\documentclass[twocolumn]{aastex631}

\usepackage[capitalise,nameinlink]{cleveref}
\usepackage{CJKutf8}
\usepackage [english]{babel}
\usepackage [autostyle, english = american]{csquotes}
\MakeOuterQuote{"}

\makeatletter
\usepackage{etoolbox}
\patchcmd\H@refstepcounter{\protected@edef}{\protected@xdef}{}{}
\makeatother

\crefname{section}{Sec.}{Secs.}

\accepted{April 5, 2024}

\shorttitle{Katachi \begin{CJK*}{UTF8}{min}(形)\end{CJK*}: The morphology-SFH connection}
\shortauthors{Alfonzo et al.}
\begin{document}


\title{Katachi \begin{CJK*}{UTF8}{min}(形):\end{CJK*}Decoding the Imprints of Past Star Formation on Present Day Morphology in Galaxies with Interpretable CNNs
\footnote{ \href{mailto:juanpabloalfonzo@astr.tohoku.ac.jp}{juanpabloalfonzo@astr.tohoku.ac.jp} \\ \href{https://github.com/juanpabloalfonzo/MaNGA-CNN}{Github Repo} \\ $\dag$ NASA Hubble Fellow.}}

\author[0009-0007-3423-1332]{Juan Pablo Alfonzo*}
\affiliation{David A. Dunlap Department of Astronomy $\&$ Astrophysics, University of Toronto, 50 St George Street, Toronto, ON, M5S 3H4, Canada}
\affiliation{Astronomical Institute, Tohoku University, 6-3, Aramaki, Aoba-ku, Sendai, Miyagi, 980-8578, Japan}

\author[0000-0001-9298-3523]{Kartheik G. Iyer $\dag$}
\affiliation{Dunlap Institute for Astronomy $\&$ Astrophysics, University of Toronto, 50 St George Street, Toronto, ON, M5S 3H4, Canada}
\affiliation{Columbia Astrophysics Laboratory, Columbia University, 550 West 120th Street, New York, NY 10027, USA}

\author[0000-0002-2651-1701]{Masayuki Akiyama}
\affiliation{Astronomical Institute, Tohoku University, 6-3, Aramaki, Aoba-ku, Sendai, Miyagi, 980-8578, Japan}


\author[0000-0003-2630-9228]{Greg L. Bryan}
\affiliation{Columbia Astrophysics Laboratory, Columbia University, 550 West 120th Street, New York, NY 10027, USA}

\author[0000-0002-9217-1696]{Suchetha Cooray}
\affiliation{National Astronomical Observatory of Japan, 2-21-1 Osawa, Mitaka, Tokyo 181-8588, Japan}

\author[0000-0003-0888-0789]{Eric Ludwig}
\affiliation{CUNY College of Staten Island, 2800 Victory Blvd, Staten Island, NY, 10314, USA}

\author[0000-0002-8530-9765]{Lamiya Mowla}
\affiliation{Dunlap Institute for Astronomy $\&$ Astrophysics, University of Toronto, 50 St George Street, Toronto, ON, M5S 3H4, Canada}
\affiliation{Whitin Observatory, Department of Physics and Astronomy, Wellesley College, 106 Central St., Wellesley, 02481, MA, USA}

\author[0000-0002-8432-6870]{Kiyoaki C. Omori}
\affiliation{Division of Particle and Astrophysical Science, Nagoya University, Furo-cho, Chikusa-ku, Nagoya 464–8602,
Japan}

\author[0000-0003-4196-0617]{Camilla Pacifici}
\affiliation{Space Telescope Science Institute, 3700 San Martin Drive, Baltimore, MD 21218, USA}

\author[0000-0003-2573-9832]{Joshua S. Speagle \begin{CJK*}{UTF8}{min}(沈佳士)\end{CJK*}}
\affiliation{Department of Statistical Sciences, University of Toronto, 9th Floor, Ontario Power Building, 700 University Avenue, Toronto, ON, M5G 1Z5, Canada}
\affiliation{Data Sciences Institute, University of Toronto, 17th Floor, Ontario Power Building, 700 University Avenue, Toronto, ON, M5G 1Z5, Canada}
\affiliation{David A. Dunlap Department of Astronomy $\&$ Astrophysics, University of Toronto, 50 St George Street, Toronto, ON, M5S 3H4, Canada}
\affiliation{Dunlap Institute for Astronomy $\&$ Astrophysics, University of Toronto, 50 St George Street, Toronto, ON, M5S 3H4, Canada}

\author[0000-0002-5077-881X]{John F. Wu}
\affiliation{Space Telescope Science Institute, 3700 San Martin Drive, Baltimore, MD 21218, USA}
\affiliation{Department of Physics $\&$ Astronomy, Johns Hopkins University, Baltimore, MD 21218,USA}



\begin{abstract}
The physical processes responsible for shaping how galaxies form and quench over time leave imprints on both the spatial (galaxy morphology) and temporal (star formation history; SFH) tracers that we use to study galaxies. While the morphology-SFR connection is well studied, the correlation with past star formation activity is not as well understood. To quantify this we present Katachi \begin{CJK*}{UTF8}{min}(形),\end{CJK*} an interpretable convolutional neural network (CNN) framework that learns the connection between the factors regulating star formation in galaxies on different spatial and temporal scales. Katachi is trained on 9904 galaxies at 0.02$<$z$<$0.1 in the SDSS-IV MaNGA DR17 sample to predict stellar mass (M$_*$; RMSE 0.22 dex), current star formation rate (SFR; RMSE 0.31 dex) and half-mass time (t$_{50}$; RMSE 0.23 dex). This information allows us to reconstruct non-parametric SFHs for each galaxy from \textit{gri} imaging alone. To quantify the morphological features informing the SFH predictions we use SHAP (SHapley Additive exPlanations). We recover the expected trends of M$_*$ governed by the growth of galaxy bulges, and SFR correlating with spiral arms and other star-forming regions. We also find the SHAP maps of D4000 are more complex than those of M$_*$ and SFR, and that morphology is correlated with t$_{50}$ even at fixed mass and SFR. Katachi serves as a scalable public framework to predict galaxy properties from large imaging surveys including Rubin, Roman, and Euclid, with large datasets of high SNR imaging across limited photometric bands.
\end{abstract}



\section{Introduction}


Morphology and star formation activity are the two primary ways in which we study galaxies to understand how they form and evolve. A rich body of literature, dating back to Hubble's original classification \citep{1936rene.book.....H}, have categorized galaxies by morphological type and found it to be correlated with present-day star formation (\citealt{1998ApJ...495..139M, 2001AJ....122.1861S, 2005ApJ...629..143B, 2009ApJS..182..216K}, also see reviews by \citealt{1994ARA&A..32..115R, 1998ARA&A..36..189K, 2014ARA&A..52..291C} and references therein). More data from observatories that peer out to higher redshifts and have better spatial resolution have further refined and augmented this relation \citep{2008MNRAS.389.1179L, 2019MNRAS.488.3929C, 2023ApJ...951..115E} and linked the morphological structure of galaxies to their colors, environments, clustering, gas content, sizes and kinematics \citep{1998ApJ...495..139M, 2005ApJ...629..143B, 2005ApJ...630....1Z, 2011MNRAS.413..813C, 2023MNRAS.521.1292P}.

This leads to two distinct approaches toward understanding galaxy evolution - a `spatial' approach studying morphology and dynamics, and a `temporal' approach studying star formation activity and other time-dependent phenomena. A key advantage of the `spatial' approach is that it can be studied with single-band data, providing a visually intuitive classification of galaxy populations and an understanding of how galaxy structure evolves over redshift. In contrast, the rise of multiwavelength surveys like GAMA, CANDELS/3D-HST, CEERS and integral field spectroscopic (IFS) surveys like SDSS-IV MaNGA, CALIFA and SAMI provide exquisite information about a galaxy's spectral energy distribution (SED) over a wide wavelength baseline. Since the spectra of galaxies contain contributions from the different stellar populations they are comprised of, careful analysis lets us infer the stellar populations of individual galaxies, and even reconstruct their star formation histories (SFHs) over cosmic time. Thus, while the spatial view leads to morphological populations of spiral, elliptical, barred, ring, compact vs extended, interacting, and irregular galaxies, the temporal one classifies galaxy populations as early- vs late-type galaxies, starburst, main-sequence, green valley, post-starburst, quiescent, rejuvenating, late bloomer, and other galaxy types based on their star-forming activity.

Previous studies that link star formation activity and morphology \citep{2015ApJ...811L..12W, 2016MNRAS.455..295H, 2018MNRAS.480.2544R, 2021MNRAS.500L..42P, 2021ApJ...923..205Y} have found that they correlate with the different phases of galaxy evolution, since the same underlying physical processes are responsible for both morphological transformation and changes in star formation activity \citep{Hubble,Morph_1,Morph_2,Morph_3,Morph_4,Morph_5,Morph_6} \citep{Morph_time1,Morph_time2,Morph_time3,Morph_time4,Morph_time5, Morph_time6,Morph_time7}. Both the spatial and temporal pictures offer unique insights into how galaxies evolve over time, with both morphology and star formation regulated by gas inflows and mergers on long spatial and timescales, and dynamical processes, feedback, etc on shorter scales \citep{gal_mech1,gal_mech2,gal_mech3,gal_mech4,gal_mech5,gal_mech6, gal_mech7, gal_mech8,gal_mech9,gal_mech10,gal_mech11}.

It is well established that the present-day star-forming rate (SFR) of a galaxy can give us insights into the internal physical phenomenons occurring in a galaxy. It is also well known that the morphology of galaxies is directly linked to their location on the SFR-M$_*$ plane. Furthermore, a galaxy's SFH is known to be connected to the tight relation that exists between its stellar mass (M$_*$) and SFR (from hereon referred to as the star-forming main sequence (SFMS) or the SFR-M$_*$ correlation). \citep{Morph-SFH-6, Morph-SFH-1, Morph-SFH-2, Morph-SFH-3,Morph-SFH-4,Morph-SFH-5}.
In addition, \cite{Morph-SFH-7} and \cite{galaxy-manifold} demonstrated that there is a strong dependence between SFR, M$_*$ and galaxy structure in the local universe. Both of their conclusions support the idea that the SFMS correlates very strongly with galaxy morphology. While we know that the recent SFR and morphology of galaxies are correlated, the extent to which these morphological imprints extend backwards in time is unclear and difficult to quantify.

Part of the reason for this is that much of the current literature uses summary statistics to quantify morphology such as sersic index, concentration, Gini, and M20 \citep{Morph_6, gini,concentration}. These, however, miss more fine details such as bars, spiral arms, star forming clumps, and the complex structures in starbursting and interacting galaxies in an attempt to summarize the entire morphological structure in one number. This fails to account for the unique morphological structures of galaxies which are often direct results of their evolutionary process. There have been many approaches to dealing with galaxy morphology via machine learning and more recently using deep learning \citep[see review by][]{Deep_Learning_Galaxy_Review}. However, other works in the literature have tried to address this issue by using deep learning to categorize galaxies into discrete classes or categories \citep{Morph-class-1,Morph-class-2,Morph-class-3,Morph-class-4,Morph-class5}. 

In this work, we aim to move away from the thinking that galaxy morphology is a discretizable parameter, and rather turn to interpretable CNNs to get a full encapsulation of galaxy morphology. Instead of dividing the galaxy population into discrete classes, we train a CNN to predict continuous galaxy properties (i.e. the star formation history) from 3-band images and then use explainable AI (XAI) methods to determine the relevant features of the image that are responsible for the network's prediction.

The goal of this work is to characterize galaxy morphology as a function of its physical properties, rather than the typical classification. Similar work has been done by \cite{Wu2020} to predict galaxy spectra from their images, and using RGB images to predict galaxy parameters in \cite{Wu_2020_2} and \cite{Wu_2019}. However, these approaches do not introduce the ability to pick out morphological features of importance like the method presented in this paper.To make a more detailed assessment of the morphology-SFH connection we introduce Katachi \begin{CJK*}{UTF8}{min}(形)\end{CJK*}, a non-parametric approach which uses interpretable CNNs to predict: M$_*$ and SFRs, and the 4000 Å break strength (hereafter D4000 or D4000 break strength) of galaxies using images that contain their full morphological information as an input.

Rather than training a `black box' CNN to make predictions, the main goal of the paper is to create a robust CNN interpretation framework. In doing this, we require that the network is learning something about the underlying physical structure of galaxies and not simply memorizing patterns in the images.
Many methods have been used to interpret the results of CNN predictions that range from creating heat maps on the input images to see which parts of the image the network focuses on to make predictions (known as saliency maps), to having the network generate a textual explanation to justify its predictions \citep[see references in][]{XAI-texbook}. Saliency maps have grown immensely in popularity because they provide a simple way to visually interpret what image features are the most important for the network for making its prediction for any given image. This lends itself nicely to galaxy morphology problems as we are particularly interested in seeing if certain groups of galaxies that we know have similar physical characteristics also share morphological properties. A saliency map would make such patterns easy to spot by for example highlighting the use of long pronounced spiral arms to predict a galaxy as being highly star-forming. For this reason, we see in the literature that there is already widespread use of saliency maps, the XAI tool of choice, for picking apart networks that focus on predicting galaxy morphology \citep{Bhambra,Peruzzi}. 

We aim to use Katachi \begin{CJK*}{UTF8}{min}(形)\end{CJK*}to form a better understanding of how the morphology of a galaxy is ultimately tied to its formation/evolution than what is currently available in the literature. This is the first study to directly look at the imprints of past star formation on current galaxy morphology in a way that quantifies the effect of the underlying physical processes that affect both of these observables. The main conclusion of this paper is to establish a correlation between stellar populations of different ages and the morphological features of galaxies. With this conclusion, we get a clearer picture as to how certain morphological features relate to the physical phenomena that occur in the formation histories and evolutionary tracks of given galaxies.

The goal of this paper is to develop a framework that relates the spatial and temporal aspects of a galaxy without the need for parametric assumptions about morphology or star formation history. Since the same physical processes are responsible for shaping galaxies and regulating their star formation, this will enable us to study both on a common footing and let us answer the question - What are the imprints of past star formation on a galaxy's present-day morphology? Beyond this, the framework also predicts galaxy SFHs using only 3-band \textit{gri} images at a level comparable to current SED fitting codes \citep{2024arXiv240112300W}, a fact that will be extremely useful to future surveys with Rubin, Roman, and Euclid that will cover a wide area with a limited set of filters. 
In \cref{sec:data} we lay out the data used to train the networks. \cref{sec:methods} describes all the methods used in this work, most critically it outlines the Katachi framework. In \cref{sec:results} we lay out the main results found between galaxy morphology and SFH. In \cref{sec:discussion} we explore the results further and provide context and meaning for them by considering their significance to the galaxy evolution paradigm. In \cref{sec:conclusion} we summarize our results and state our conclusions. Throughout this paper, we adopt a standard $\Lambda$ cold dark matter($\Lambda$CDM) cosmology with $\Omega_m = 0.3, \Omega_{\Lambda} = 0.7,$ and $H_0 =70$ km Mpc$^{-1}$ s$^{-1}$ .

\section{Data} \label{sec:data}

\begin{figure*}
    \centering
    \includegraphics[width=\textwidth,trim={2cm 0.8cm 0 0},clip]{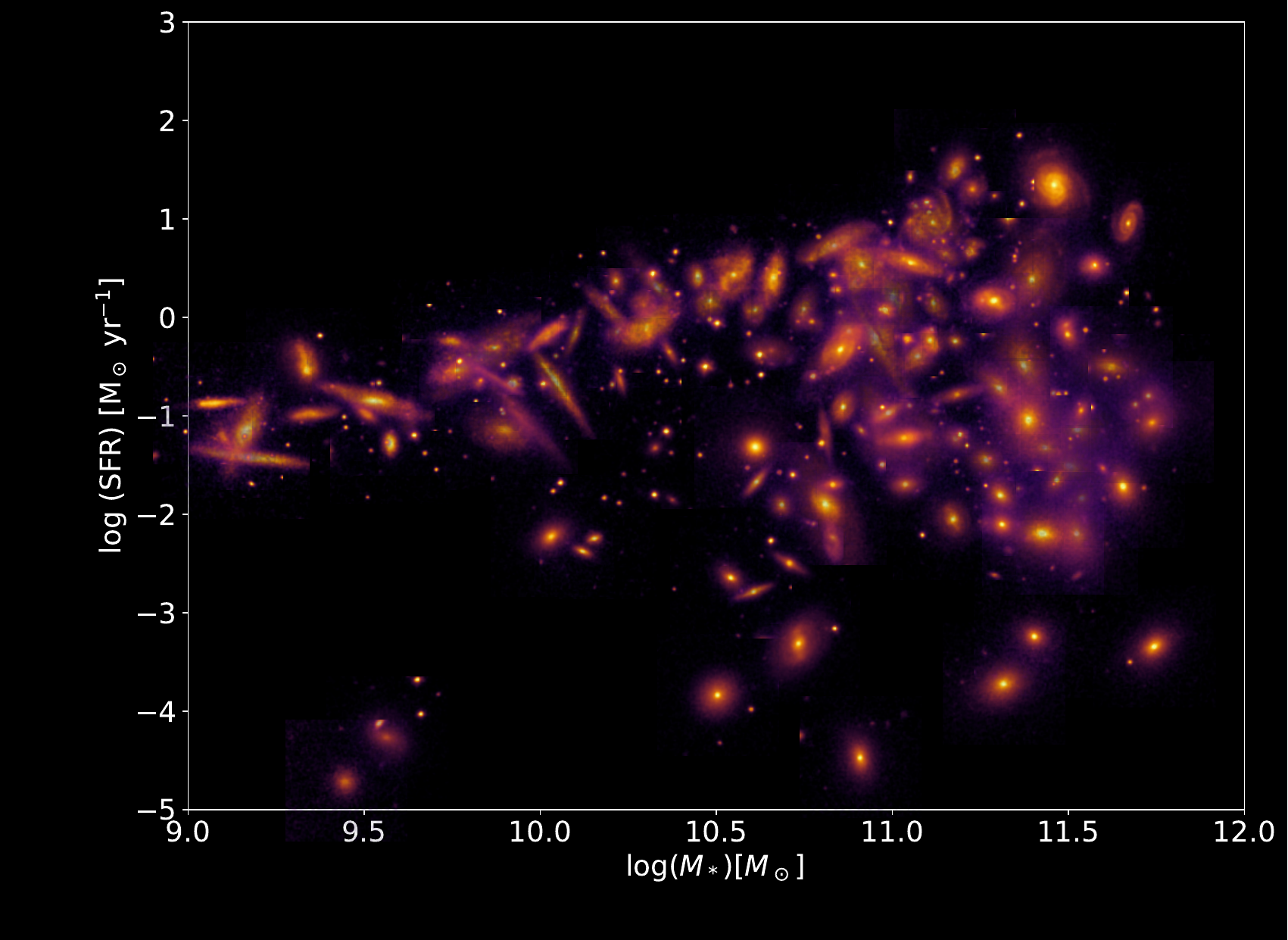}
    \caption{A subset (N=111) of the images (N = 9904) used from the SDSS SDSS-IV MaNGA catalogue (N=10,081) with stellar mass and SFR values taken from the PIPE3D VAC. The  M$_*$-SFR main sequence (SFMS) is visible here as well as the morphological trend for galaxies on and off this main sequence. On the MS we see more disk/spiral galaxies and off the SFMS we see more elliptical galaxies. This a well-known and expected trend.}
    \label{fig:main sequence}
\end{figure*}

The SDSS-IV MaNGA (Mapping Nearby Galaxies at APO (Apache Point Observatory), \citet{MaNGA}) survey gives spectrographic data on low red-shift galaxies. Unlike previous Sloan Digital Sky Survey (SDSS) surveys which obtained spectra only at the centers of target galaxies, SDSS-IV MaNGA enables spectral measurements across the face of each of $\sim$10,000 nearby galaxies thanks to 17 simultaneous "integral field units" (IFUs), each composed of tightly-packed arrays of optical fibers. The data release 17 (DR17) of SDSS-IV MaNGA was used for this study \citep{DR17}. Of all the galaxies in DR17 (N=10,081) we used only the ones which satisfied the conditions $10^8 M_\odot<M_*<10^{12} M_\odot$ and $10^{-4}<$ SFR $ (\frac{M_\odot}{yr})<10^{2}$ (N=9904) to avoid any galaxies that had unphysical parameters due to bad spectral fitting. We used the PIPE3D \citep{PIPE3D} data reduction pipeline's value-added catalog (VAC)\footnote{\href{https://www.sdss4.org/dr17/data_access/value-added-catalogs/?vac_id=manga-pipe3d-value-added-catalog:-spatially-resolved-and-integrated-properties-of-galaxies-for-dr17}{PIPE3D Data Model}} to acquire the resolved galaxy properties to make this data split. After we had this selected sample we pulled 256 by 256 pixel image cutouts of the SDSS \textit{gri} broadband images of these galaxies. These images were converted to RGB using the pipeline described in \cite{RGB_Images} which preserves relative flux. These RGB images were acquired from SDSS SkyServer through the Marvin API \citep{marvin}. A subset of the selected sample is shown in \cref{fig:main sequence}. In this subset we can clearly see there is a morphological trend for galaxies on the  M$_*$-SFR main sequence and those below it. We see more disk/spiral-type galaxies along the main sequence and more elliptical-type galaxies below it. This is as we expected based on the literature that exists documenting the connection between the morphology of galaxies and their star formation rates. \citep{2015ApJ...811L..12W, 2016MNRAS.455..295H, 2018MNRAS.480.2544R, 2021MNRAS.500L..42P, 2021ApJ...923..205Y} 

From the PIPE3D VAC, we pulled the M$_*$, the SFR, and the D4000 break strength for each of the galaxies. These are derived on the spaxel level and summed up to give the value for the entire galaxy. These values were set as the truth values for the training of the CNNs described in \cref{sec:training}. Additionally, the minimum and maximum values of t$_{50}$, metallicity, redshift, and dust attenuation from the PIPE3D VAC are used to generate mock spectra (discussed in \cref{sec:t50 model}). We use the Galaxy Zoo VAC\footnote{\href{https://www.sdss4.org/dr17/data_access/value-added-catalogs/?vac_id=galaxy-zoo-classifications-for-manga-galaxies}{Galaxy Zoo Data Model}} \citep{Galaxy_Zoo,Galaxy_Zoo_2} that contains various morphological feature flags in the discussion section of this paper for galaxies in our sample where these values were available. We use the PyMorph VAC\footnote{\href{https://www.sdss4.org/dr17/data_access/value-added-catalogs/?vac_id=manga-pymorph-dr17-photometric-catalog}{PyMorph Data Model}} \citep{PyMorph} to the same end as we used the Galaxy Zoo VAC as it provides additional morphological features. These include things like the presence of spiral arms, bulges, and merger events.  

\section{Methods} \label{sec:methods}
This section will outline all the methods used in this paper, which together make up the Katachi \begin{CJK*}{UTF8}{min}(形)\end{CJK*} framework. In \cref{sec:arch} we briefly introduce and explain what CNNs are, then continue to explain the specific architecture used in this paper. In \cref{sec:training} we describe how the CNNs were trained using the data outlined in \cref{sec:data}. In \cref{sec:transfer learning} we introduce and explain the concept of transfer learning as well as motivate the reason we used it in this work. In \cref{sec:data pertubation} we discuss the data perturbations that were done when training the CNNs to ensure the statistical robustness of the method. In \cref{sec:t50 model} we describe how we acquired t$_{50}$ values for the galaxies in our sample given their D4000 through the use of a simple linear network.  In \cref{sec:dense basis} we describe the dense basis method for generating SFHs for all the galaxies in our sample. In \cref{sec:shap map methods} we introduce the concept of SHAP maps and explain how they were used in this work to study galaxy morphology. In \cref{sec:radial profiles method} we describe the creation of radial dependence plots for the 3 sets of SHAP maps that are generated for each galaxy in our sample.

\subsection{Predicting M$_*$,SFR, and D4000 From Images}

\subsubsection{CNN Architecture}\label{sec:arch}

We use convolutional neural networks (CNNs) on RGB images of galaxies to gain insight into their morphological structure. This is because CNNs are well-documented to be good tools in computer vision tasks. They have recently had unprecedented breakthroughs in image classification, object detection, semantic segmentation, and many more. CNNs are a category of deep learning models specifically engineered for the analysis of grid-like data, particularly images and videos. CNNs operate through a series of layers, with convolutional layers being central to their functioning. In these layers, learnable filters slide over the input data, calculating dot products at each location. This process produces feature maps that highlight patterns like edges, corners, and textures. Multiple filters are used in each convolutional layer to capture diverse features. Subsequently, pooling layers like max-pooling or average-pooling are employed to reduce spatial dimensions and retain critical information. Activation functions, such as the Rectified Linear Unit (ReLU), introduce non-linearity to capture complex relationships. After several convolutional and pooling layers, CNNs often include fully connected layers for high-level representation learning and predictions. Training CNNs involves backpropagation and gradient descent, where the model adapts its parameters to minimize a predefined loss function. Overall, CNNs excel in computer vision tasks by automatically extracting hierarchical features from input data, making them a cornerstone of image analysis in deep learning. 

In the realm of astronomy, CNNs have also been effective in dealing with photometry and spectrographic data. In particular, there is already some work in the literature that suggests that CNNs are very effective at recognizing galaxy morphology \citep{Wu2020,cheng,Bhambra,nn_galmorph,deep_galaxy_zoo}. In this paper, we adopt a unique approach that strings 3 CNNs that each use the well-known ResNet50 architecture. ResNet-50, short for "Residual Network with 50 layers," is a significant deep convolutional neural network (CNN) architecture introduced in 2015. It was designed to tackle the challenge of training very deep neural networks effectively by addressing the vanishing gradient problem. The architecture begins with an initial convolutional layer followed by max-pooling, reducing the spatial dimensions of the input image. The core innovation lies in the use of residual blocks, each containing multiple convolutional layers with skip connections, allowing gradients to flow more easily during training. ResNet-50 consists of 16 residual blocks organized into different stages, with varying numbers of blocks in each stage. After the last residual block, global average pooling is applied to convert feature maps into a vector. A modified fully connected layer with a single neuron is employed to generate a continuous numerical output, directly addressing the regression task. During training, the network aims to minimize the Mean Squared Error (MSE) loss, adjusting its parameters through backpropagation and gradient descent. \cite{ResNet50} introduced residual layers in neural networks (ResNets), which advanced optimization and influenced the design of deep CNNs. Its innovative use of residual blocks and skip connections remains a fundamental concept in deep neural network design \citep{ResNet50_2}. We call our new technique of stringing 3 ResNet50 architecture networks together a "chain architecture". This section will describe the design of this chain architecture as well as the motivation behind its design. For further discussion as to why ResNet50 was chosen as the architecture for each of the networks in the chain please refer to \href{sec:transfer learning}{Sec. 3.3}. 

\begin{figure*}
    \centering
    \includegraphics[scale=0.5,trim= 0 11.5cm 0 0,clip,width=\textwidth]{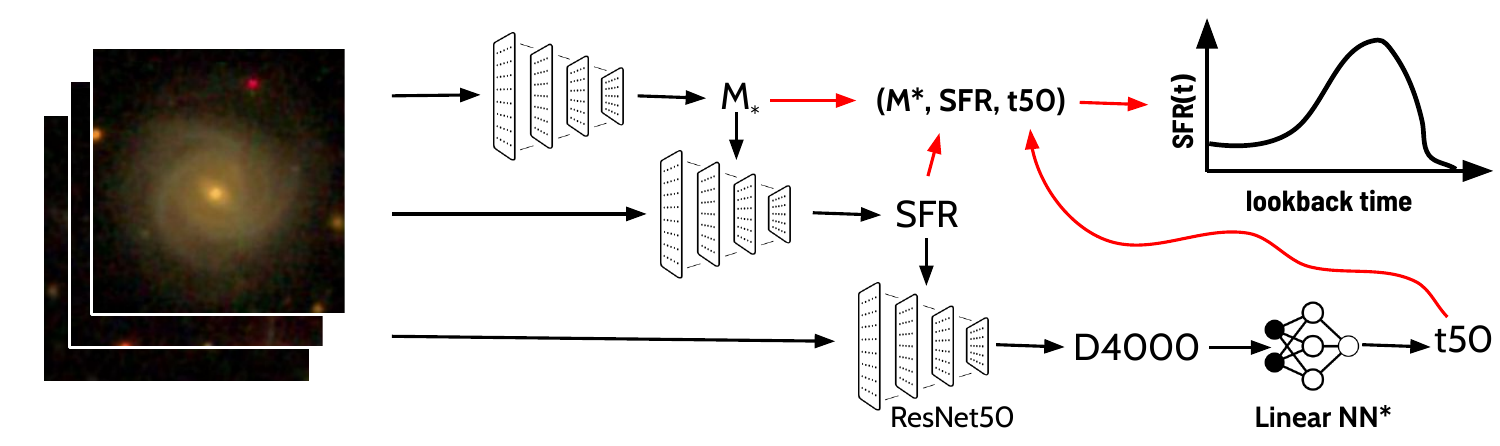}
    \caption{Chain architecture overview, where the output of the previous network in the chain gets appended to the original input and then fed to the next network down the chain. The final output of the whole chain is then a value for M$_*$, SFR, and D4000 break strength of a galaxy just from its \textit{gri} image.
    \\$^*$While not part of the chain architecture this linear neural network is used to convert D4000 values to t$_{50}$ values which allows for the reconstruction of SFHs via the use of the dense basis method.}
    \label{fig:chain arch}
\end{figure*}

The chain architecture is laid out in such a way that each network down the chain takes the output of the previous network as part of its input. This means it takes the output of the previous network, appends it to the input of that previous network, and the appended output in addition to the original input becomes the input for the next network. For the purpose of this paper, the chain network is three networks long. Each of the three networks is responsible for predicting one of the three parameters (M$_*$, SFR, and D4000 break strength) we need to reconstruct the SFH of the galaxy. In the setup used for this paper, the first network in the chain predicts the stellar mass of the galaxy using an image of the galaxy with the specifications outlined in \href{sec:data}{Sec. 2.1}. The predicted stellar mass is then appended to the image and this becomes the new input for the next network down the chain which predicts the SFR of the galaxy given its image and the predicted mass. Finally, we repeat this process and append the predicted stellar mass and predicted stellar SFR to the image and this becomes the input for the final network which predicts the strength of the D4000 break for the galaxy spectra. A visual representation of the chain network is shown in \cref{fig:chain arch}. 

The motivation behind this architecture is to allow networks further down the chain to have more data to learn from than those at earlier parts of the chain. The idea behind this came about as we noticed that just from the image a ResNet50 architecture could predict stellar mass much better than SFR and the D4000 break strength (for more details and to see the improvement in prediction performance refer to \cref{sec:solo vs chain}). The biggest improvement is seen in SFR prediction once a predicted stellar mass is added to the image, this serves to confirm the well-established M$_*$-SFR relation (\cite{Mass-SFR-Speagle}, \cite{Mass-SFR-Pan}, \cite{Mass-SFR-Chang}). For further discussion regarding this refer to \cref{sec:discussion galaxy}. 

\subsubsection{Training of Networks}\label{sec:training}

In this study, we use a 90/10 train/test split of our sample. The networks are trained on 90$\%$ of the data, meaning the network knows the true value of the targets (M$_*$, SFR, and D4000 break strength) and minimizes a loss function to improve its prediction accuracy. The loss function measures how accurate the predictions of the network are. In the case of this paper since we had three different networks each has its own loss function to minimize as the training loop is happening. This means that each network is trained independently of each other. This is to say that each one minimizes its own loss function without knowing the loss function value of any of the other networks. This leads to each network minimizing its own error in prediction and not being affected by errors the other networks in the chain might be making. The only way error can be propagated down the chain is if the predictions of a network up the chain are incorrect, but this is precisely what the networks are independently set up to minimize by having their own loss functions. This is why we preferred to take this approach rather than have the chain minimize one final loss function that took into account all the errors in the three predicted parameters. We used the mean square error (MSE) as our loss function for training all three networks and used the Adam optimizer \citep{Adam-Optimizer} to minimize these loss functions. We used a learning rate of $10^{-3}$ for all three networks and trained with batch sizes of 32. We used the PyTorch function \texttt{optim.lr$\_$scheduler.ReduceLROnPlateau} to reduce the learning rate automatically when the loss value had not decreased in more than 3 epochs. Additionally, if the training loss had not been reduced by more than $10^{-3}$ in 10 epochs the training was stopped as it implied the model was no longer learning, and continuing the training could lead to overfitting of the data. All training and evaluation of the models was done using PyTorch \citep{Pytorch}. 

The networks were then tested after being trained to assess for over-fitting of the data. This is done by providing a trained network with images, not in its training set (i.e images in the test set) and assessing how accurately it can predict the targets for those images. If we get similar amounts of accuracy from the model with both the training and test data sets then we can be confident that the network is not over-fitting. Similar to the training of the networks, the testing was done independently for each of the three networks in the chain. For further details on how the robustness of the network was verified refer to \cref{sec:data pertubation}.

\subsubsection{Transfer Learning Approaches} \label{sec:transfer learning}

In machine learning, there exists a process known as "transfer learning" which can be beneficial to solving complex machine learning tasks, while also reducing computational loads and times. The idea of transfer learning is to leverage the "knowledge" of previously trained models and apply it to solve the problem you are currently interested in. In cases involving CNNs specifically, it is common to use deep architecture models that have been trained on massive data sets as a starting point for your own training. This allows the network to learn aspects specific to your data set with much higher performance and with a relatively small amount of data. This is because the pre-trained network knows how to classify objects, for example, a network that can tell what animal is contained in an image. To be able to classify objects the network learns general features of any objects such as edges and shapes in images. These general features will be useful in any image classification problem, even if the network has never seen images like the ones you intend to train it on. The idea with transfer learning then with CNNs is to leverage this knowledge the network has about general features in images and make it learn just specific features unique to the data set it's being trained on. In the case of this paper for example the use of a pre-trained network might mean the network can quickly pick out where the galaxy is in the image and the final layers that are trained can focus on the fine morphological details of the galaxy. 

This is precisely the approach we took in this paper, utilizing a very popular pre-trained model known as ResNet50, which has been trained on more than one million images from the ImageNet database \citep{ResNet50,ResNet50_2,imagenet}. We used the pre-trained ResNet50 model from the PyTorch library. This model was pre-trained to recognize many commonplace objects and the goal of the paper was to leverage its knowledge of general object features to create a model that was effective at recognizing galaxy morphology. We used this transfer learning approach on all three of the networks that make up the chain architecture as outlined in \cref{sec:arch}. All the layers of this pre-trained model remained frozen, and we trained the final convolutional and linear layers of the model using the process that was described in \cref{sec:training}. This allowed for the networks to achieve a higher accuracy in much less training time when compared to not using the pre-trained weights.

ResNet50 is not the only architecture that is used to create pre-trained transfer learning models. There is an abundance of popular model architectures used for transfer learning tasks, a few of which were also explored during this study. The results of using the other pre-trained architectures made it clear that in this particular case, ResNet50 gave the most accurate results of all the pre-trained models that were tried. Additionally, previous literature has shown that ResNet50 is very capable in galaxy morphology problems \citep{ResNet50_Morph}. As such, it was the one that was selected to carry on the bulk of the analysis of this paper. However, it is worth noting that with different galaxy image data sets (i.e not SDSS) some groups have had success with transfer learning to predict galaxy morphology using different model architectures (\cite{VGG16_Morph}, \cite{Xception_Morph}, \cite{PixelCNN_Morph}). 

We additionally explored the use of Zoobot \citep{Zoobot} which is based on the EfficentNet \citep{EfficientNet} architecture and was trained on over 314,000 galaxies from SDSS DR8 and using Galaxy Zoo DECaLS \citep{decals} collected labels. This network was specifically trained on galaxy images and so the hope was to leverage the trained weights of this architecture and fine-tune the output layer to match our specific morphology problem. Since Zoobot was trained on galaxy images not general ImageNet images the idea is that its weights would be better equipped to solve galaxy morphology problems. This however did not turn out to be the case, and we saw comparable performance from the networks when using Zoobot compared to the ImageNet weights while maintaining the architecture (EfficentNet) constant. Zoobot did not significantly outperform the other transfer learning approach, which was unexpected. It is worth keeping in mind that this comparison and analysis was done in January of 2023 and Zoobot has been significantly updated and improved since then. We aim to repeat this comparison with the updated Zoobot model in future works. 

It is worth noting however that the performance across all architectures was only marginally different. This is to say that all the architectures we perform roughly are in the same ballpark accuracy-wise. This is expected as the wider literature has found the same to be generally true \citep{Deep_Learning_Galaxy_Review,arch-comp-1,arch-comp-2}. The commonly accepted idea in the community is that this is the case since astronomical images generally present less diversity than natural images, where different CNN architectures have had varying success. The lack of diversity means that simpler/less advanced CNN architectures generally fare as well as more intricate or advanced architectures. 

\subsubsection{Data Augmentation} \label{sec:data pertubation}

The first set of data augmentation was done on the images that comprise the training and test sets. When these images are packaged into data loaders they are randomly perturbed by having vertical/horizontal flips applied to them, as well as a random rotation by some degree about its origin. This is done to ensure the network looks for patterns in the structures of each galaxy and does not develop a dependence on having to see galaxies from a certain angle/perspective to be able to identify the physical structure of the galaxy. This also aids the networks in learning approximate symmetries \citep{nn_galmorph}. For a discussion on the effects of age-metalicity degeneracy from dust attenuation we refer the reader to \hyperref[app:age morph]{Appendix F}.

\subsection{From M$_*$, SFR and D4000 to SFHs}

\subsubsection{D4000 to t$_{50}$ Model}\label{sec:t50 model}

Given that we are interested in studying the SFH of the galaxies we needed a quantify the ages of the galaxies to derive their SFH. The D4000 break is well documented to be directly tied to the age of a galaxy as it is an indicator of how metal-rich the galaxy is. Usually, galaxies that are more metal-rich are older as they are populated with older stars that are more metal-rich. \citep{D4000_Age,D4000_Age2,D4000_Age3,D4000_Age4} However, to reconstruct the SFH of the galaxies using a SED fitting algorithm we needed a more solid definition of the galaxy's age. For this, we trained to try to acquire the t$_{50}$ of a galaxy, which is the time at which the galaxy formed half of its total stellar mass. Since we expect the D4000 break strength and the t$_{50}$ of a galaxy to be correlated we created a model that could predict the t$_{50}$ of a galaxy given its M$_*$, SFR, and D4000 break strength. The model is a simple 5-layer linear neural network with a ReLU activation function used in between each layer. It was trained using a train/test split of 80/20 with batch sizes of 32. The Adam optimizer was used to train the model parameters for 400 epochs. 

The data used to train this model was simulated using the dense$\_$basis SED fitting algorithm \citep{dense_basis}. More details about the specifics of this SED fitting algorithm are shown in \cref{sec:dense basis}. The dense$\_$basis algorithm allows us to simulate galaxy spectra given the following parameters: t$_{50}$, M$_*$, SFR, dust attenuation, metallicity, and redshift. Using PIPE3D we found the max and min values of all these parameters for galaxies present in the SDSS-IV MaNGA survey. We then generated 10,000 values for each of these parameters following a uniform distribution given the min and max values for each parameter in the SDSS-IV MaNGA dataset. Each set of generated values would correspond to an SDSS-IV MaNGA-like galaxy. We then generated what these spectra would look like and measured the strength of the D4000 break in each of these spectra. This simulated data set was then used to train the network described above.

After we trained this network we used the predicted values of stellar mass, SFR, and D4000 break strength from our chain model to generate the t$_{50}$ values of the galaxies in our sample. Combining this network with the chain model previously described we now have a final product that could take in the galaxy images and give us the galaxy's: M$_*$, SFR, D4000 break strength, t$_{50}$, and SFH.

\subsubsection{Generating SFH Using dense$\_$basis} \label{sec:dense basis}

The Dense Basis method (\cite{dense_basis_2},\cite{dense_basis}) provides a non-parametric description of a galaxy's star formation history using its stellar mass, SFR, and a set of $t_X$ values that specify when a galaxy formed X\% of its total mass and provide constraints on the shape of the SFH. These values are used by a Gaussian process framework to create flexible SFHs that satisfy the integral constraints imposed by the $t_X$ values. While the method is generally used for spectral energy distribution (SED) modeling and fitting, it can also be used to reconstruct the SFHs of galaxies given their (M$_*$, SFR, $t_X$) tuples \citep{AGN_Feedback}, as we have from the PIPE3D VAC or from our neural network predictions. For this paper, we state $t_X$ values in cosmic time, unless otherwise stated.

\subsection{Interpreting the Learned Behavior}

\subsubsection{SHAP Maps} \label{sec:shap map methods}

A big focus of this work was ensuring the networks trained were interpretable. It is often the case in deep learning that the effectiveness of the model is put first, and understanding why the model predicts what it predicts is often left as an afterthought. However, in conducting scientific research we know that this is not good enough. If deep learning is to play a significant role in the field we must understand what it is fundamentally doing if we hope to extract new scientific findings from them. In other words, it's not a matter of just ensuring the model is making correct predictions, but rather \textit{how} it is making these predictions that matter most. We focused on the interpretability of our trained network using saliency maps based on SHapley Additive exPlanations (SHAP) values \citep{SHAP}. 

SHAP values are a method to explain the output of any machine learning model. They are based on the concept of Shapley values from cooperative game theory and aim to fairly distribute the contribution of each feature to the model's output for a specific prediction. The SHAP values of a prediction are calculated as a linear combination of the Shapley values for all possible coalitions of features. The contribution of each feature can be positive or negative, indicating whether it increases or decreases the prediction compared to the baseline. The baseline can be the expected value of the prediction or the average prediction over the training dataset. The sum of all SHAP values is equal to the difference between the prediction and the baseline. In all the following plots, a red (positive) value for the SHAP indicates that increased flux in the image at that location will increase the estimate of the associated physical property (increased mass, SFR or younger ages), and a blue (negative) value indicates that increased flux will result in a decreased physical estimate (reduced mass, SFR or older ages). 

The computed SHAP values for the predictions are used to generate SHAP maps, which are a type of saliency map or "heat map". To create a SHAP map, we first define a baseline, which serves as a reference point for the SHAP value calculation. The baseline can be the average prediction of the model over the training dataset or any other meaningful value. Next, we compute the SHAP values for each feature in the model for each prediction. This process involves using a SHAP explainer, in our case GradientExplainer, to determine the contribution of each feature to the model's predictions. Once the SHAP values are calculated for all predictions, we aggregate them across the dataset to gain an overview of each feature's impact on the model's predictions. This aggregation can be done by taking the mean or sum of SHAP values for each feature. The resulting SHAP map is a visualization presented in a heat map format. The color of each cell in the heat map indicates the magnitude and direction (positive or negative) of the SHAP value for that feature and prediction. These saliency maps allow us to see what parts of galaxy images the network is focusing on to make its predictions. In our case, a positive SHAP value (red on the heat map) means that increasing the flux in these pixels makes the network overpredict the target value. Likewise, a negative SHAP value (blue on the heat map) means increasing the flux in these pixels makes the network underpredict the target value. With this, we explored which morphological features of galaxies have the greatest impact on predicted star formation history parameters. This allows us to gain insights into the links between the underlying physical processes regulating star formation in galaxies and imprinted morphological features on the galaxy. This means that for any given galaxy in the sample we generated 3 distinct SHAP maps. Each map corresponds to one of the three networks in the chain, and therefore to one of the three predicted parameters (M$_*$, SFR, D4000 break strength). The results of this analysis are shown in \cref{sec:shap maps} and the discussion of these results in \cref{sec:discussion galaxy}. 

In addition to SHAP maps, we also used other saliency map methods such as Grad-CAM and Eigen-CAM. These methods are designed to produce saliency maps, but instead of perturbing the input as SHAP does, these methods instead focus on seeing how the gradients of the last convolutional network of the model change as the input is passed through it. \citep{GradCAM} The main difference between Grad-CAM and Eigen-CAM is that Eigen-CAM takes the first principle component of the 2D activations from the final convolutional layer to create a CAM, while Grad-CAM uses the gradients of the selected target in the final convolutional layer \citep{Eigen-CAM}. 

However, we found that gradient-based saliency maps were inconsistent with our data and network and would often produce maps that were difficult to interpret. For this reason, we decided not to continue further with these and used the SHAP saliency maps which made more physical sense and were more consistent.

\subsubsection{Generating Radial Profiles of SHAP Maps} \label{sec:radial profiles method}
Once we generated the SHAP maps described in the previous section we created radial profiles based on these maps to investigate the areas of focus of the network. These areas of focus serve as a proxy of galaxy morphology and allow us to seek trends in the population of galaxies present. This section describes the creation of these radial profiles.

The radial profiles are designed to give us the average SHAP value for a given galaxies at some radius $r$. The radius is measured from the galaxy's center and is measured in kpc. To achieve this the physical size of the galaxies must be calculated from their images. This is quite straightforward to do using SDSS-IV MaNGA as we know the relation between spaxel size, angular size, and physical size. This then becomes the x-axis of the radial plots. The y-axis is then the average value of the SHAP map at this given radius from the center. To calculate this we simply move out from the center of the SHAP map in discrete radial steps and average all the SHAP map values that fall into each radial bin.

Having these radial plots we also define a component of them to summarize the importance of the morphological profiles of the galaxies for each of the predicted parameters. We define a "gradient" which is the slope of a line that runs from the value the radial plot has at $r=0$ kpc to $r=20$ kpc. Since we find this gradient for all galaxies and therefore all would have the same $\Delta r$, we take the $\Delta y$ (the difference in the average SHAP values at $r=20$ kpc and $r=0$ kpc) to be the "gradient". We calculate these gradients for all 3 sets of SHAP maps that we have for each galaxy in the sample (each corresponding to one of the predicted parameters). These gradients become a helpful summary statistic to describe the morphological dependence of the galaxy on each of the predicted parameters as is shown in \cref{sec:radial plots results} and is discussed further in \cref{sec:discussion galaxy}. 

We chose 20 kpc as the point to take the slope as for galaxies in the local universe we expect spiral arms of galaxies to be in the 15-25 kpc range \citep{spiral_arm_distance}. Therefore, this slope encompasses the area of most morphological diversity of the galaxies in the sample. It is also worth mentioning that we tried to take this same gradient from the center of the galaxy to its effective radius. We found that the gradient distribution was roughly the same as choosing 20 kpc as our outer point. We believe this is because the galaxies span a relatively narrow range of redshifts and stellar masses, so the variation in effective sizes is not large. This does not mean that re-scaling to effective radius does not improve the information content, just that the gain in doing so does not significantly affect the qualitative interpretation of the radial trends analyzed in this paper. Therefore, we decided to use 20 kpc.

\subsection{A Recap of the Katachi Framework}
In this subsection, we will briefly recap the broad overview of the Katachi method, and its uses in this work, as the fine details have all been laid out. (1) Train a chain CNN on galaxy images (the number of bands is flexible, in this work we only use \textit{gri} bands). (2) Use SHAP to understand feature importance for predicting parameters of interest (in this work M$_*$, SFR, and D4000). (3) Generate SHAP maps for each galaxy and for each parameter of interest. (4) Measure and fit SHAP radial gradients. (5) Compare best-fit SHAP radial gradients to position in M$_*$-SFR plane. (6) Fit star formation histories, and repeat (5) except using \textit{past} positions in M$_*$-SFR plane. (7) Variations of (5) and (6) using the different parameters (M$_*$, SFR, D4000) to see their connection to galaxy morphology.

\section{Results} \label{sec:results}
The main aim of this paper is to introduce the Katachi framework as a new and unique way to look at galaxy morphology. The results of the methods outlined in \cref{sec:methods} are outlined below. In \cref{sec:solo vs chain} we evaluate the performance of the CNNs to predict M$_*$, SFR, and D4000. In \cref{sec:val test} we discuss the results of our validation test. In \cref{sec:t50 performance} we evaluate the performance of the simple linear network used to predict the t$_{50}$ values of the galaxies in the sample. In \cref{sec:SED comp} we compare the predictions of our chain CNNs to SED fitting. In \cref{sec:shap maps} we use SHAP to understand feature importance for predicting M$_*$, SFR, and D4000. In \cref{sec:radial plots results} we show the radial plots of the SHAP maps, and define a summary statistic for them which we call "gradients". In \cref{sec:color} we examine the effect of the galaxy colors on the SHAP maps. In \cref{sec:mass divide} we examine the divide in gradients around M$_*$ and D4000. In \cref{sec:age similar} we compare the SHAP maps for D4000 to the SHAP maps for M$_*$ and SFR. In \cref{sec:physical significance of gradients} we examine the physical significance of the gradients. In \cref{sec: SFH and morphology} we use the reconstructed SFHs of the galaxies in our sample to construct a link between the galaxy's present morphology and its SFH.

\subsection{How Well Can We Predict SFHs from Images?}

\subsubsection{Prediction Accuracy of CNNs} \label{sec:solo vs chain}
\begin{figure}
    \centering
    \includegraphics[width=0.49\textwidth, trim={3cm 0 5cm 0},clip]{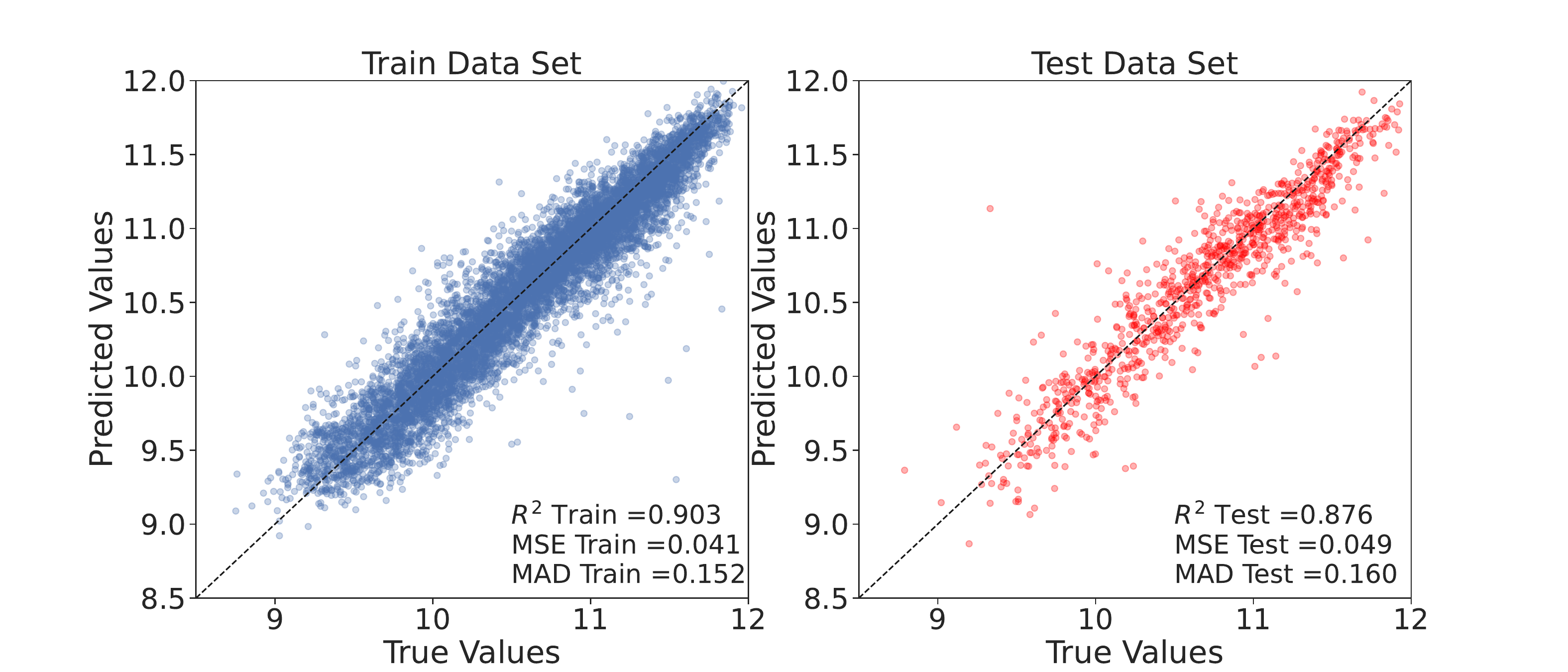}
    \includegraphics[width=0.49\textwidth, trim={3cm 0 5cm 0},clip]{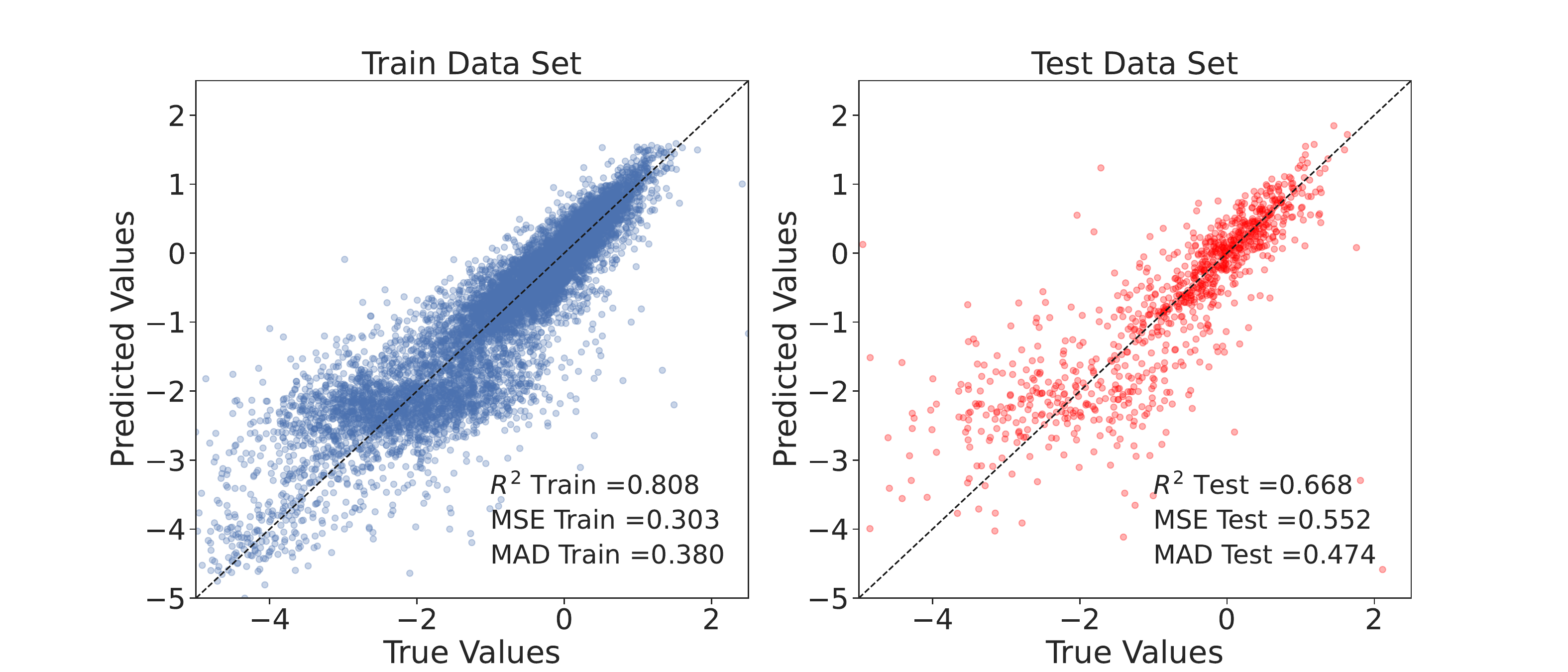}
    \includegraphics[width=0.49\textwidth, trim={3cm 0 5cm 0},clip]{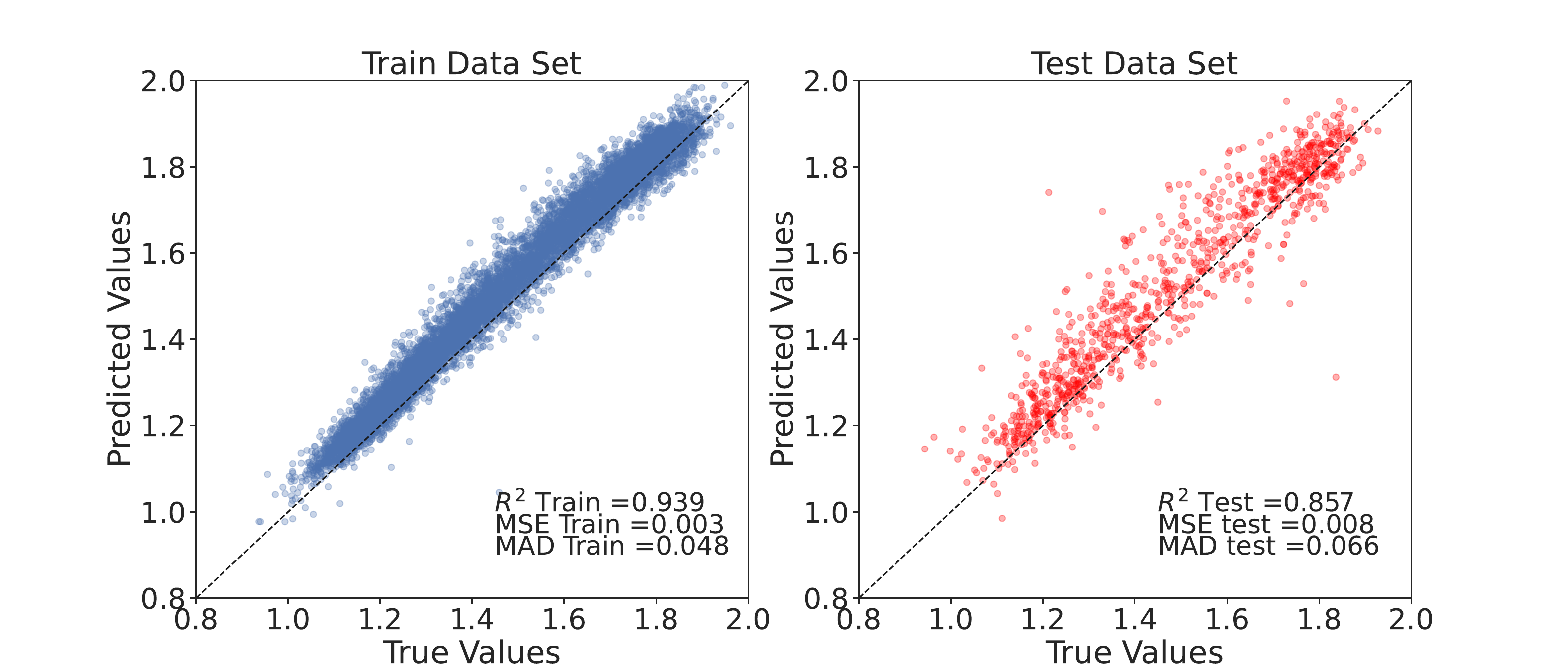}
    \caption{\textbf{Top:} The accuracy of predicting log M$_*$ of a galaxy using the RGB image of it as an input using the ResNet50 Architecture. The left panel shows the accuracy of the training data set and the right panel shows the accuracy of the test data set. The training set performs slightly more accurately than the test set (RMSE of 0.20 dex vs 0.22 dex respectively). \\
    \textbf{Middle:} The accuracy of predicting log SFR of a galaxy using the RGB image of it as an input in addition to the predicted M$_*$ from the previous network in the chain. The left panel shows the accuracy of the training data set and the right panel shows the accuracy of the test data set. The training set performs slightly more accurately than the test set (RMSE of 0.55 dex vs 0.75 dex respectively). The higher than expected RMSE values are due to the network having trouble predicting how quenched a galaxy is. Filtering out galaxies with sSFR $<-10.8$ improves the test RMSE value to 0.31 dex. \\
    \textbf{Bottom:} The accuracy of predicting the D4000 break strength of a galaxy using the RGB image of it as an input in addition to the predicted M$_*$ and SFR from the previous networks in the chain. The left panel shows the accuracy of the training data set and the right panel shows the accuracy of the test data set. The training set performs slightly more accurately than the test set (RMSE of 0.17 dex vs 0.28 dex respectively).}
    \label{fig:network accuracy}
\end{figure}

The accuracy of the network is measured by observing how close the predicted value for each of the three target parameters (M$_*$, SFR, D4000 break strength) is to the real values (from the PIPE3D VAC). The results for the M$_*$ prediction network are shown in the top row of \cref{fig:network accuracy}, where we compare the real values (x-axis) to the predicted values (y-axis) and draw a 1 to 1 line to indicate what a perfect regression would look like. The figure has 2 panels to compare the results from the training and test data sets independently of each other. The plot also includes the $R^2$, MSE, and mean absolute deviation (MAD) values to measure the accuracy of the predictions for each of the data sets.  

For the other two parameters predicted by the chain network (SFR and D4000 break strength) the results for their accuracy are shown in the second row and bottom row of \cref{fig:network accuracy} respectively. These plots are the same as was described above for the top row of \cref{fig:network accuracy}. For a comparison of predictions between the chain architecture and the ResNet50 architecture outside of the chain please refer to \cref{fig:solo vs chain} in \hyperref[app:chain network appendix]{Appendix A}. 

Note that for the M$_*$ there is no comparison of in and out of chain architecture since it is the first network in the chain and therefore does not take any additional input but the image of the galaxy. Due to this, the performance of this network is completely unaffected by the chain architecture. 

\subsubsection{Validation Tests} \label{sec:val test}

Validation tests were done to verify how robust the SHAP maps generated were. The two main validation tests conducted tested how SHAP maps would look for just random noise images, and for galaxy images that were rotated. The random noise images produced essentially equally random noise SHAP maps which is to be expected as there is no real information for the network to pull from to make an accurate prediction. As for the rotated galaxy images we tested how the SHAP maps would look for a galaxy after being rotated by some random angle. We find that the SHAP maps generated are rotated by the same angle, meaning the SHAP maps are picking up on the same morphological feature regardless of image orientation. This gives us more confidence in the SHAP maps as we can trust they are following the galaxy's morphological features and that they are rationally invariant. 

To test the robustness of our chain network and t$_{50}$ network, and to quantify the error of our whole method, we performed bootstrapping on the entire chain shown in \cref{fig:chain arch}. We did this by taking the standard deviation of the pixels within a 5-pixel border of all the galaxy images in the test set. We then added Gaussian noise to each image setting the standard deviation to the one we derived from the border pixels. This was done so that the Gaussian noise was similar in magnitude to the sky background for each image in the test set. We performed this bootstrapping over 100 iterations for each image in the test set and recorded the predicted values for M$_*$, SFR, D4000, and t$_{50}$. We then took the standard deviation of the predictions of each parameter for each galaxy image in the test set across the different iterations. Finally, we took the mean of the standard deviation between runs for each of the galaxy images in the test set. The values are as follows: 0.023 for M$_*$, 0.096 for SFR, 0.012 for d4000, and 0.021 for t$_{50}$.

We checked the performance of the chain network on images of lower quality by pixelating our original images by reducing their resolution and then upscaling them to the original resolution (256 by 256 pixels). We find that all three networks have very little degraded prediction accuracy up to a degrading factor of 4 (image resolution reduced to 64 by 64 pixels). Degradation of the images beyond this point leads to significantly worse prediction accuracy across all three networks in the chain.

\subsubsection{t$_{50}$ Model Performance} \label{sec:t50 performance}
\begin{figure}
    \centering
    \includegraphics[scale=0.23]{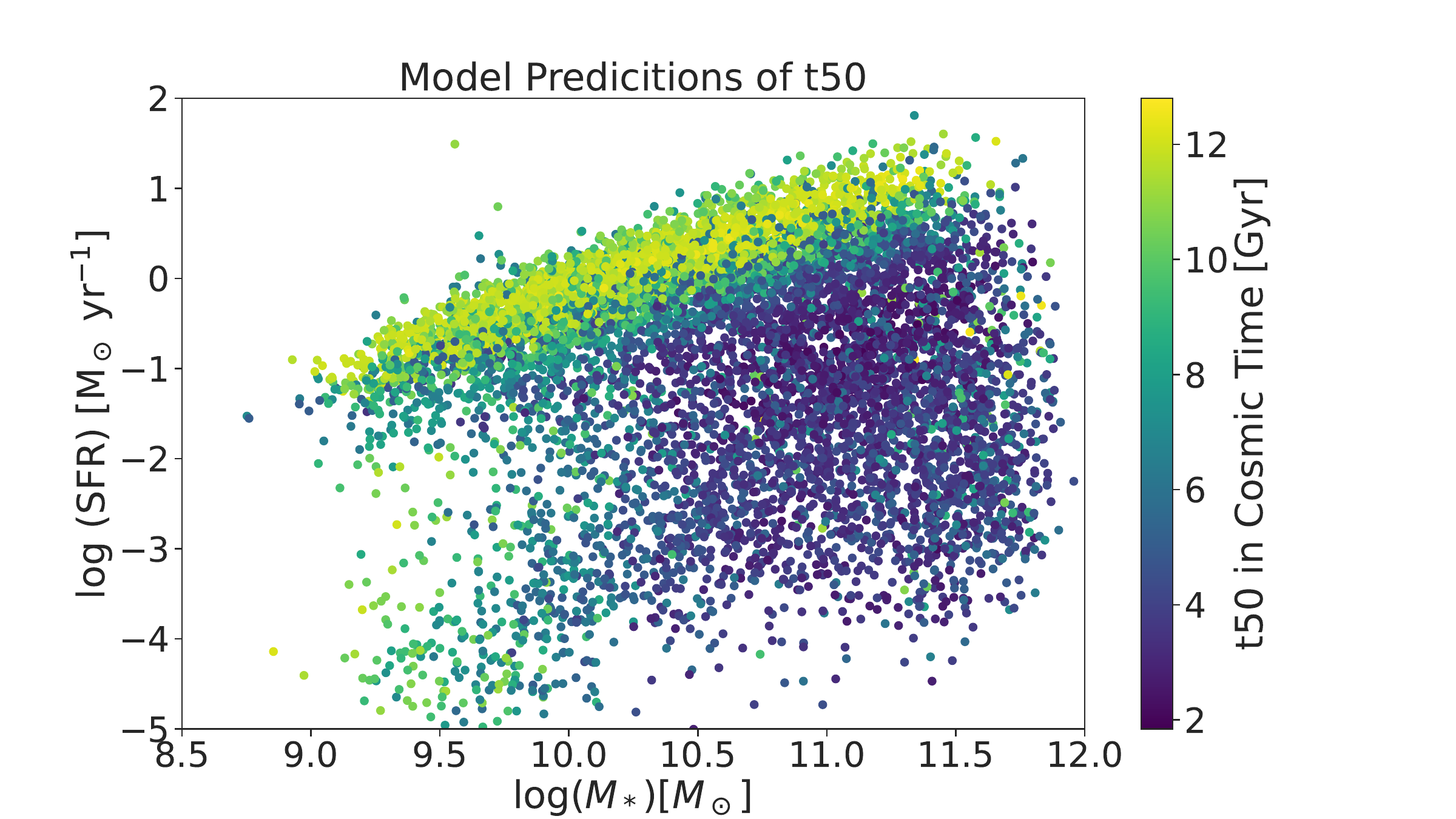}
    \caption{The t$_{50}$ values are calculated for each galaxy using a simple linear neural network. The networks take a galaxy's M$_*$, SFR, D4000 break strength, dust attenuation, metallicity, and redshift as inputs and provide the t$_{50}$ value as an output. As expected we see the SFMS galaxies having the youngest t$_{50}$ and the quiescent galaxies below the SFMS having the oldest t$_{50}$ values.} 
    \label{fig:t50 model}
\end{figure}

The t$_{50}$ values were calculated for each galaxy using a simple linear neural network as described in \cref{sec:t50 model}. The network outputs the t$_{50}$ value and takes the galaxy's M$_*$, SFR, D4000 break strength, dust attenuation, metallicity, and redshift as inputs. The results of the predictions can be visually seen in \cref{fig:t50 model}. We see that the network correctly predicts that galaxies along the SFMS have produced most of their stars more recently, meanwhile, galaxies below the SFMS which are mostly quenched have produced most of their stars in the far past. A comparison to the values found in the PIPE3D VAC is shown in \cref{fig:t50 model vs PIPE3D} in \hyperref[app:t50 appendix]{Appendix B}.

\subsubsection{Comparison to SED Fitting Approaches} \label{sec:SED comp}

In this section, we compare how our CNN-based approach compares with popular SED fitting codes. \cite{SED_Comp} compare a vast number of SED fitting suites to compare how the model assumptions impact the predicted values given the same spectral data. They report a median uncertainty of 0.12 dex for M$_*$ and 0.27 dex for SFR. Both of which are lower than the errors we measure from our networks (0.22 dex and 0.75 dex). The paper does not assess the accuracy of extracting age information. In \citealt{2024arXiv240112300W}, different SDSS-IV MaNGA SED fitting codes were tested, including pyPIPE3D, on mock spectra to compare their results. They report a scatter of 0.11 dex for M$_*$ and 0.14 dex for age. The paper does not assess the accuracy of extracting SFR information. Of course, since our train values are taken from the pyPIPE3D fitting suite we are also inheriting any model assumption bias present in their fitting pipeline  Future work is needed to see how a network trained without the inherent bias of a previous SED fitting model would perform.

\subsection{The Information Content of Galaxy Images}
\subsubsection{SHAP Maps}\label{sec:shap maps}
\begin{figure*}[ht]
    \centering
    \hbox{\includegraphics[width=\textwidth, trim={5cm, 0 , 5cm, 0}]{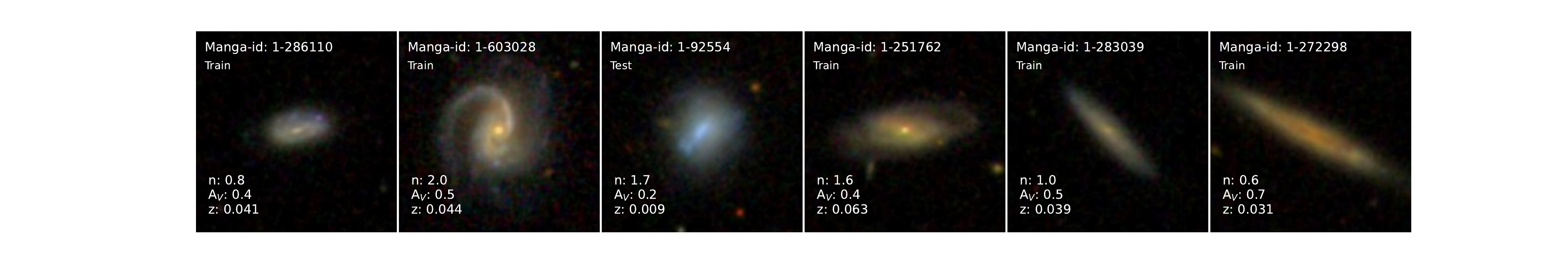}}
    \vspace{1.1cm}
    \hbox{\includegraphics[width=\textwidth, trim={5cm, 0 , 5cm, 0}]{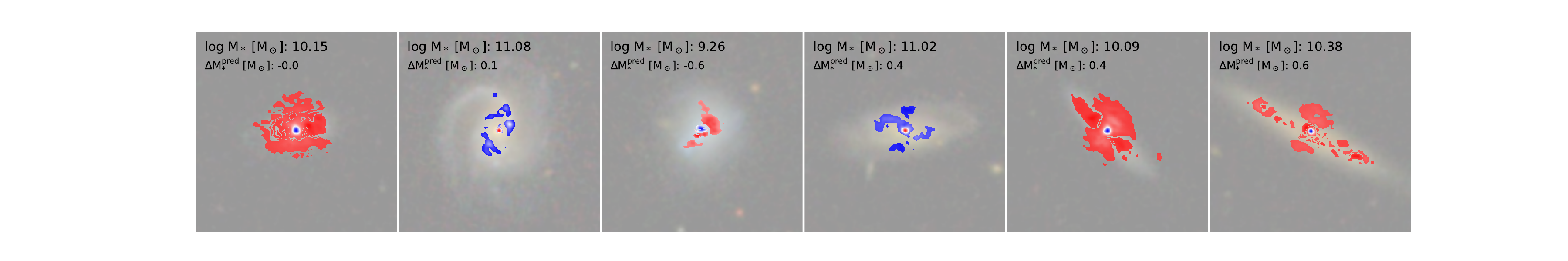}}
    \vspace{1cm}
    \hbox{\includegraphics[width=\textwidth, trim={5cm, 0 , 5cm, 0}]{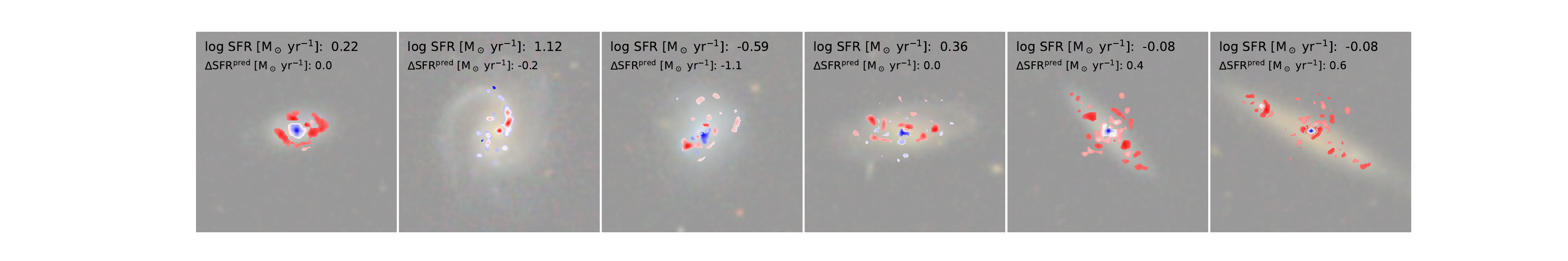}}
    \vspace{1cm}
    \hbox{\includegraphics[width=\textwidth, trim={5cm, 0 , 5cm, 0}]{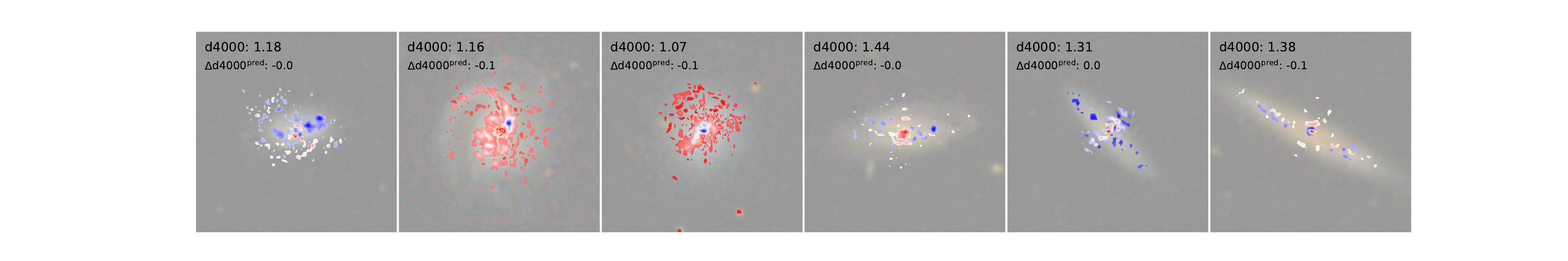}}
    \caption{SHAP maps and images for a selected sample of 5 galaxies. The first row shows the RGB images of the galaxies that the chain networks take as input. The second row, third row, and fourth row show the SHAP map generated by the M$_*$, SFR, and D4000 prediction network along with the true and predicted M$_*$, SFR, and D4000 values respectively. The red pixel indicates a positive SHAP value and the blue pixels indicate a negative SHAP value. These saliency maps tell us which morphological features of the galaxy are important for the prediction of the respective parameter.}
    \label{fig:selected shap maps}
\end{figure*}

A visual representation of the SHAP maps generated for each of the 3 parameters predicted by the chain networks is shown in \cref{fig:selected shap maps}. The figure shows the RGB image of the galaxy along with its 3 corresponding SHAP maps and includes information about the galaxy itself as well as about the prediction accuracy of the network in the given image. The red pixel on the SHAP maps indicates a positive SHAP value and the blue pixels indicate a negative SHAP value. A positive value indicates that increasing the value of this pixel leads to the model overpredicting the target variable, whereas a negative value indicates that increasing the value of the pixel leads to an underprediction of the target variable. A pixel that lacks any color has a SHAP value of around 0 meaning increasing the pixel value has little to no effect on the prediction.

The plot includes identification information and parameters taken from the PIPE3D VAC on the image to provide some additional information about the galaxy. A version of this plot with 10 randomly selected galaxies is shown in \cref{fig:shap map random} in \hyperref[app:SHAP map appendix]{Appendix C}. 

\subsubsection{Radial Plots and Their Gradients} \label{sec:radial plots results}
\begin{figure}
    \centering    
    \includegraphics[scale=0.3,trim={7cm 0cm 7cm 0cm},clip]{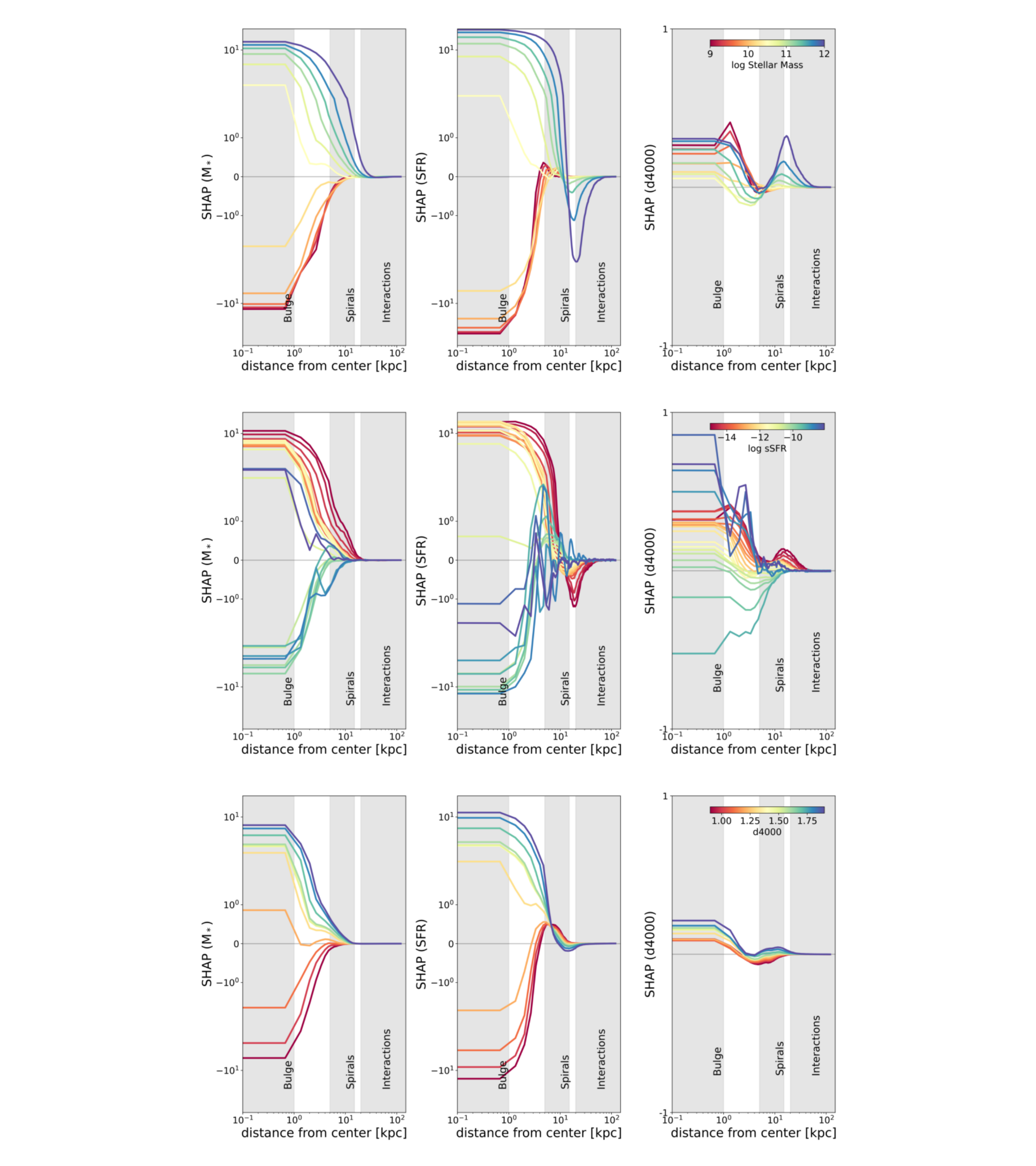}
    \caption{The radial plots of all galaxies that fall within the respective bins as indicated by the color bar on each row. Each row represents the radial plots across the three parameters for the same binning scale. The plot has three different sections that label where roughly most galaxies in the sample have their bulge, spiral arms, and interactions with neighboring galaxies. This plot allows us to see how galaxies in different populations have on average different morphological features linked to the predicted parameters.}
    \label{fig:primary radial plots}
\end{figure}

To create a comprehensive summary of the SHAP maps discussed in \cref{sec:shap maps} we generated radial plots of these SHAP maps as described in \cref{sec:radial profiles method}. The resulting plots which tell us the average SHAP value at a given distance from the galaxy center for a group of galaxies are shown in \cref{fig:primary radial plots}. In this plot the first row bins the galaxies by M$_*$, the second row bins the galaxies by sSFR (specific SFR, meaning normalized by M$_*$), and the third row by D4000 break strength. This plot allows us to see the average morphological profile for galaxies with similar physical properties. 

As described in \cref{sec:radial profiles method} gradients of the radial plots were taken for each galaxy and each of the three parameters. These gradients serve as a summary of the galaxy's morphological structure. Essentially the gradient is simply the difference in the $y$ values (the average SHAP value at a given radial distance from the center of the galaxy) between the start of the plot and halfway through the "spirals" box. While \cref{fig:primary radial plots} shows the radial profile of a collection of galaxies, it is worth remembering that a radial profile is generated for each galaxy for each of the three parameters (M$_*$, SFR, D4000). This means that each galaxy has three gradients, each one describing the connection between the morphological profile of the galaxy and the predicted parameter. We add shaded regions to \cref{fig:primary radial plots} to show where we expect the bulge, spirals, and interactions would be for the galaxies in our sample. We expect bulges to be in distances of 0-5 kpc from the center \citep{bulge_distance}, spiral arms to be in distances of 15-25 kpc \citep{spiral_arm_distance}, and interactions to be in distances of over 130 kpc (\cite{interaction_distance1},\cite{interaction_distance2}). 

\begin{figure}[ht]
    \centering
    \includegraphics[scale=0.33]{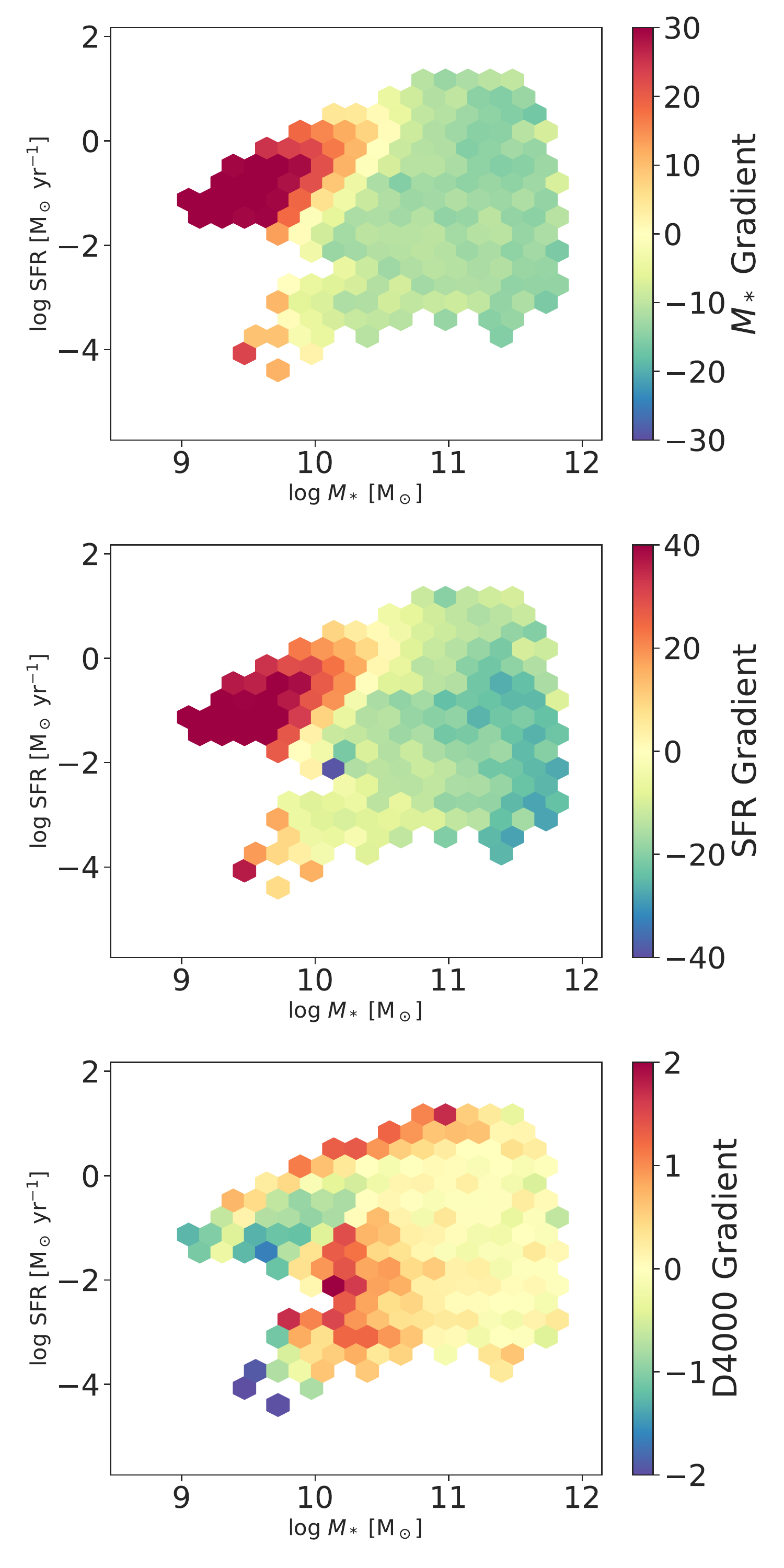}
    \caption{A visualization of the gradients across the 3 parameters for each galaxy in the M$_*$-SFR plane. The plots show the M$_*$, SFR, and D4000 gradients for each galaxy respectively. A red color (positive value) means the network is finding morphological features that are positively correlated for the prediction of the parameter further out from the galaxy center. This could be things like spiral arms. It could also mean that it finds morphological features in the galaxy center to be negatively correlated with the predicted parameter. A blue color (negative value) means the network is finding morphological features that are positively correlated for the prediction of the parameter near the galaxy center. This could be things like the bulge.}
    \label{fig:hexbin gradients}
\end{figure}

The 3 sets of gradients calculated for each galaxy can be visualized in \cref{fig:hexbin gradients}. Here the galaxies are plotted according to their position on the M$_*$-SFR plane and are colored by the gradient for each of the three parameters. The plot highlights how the gradient profiles vary across the same galaxy population depending on what the prediction target parameter is. Recalling how the gradients are calculated we can infer that a positive gradient means that as we move away from the center of the galaxy the SHAP values generally increase. On the other hand, a negative gradient means that as we move away from the center of the galaxy the SHAP values generally decrease. The magnitude of the gradient then tells us how fast this increase is as we move out from the center of the galaxy (for positive gradients), and how much of a drop-off we have as we move from the center of the galaxy outward (for negative gradients).

 For simplicity since we have three sets of gradients (one for M$_*$, one for SFR, and one for D4000), we refer to them as $\nabla M_*$, $\nabla$SFR, and $\nabla$D4000) respectively. $\nabla M_*$ and $\nabla$SFR are noted to have very similar behaviors while $\nabla$D4000 is very distinct to both of them. This trend is shown qualitatively in \cref{fig:hexbin gradients} where both $\nabla M_*$ and $\nabla$SFR have a clear bi-modality about $M_*=10^{10.5}M\odot$ yet $\nabla$D4000 has a less organized split in the M$_*$-SFR plane. From this plot, it is also clear that while galaxies tend to have very similar $\nabla M_*$ and $\nabla$SFR values, their $\nabla$D4000 are quite different and in some cases even the exact opposite. 

These trends lead us to conclude that the morphological features that are important for predicting M$_*$ and SFR are very different from the ones that are important for predicting D4000. This leads us to the idea that morphological properties of galaxies associated with the galaxy's age are perhaps somewhat unique and directly related to events in its evolutionary history. 

\cref{fig:morph-sfh} shows how the current morphologies of galaxies are connected to their SFHs. The plot shows where the galaxies would lie in the M$_*$-SFR plane at 0.1 Gyr and 12 Gyr before the time of observation. We were able to derive these M$_*$ and SFR values using the SFH curves generated using the methods described in \cref{sec:dense basis}. What we observe is that galaxies with the same important morphological features stay clustered in the M$_*$-SFR plane as we move back in time. This is a very promising result as it indicates that galaxies with similar significant morphological features have a similar SFH. This points to the idea that a common galactic evolutionary path leaves the same imprints on the galaxy's morphology. In other words, the current morphological features of galaxies are a fossil record of their formation, interactions, and evolution.

\subsubsection{The Effects of Galaxy Color}
\label{sec:color}

\begin{figure*}[ht]
    \centering
    \includegraphics[width=0.45\textwidth]{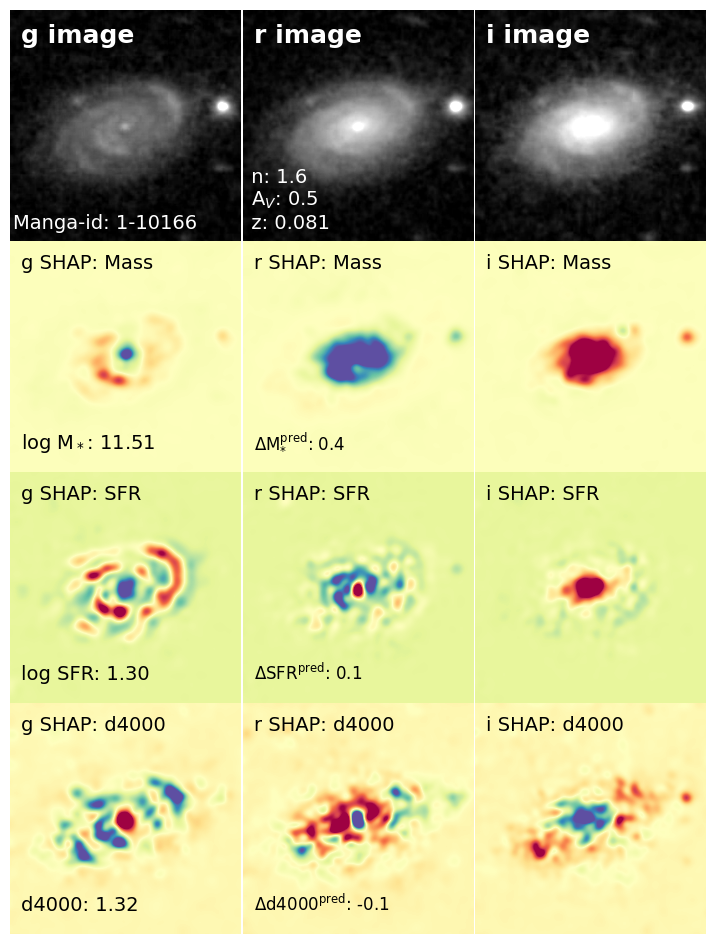}
    \includegraphics[width=0.45\textwidth]{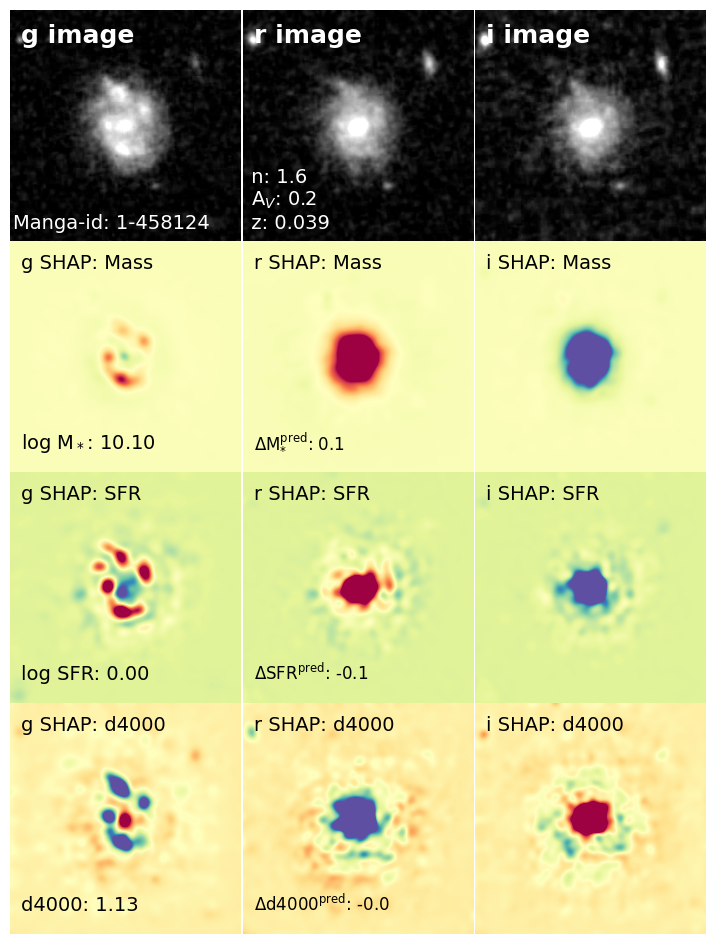}
    \includegraphics[width=0.45\textwidth]{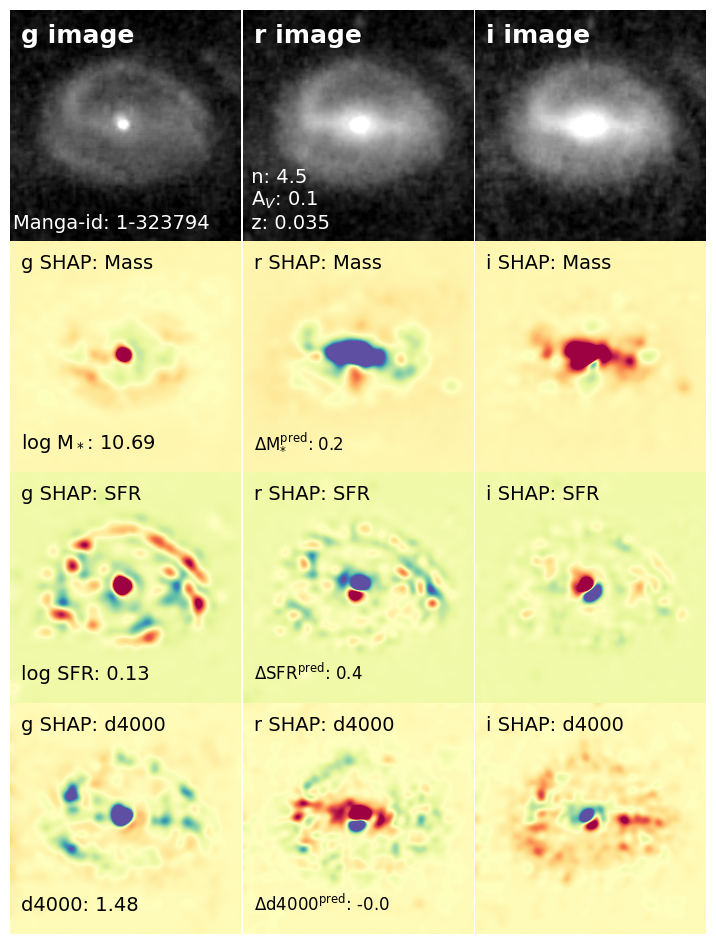}
    \includegraphics[width=0.45\textwidth]{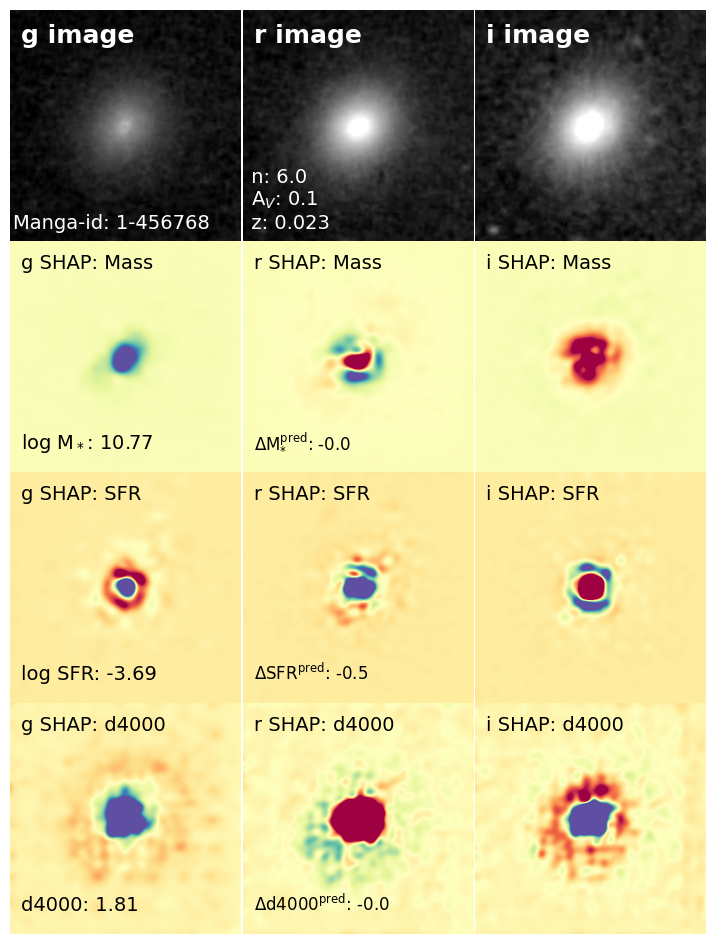}
    \caption{Single color SHAP maps for four galaxies in our sample, along with greyscale images of the input galaxies. }
    \label{fig:spatial_shap_examples}
\end{figure*}

We have so far focused on the spatially resolved SHAP maps averaged across the three channels of our input (i.e. the normalized \textit{gri} fluxes of the galaxies in our sample). Breaking this down further and studying the maps on a color-by-color basis reveals the presence of rich structure in the individual bands that maps the features at different wavelengths, from the spiral arms and star-forming clumps being much more prominent in the \textit{g}-band SHAP maps to the bulges of galaxies being featured in the \textit{i}-band SHAP maps. A few examples of this are shown in \cref{fig:spatial_shap_examples}. 

While a full treatment of this effect is outside the scope of this work, we can see that the three colors tend to trace different features, leading to the CNN looking at a combination of color and morphological features for different galaxies. For example, generally, the r-i color drives the stellar mass, except for the extremely clumpy galaxy (without a noticeable bulge), where the trend inverts. SFR is driven by the presence of spiral arms or clumps, except for the elliptical galaxy where the color is more important. Finally, age (or its proxy in D4000) is set by a balance of star-forming features (clumps and spiral arms, for example), which drive the value to younger ages, contrasting with the presence of a mature bulge, which drives the age to older values. Interestingly, the bar in the lower-right example (MaNGA id: 1-323794) seems to drive the age to younger values, but the lack of a statistically large sample means that this effect can't be investigated at present. 

\begin{figure*}[ht]
    \centering
    \includegraphics[width=0.32\textwidth]{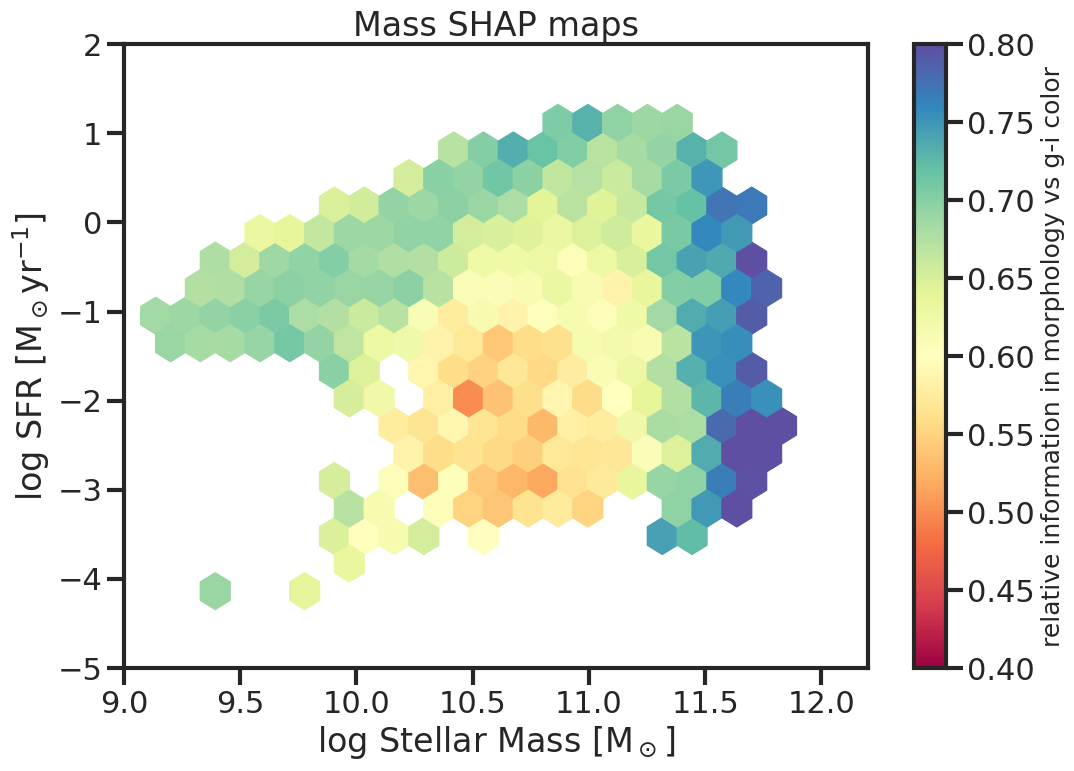}
    \includegraphics[width=0.32\textwidth]{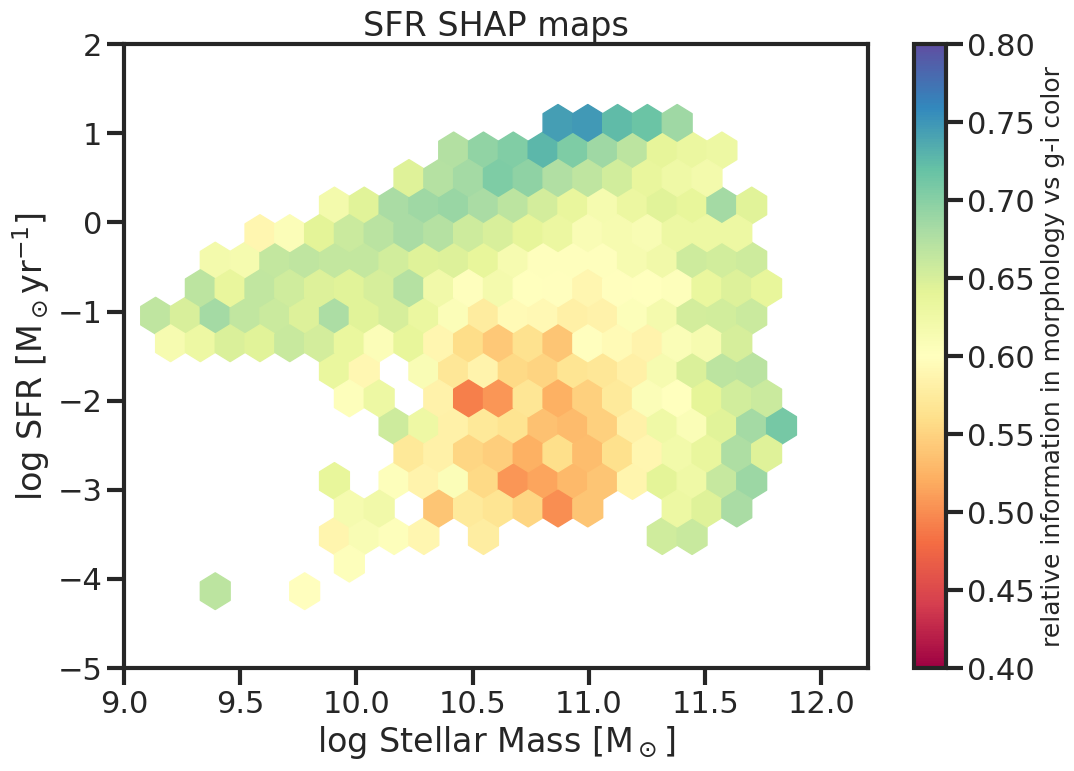}
    \includegraphics[width=0.32\textwidth]{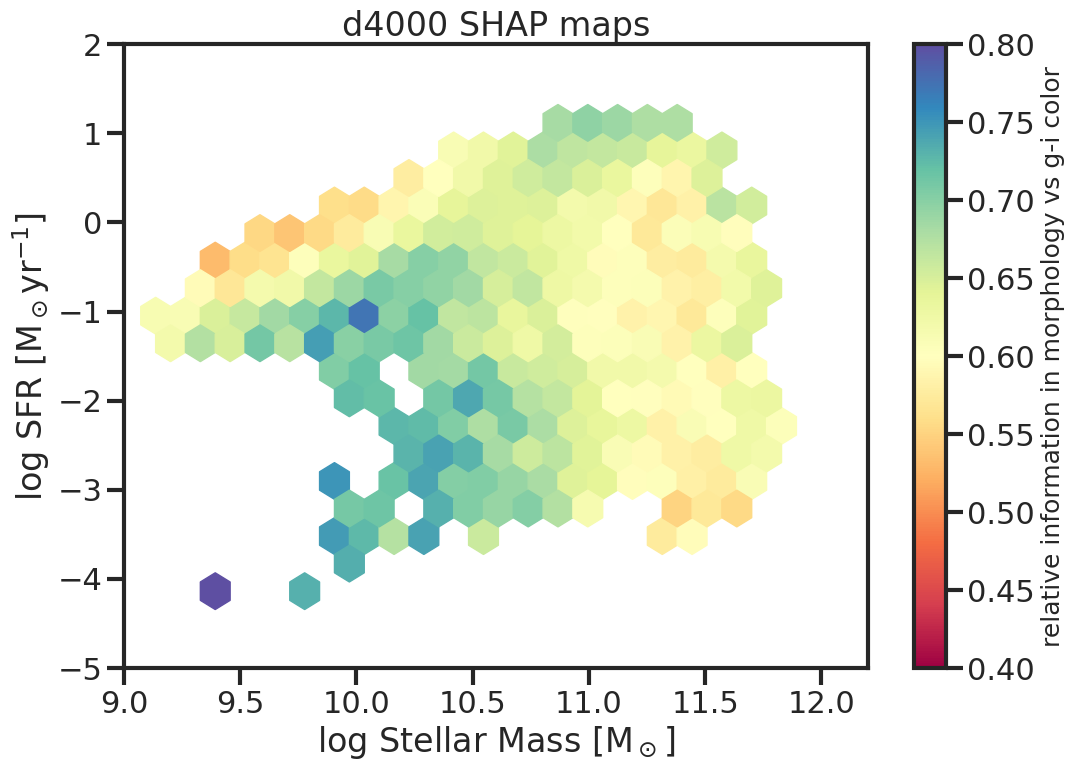}
    \includegraphics[width=0.32\textwidth]{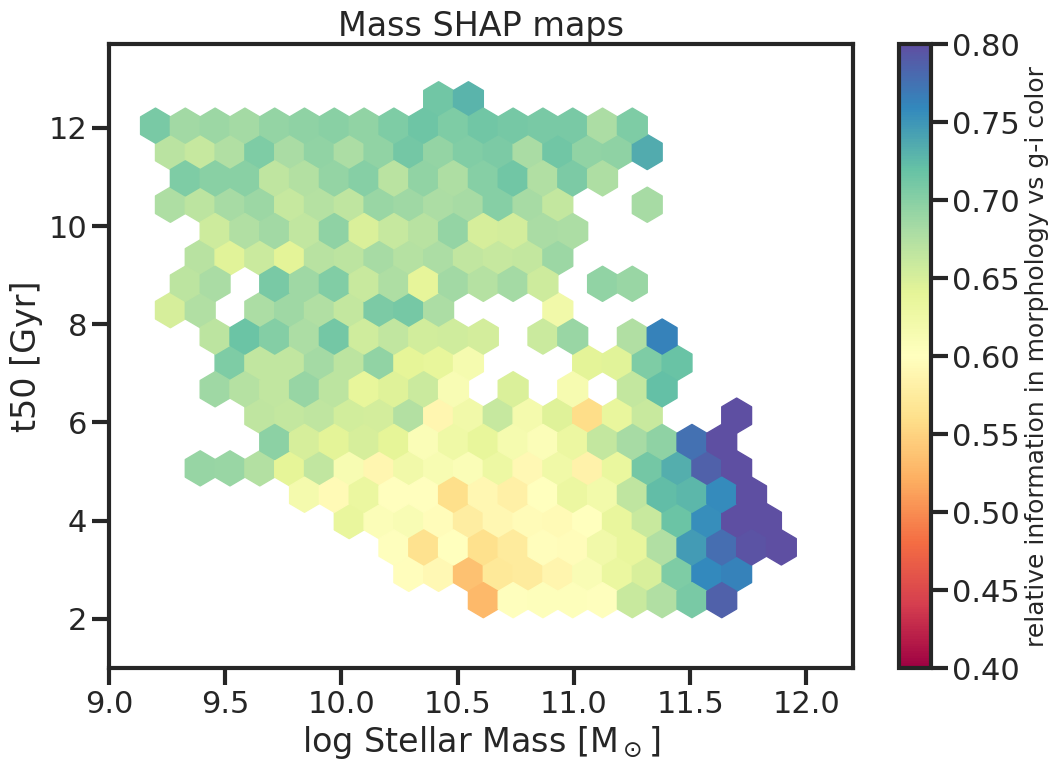}
    \includegraphics[width=0.32\textwidth]{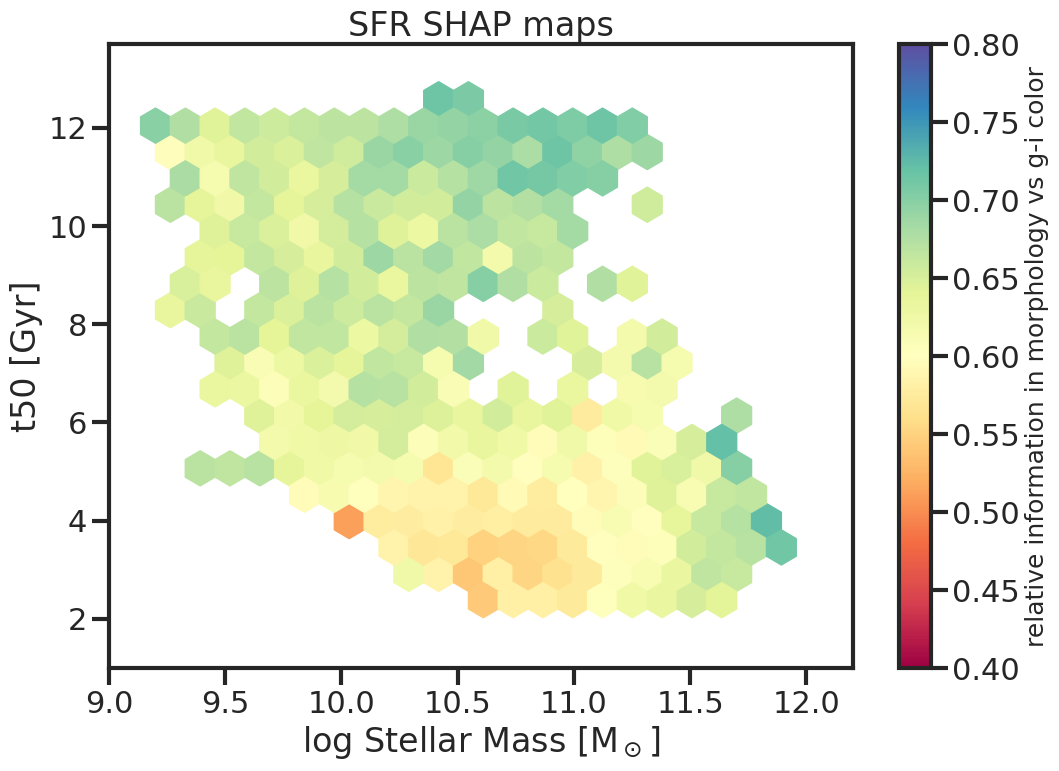}
    \includegraphics[width=0.32\textwidth]{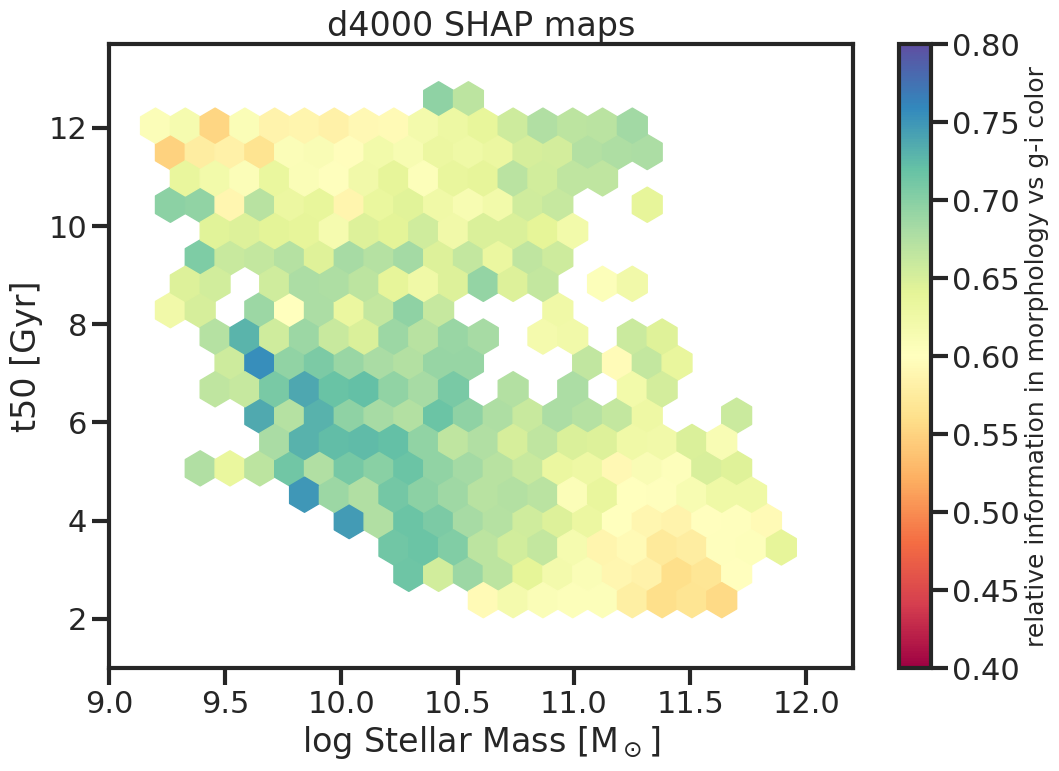}
    \caption{Relative amount of information contained in the g-i color compared with the morphology, indicated by the color bar. The plots show the average contribution across the stellar mass-SFR plane (top) and stellar mass-t$_{50}$ plane (bottom) for predicting the stellar mass (left), SFR (middle) and d4000 (right). Lower values imply that the color drives the prediction while higher values imply that the morphological information is of primary importance. We see that while color is important for getting the SFR of low-mass galaxies or quenching right, getting the masses of high-mass galaxies, the SFR of massive star forming galaxies and the ages of low-mass low-sSFR galaxies are strongly influenced by the morphology.}
    \label{fig:color_importance}
\end{figure*}

On an ensemble level, the relative balance between the amount of information in the galaxy colors and that in their morphology is studied in \cref{fig:color_importance}. 
We define the relative information as $I = \sigma_{\rm morph} / (\sigma_{\rm morph} + \sigma_{\rm color})$, where $\sigma_{\rm morph}$ is the variance of the mean SHAP map for a given physical quantity averaged over all the colors, and $\sigma_{\rm color}$ is the variance of the $g-i$ SHAP maps for a given physical quantity. This directly compares the variance in the monochromatic SHAP map (that highlights features in morphology) that contribute to the prediction against the variance in the difference between the g-band SHAP map and i-band SHAP map (which tends to be small when the different colors hold no information and large when there is a significant difference with color due to e.g., a large bulge). By construction,
low values of this number mean that the color strongly drives the prediction (of M$_*$, SFR or D4000), while high values imply that the morphology drives the prediction instead. We see that for M$_*$ this happens for high-mass galaxies at all SFRs (massive galaxies being star-forming or quenched depends more on their morphology), while for D4000 values this happens for low-mass, low SFR galaxies that tend to be bursty, and therefore contain rich morphological information that helps constrain their age. This is especially important for studies at high redshifts since it implies that morphological constraints might help better constrain the stochastic nature of star formation at these epochs. We aim to investigate better metrics to quantify this and discuss the effects of this in future works.

\subsubsection{M$_*$ Divide of $\nabla M_*$ and $\nabla$SFR} \label{sec:mass divide}
Looking at the first two rows of \cref{fig:hexbin gradients} it is noticeable that for $\nabla M_*$ and $\nabla$SFR, there is a clear divide at $M_* \approx 10^{10.5} M\odot$. If we were to draw a vertical line at $M_* = 10^{10.5} M\odot$ on these plots we can see two clear and distinct populations. On the left-hand side, we have galaxies colored mostly orange/red indicating positive gradients, while on the right-hand side, we have galaxies colored green/blueish indicating a negative gradient. Essentially this points to lower mass galaxies having the network focus on its outskirts to predict M$_*$ and SFR, whereas for higher mass galaxies the network is mostly focusing on the center and the areas around it to predict their M$_*$ and SFR. This leads us to believe that the predictions for both M$_*$ and SFR are heavily reliant on the same morphological features for galaxies above and below the $M_* = 10^{10.5} M\odot$ threshold. Additionally, in \cref{fig: hexbin d4000} we recreate the top two rows of \cref{fig:hexbin gradients} but we plot the galaxies in M$_*$-D4000 space instead of M$_*$-SFR space. In doing so we realize the same divide exists for galaxies below and above D4000 $= 1.6$. This leads us to conclude that the areas of morphological importance between low-mass galaxies are the same as for young galaxies and that the areas of morphological importance between high-mass galaxies are the same as for old galaxies. This means that low-mass and young galaxies, as well as high-mass and old galaxies, have similar morphological properties linked to their current M$_*$ and SFR. 

\subsubsection{The Relation Between SHAP Maps for M$_*$, SFR, and D4000} \label{sec:age similar}

\begin{figure*}
    \centering
    \includegraphics[width=\textwidth, trim={3cm 1cm 1cm 3cm}]{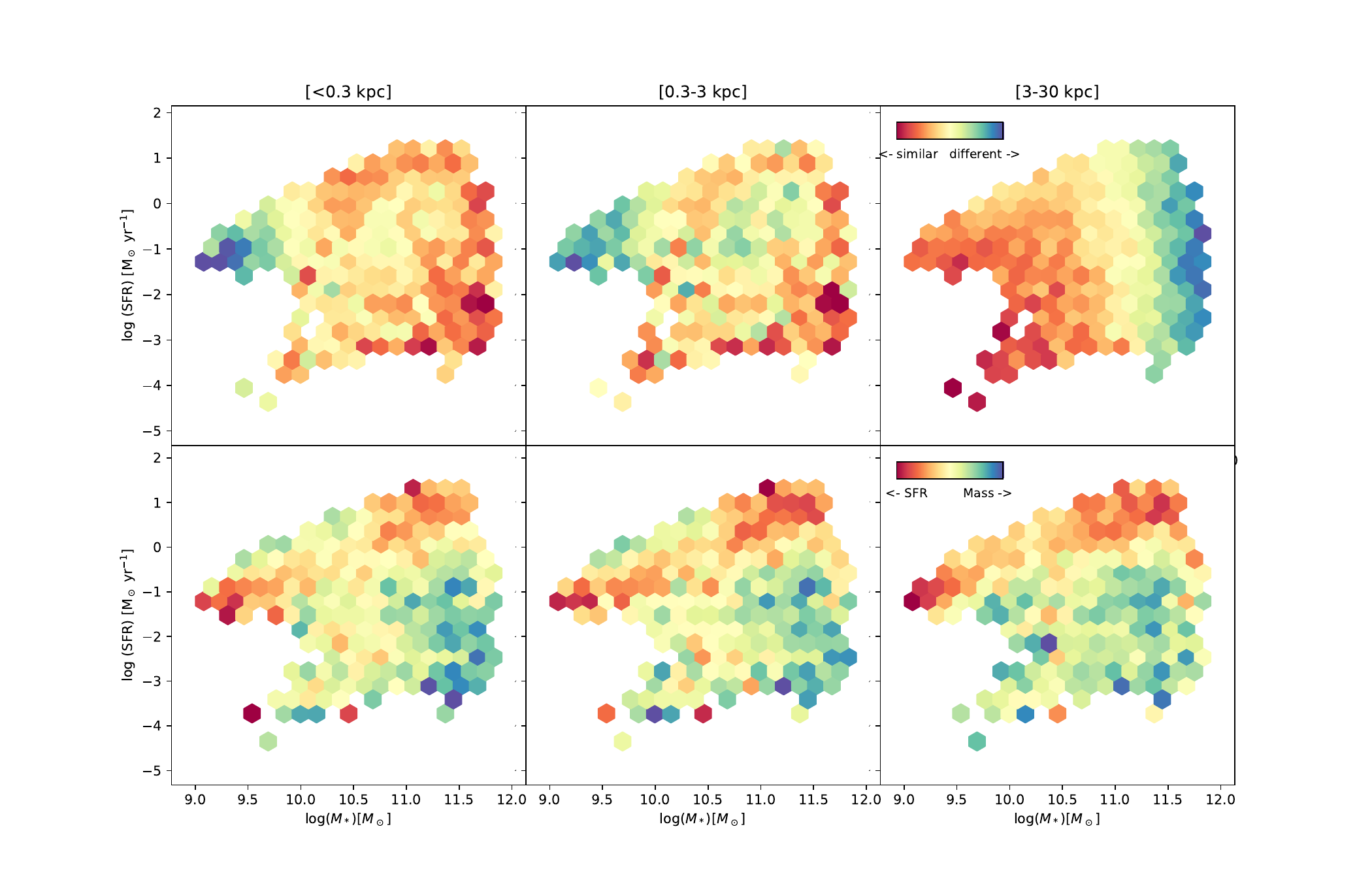}
    \caption{How similar the SHAP maps for D4000 are to the SHAP maps for M$_*$ and SFR at different radial distances (columns). The top row shows how similar the D4000 SHAP map is to the M$_*$ SHAP map, with redder colors indicating more similarity and bluer colors indicating more difference. The bottom row shows if the D4000 SHAP map is more similar to the M$_*$ SHAP map or the SFR SHAP map.}
    \label{fig:age similar}
\end{figure*}

The age of a galaxy is a function of both recent stellar populations (correlating with SFR) and older ones (correlating with M$*$). As a result, the SHAP maps for D4000 sometimes resemble those for SFR (e.g. in the outskirts of star-forming disks), and sometimes those of stellar mass (e.g. in the bulges of massive galaxies), and sometimes neither. To investigate this as a function of galaxy properties, we compute the similarity of the normalized D4000 SHAP map to the normalized M$_*$ SHAP map for each galaxy, shown in the top row of Figure \ref{fig:age similar} as a function of distance. As expected, the SHAP maps in the central regions tend to agree for massive galaxies that are bulge-dominated. As we move to the outskirts, there is an increasing trend of the D4000 and M$_*$ SHAP maps deviating from each other with increasing mass, since the D4000 maps tend to be more sensitive to the environment while the mass concentrates towards the centers. 
We also compare the normalized D4000 SHAP to see if it is more similar to the normalized mass or SFR SHAP map in the bottom row of Figure \ref{fig:age similar}. The trends here are more subtle, with high-mass star-forming galaxies generally closer to the SFR SHAP maps, since the morphological features highlight recent star formation, with the trends being most visible in the 3-30 kpc range.

\subsubsection{Physical Significance of the Gradients} \label{sec:physical significance of gradients}
The gradients, whose derivation was explained in \cref{sec:radial profiles method} and whose interpretation was explored in \cref{sec:radial plots results}, have also a physical importance which will be explored in this section. The physical importance of the gradients comes from the fact that they stand in as a summary of morphology learned by the network. The gradients highlight which areas of the galaxy are important to the prediction of each of the target parameters. This in turn means we can use them to infer the relation that exists between the morphology of the galaxy and the target parameters. To do so, however, it is important to understand how to interpret a given gradient value and what it tells us about the galaxy morphology in physical space. 

Recall that a positive gradient indicates the network positively associates the outskirts of the galaxy with the predicted parameter, or that it negatively correlates the center region of the galaxy with the predicted parameter. This is because a positive gradient indicates a higher average SHAP value at 20 kpc from the center than at the center. A higher SHAP value is associated with a positive correlation existing between that part of the image and the target parameter. Conversely, a negative gradient indicates the network positively associates the center of the galaxy with the prediction parameter. This is because a negative gradient indicates a higher average SHAP value at the center of the galaxy compared to at 20 kpc from the center. Therefore, we see that the network finds a positive correlation to exist between the center of the galaxy and the target parameter.

Now that we have examined what the sign of the gradient tells us about the physical link between the morphology and the target parameter, let us explore what the magnitude of the gradient tells us. A higher absolute value for the gradient means that the gradient has a sharper rise if it is positive and a faster drop-off if it is negative. What we can infer from this is that the more positive the gradient the more we expect the galaxy to have morphologically significant components towards its outskirts. This could be things like extended bars or long spiral arms. On the other hand, sharp declines tell us that most of the morphologically significant components of the galaxy lie in its center, and the more negative the gradient the more compact the morphologically important features are to its center. 

We expect that galaxies with similar gradients, both in magnitude and sign, have similar morphological profiles. It is important to note that this means similar gradients of the same type (i.e $\nabla $M$_*$, $\nabla$SFR, and $\nabla$D4000). Since gradients highlight the area of importance for the prediction of a given parameter it does not mean all 3 sets of gradients will agree with each other. This is because some areas of the morphological structure might be very important for predicting the M$_*$ of the galaxy, but those areas may not be so important for predicting the D4000 of the galaxy. 

\begin{figure*}
    \centering
    \includegraphics[scale=0.3]{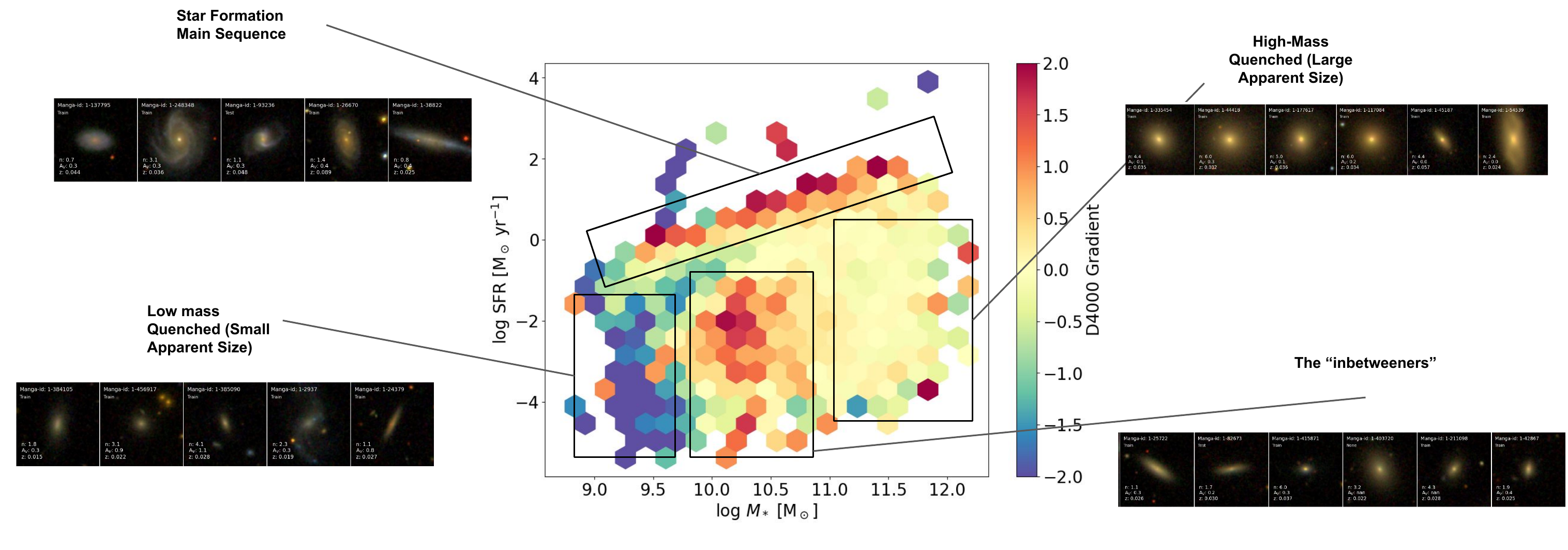}
    \caption{A visual check that similar gradients correspond to galaxies with similar morphological profiles. The galaxies shown for each defined group are selected randomly from the sample. It is also worth noting that there is a big dispersion of sersic index values for the galaxies of similar gradients, yet the galaxies have the same morphology when examined visually. This points to the gradient approach being a better indicator of morphological features than the classically used sersic index.}
    \label{fig:gradient visual}
\end{figure*}

We manually verify that similar gradients of a given parameter correspond to galaxies with similar morphology by selecting some galaxy populations with similar gradients and visually checking their images to see if they have similar morphological profiles. We demonstrate this using the $\nabla$D4000 in \cref{fig:gradient visual}. We see here that there are roughly 3 main groups of galaxies when grouped by morphology and that they all occupy a similar space in the M$_*$-SFR plane, which is to be expected. The three main groups found by grouping the gradients are labeled: star-forming galaxies, low mass quenched galaxies, and high mass quenched galaxies. There is also a 4th group called "the inbetweeners" which is made up of a diverse range of gradient types. This is expected, as we found by visual inspection that there is a wide mix of galaxy morphology structures in this area, and tends to contain a mix of galaxies in the process of quenching, low sSFR galaxies evolving secularly, and a small fraction of rejuvenating galaxies. This plot helps us confirm that indeed the network is picking up on distinct morphological populations and using this to infer what D4000 value the galaxy should have. In other words, the network is accurately finding links between morphological features and the target parameters. This gives us confidence that we can use the gradients as a stand-in for physical morphology descriptions, and hence we can use them as a new way to study galaxy morphology without relying on categorizing the morphology in discrete bins. 

Additionally, as is shown in \cref{fig:gradient visual} some of the galaxies with similar gradients have very different sersic index values, yet when examined visually the galaxies have similar morphologies. This points to the well-known shortcoming of the sersic index as a metric to study galaxy morphology namely: its lack of robustness when it comes to degeneracy (different sersic parameters can lead to the same index), its inability to capture fine-scale features such as spirals arms and bars, and difficulty handling galaxies with extended low surface brightness outer regions. These are all things the gradient approach picks up on much better than the sersic index as the approach used to generate the gradients relies on CNN and SHAP values. This approach can: avoid degeneracy problems (as it creates a unique profile map for each galaxy), pick up fine-scale features, and better capture extended low-brightness profiles (as seen in the low-mass quenched class). In addition to this, the gradients are not simply a statistical summary of the galaxy light profile, but rather the result of finding statistical trends between the physical structure of the galaxy and its corresponding physical parameter. The gradients give us a new way to study galaxy morphology and a better way to find links between morphological features and physical parameters that arise from galaxy evolution. In the following sections, we will discuss some of the trends that were found using these gradients, and their physical implications for galaxy evolution. 

\subsection{SFHs and Morphological Properties} \label{sec: SFH and morphology}
\begin{figure*}
    \centering
    \includegraphics[scale=0.17]{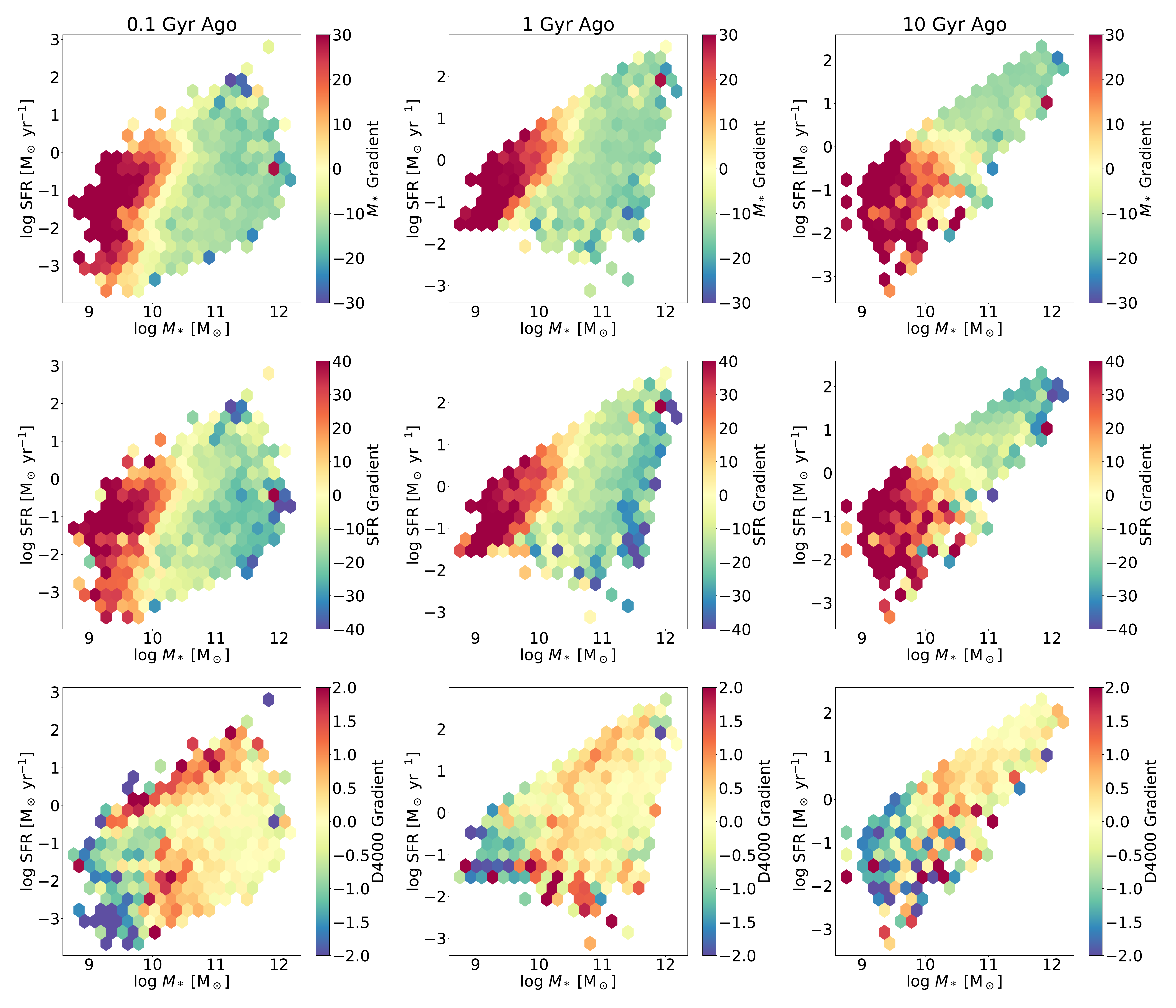}
    \caption{Galaxies on the M$_*$-SFR plane at 0.1 Gyr (left column) and 1 Gyr (middle column), and 10 Gyr (right column) before observation colored by their gradient. The rows of the figure show M$_*$, SFR, and D4000 gradients respectively. We see that galaxies with similar current morphology stick together on the M$_*$-SFR plane over cosmic time. The values of M$_*$ and SFR were derived from the reconstructed SFHs of each galaxy.}
    \label{fig:morph-sfh}
\end{figure*}

Using the dense basis method (described in \cref{sec:dense basis}) we generated the SFH of all the galaxies in the sample. This means we were able to obtain the SFR at any time from the start of star formation to the time of observation of every galaxy in the sample. We used these SFRs to generate plots similar to those found in \cref{fig:hexbin gradients}, but instead of plotting on the time of observation M$_*$-SFR plane, we make the plot on the M$_*$-SFR plane at different points in cosmic time. This allows us to see the direct trend between the galaxies' morphology (via the gradients) and their SFHs. The results of this analysis are shown in \cref{fig:morph-sfh}, where we see how the different morphological populations of galaxies shift around the M$_*$-SFR plane between 0.1 Gyr to $\sim$10 Gyr before observation. Refer to \hyperref[app:full sfh morph plots]{Appendix D} to see a similar plot but which includes more time steps between 0.1 Gyr and 12 Gyr before observation. As seen in the figure, a lot of morphological trends tend to become more well-ordered as we probe star formation further back in time. This correlates with the reduced scatter in the SFR-M$_*$ correlation at larger lookback times \citep{iyer18,sanchez18}, as well as simulations where star formation tends to be correlated over large timescales \citep{Eagle}.

\section{Discussion} \label{sec:discussion}

\subsection{Morphology and Current Galaxy Properties} \label{sec:discussion galaxy}
This section will focus on explaining relationships that were found between the morphology of the galaxies in the sample and their predicted parameters at the time of observation. Possible physically driven explanations will be given to try and explain why these trends might be occurring and how they fit into the bigger picture of galaxy evolution. 

\subsubsection{Is there a Characteristic Mass Scale in Galaxies?} \label{sec:mass age divide}

\begin{figure}
    \centering
    \includegraphics[width=0.5\textwidth]{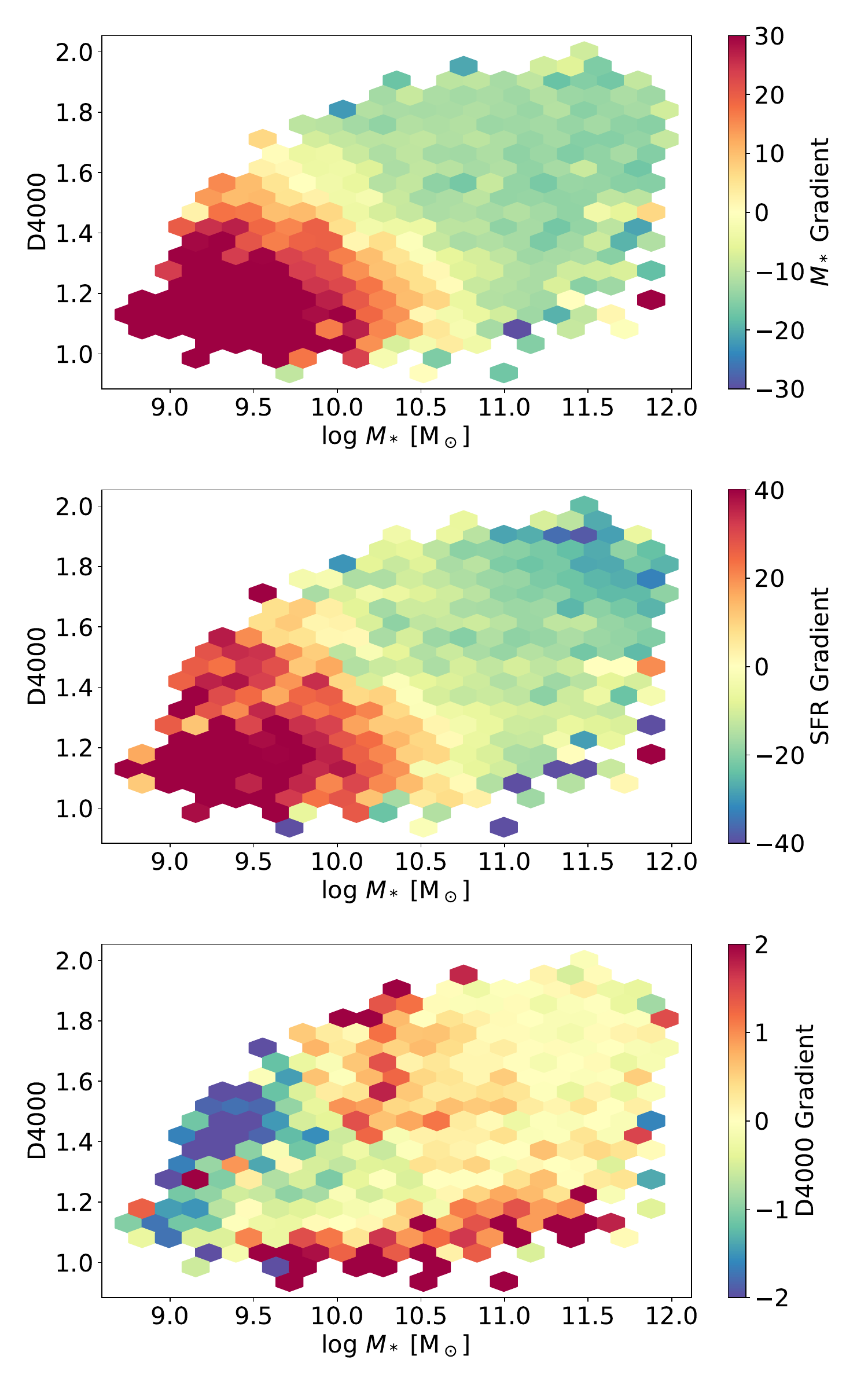}
    \caption{Similar to \cref{fig:hexbin gradients}, but the galaxies are plotted on the M$_*$-D4000 plane instead of the M$_*$-SFR plane. Here we see a similar divide for areas of morphological importance by M$_*$, but this plot further highlights that this same divide exists between galaxies with a low D4000 (young galaxies) and a high D4000 (old galaxies). The values for M$_*$ and D4000 in this figure come from the PIPE3D VAC.}
    \label{fig: hexbin d4000}
\end{figure}

Characteristic or transition mass scales have been described throughout the literature for quite a while, most notably in \cite{char_mass} and \cite{char_mass2}. This leads us to consider what physical galaxy evolution is at play that leads to low mass and young galaxies having important morphological features at their outskirts and also would cause important morphological features to be near the center of high mass and old galaxies as seen in \cref{fig: hexbin d4000}. There is a well-known galaxy "golden mass" at around $M_* = 10^{10} M\odot$ which could explain the bi-modality we see in the morphological structures around this M$_*$. The so-called golden mass of galaxies is discussed by \cite{golden-mass-1} and more thoroughly explored in \cite{golden-mass-2}. They describe two physical mechanisms that occur in galaxies just below and just above the threshold M$_*$ that lead to inside-out quenching of galaxies at cosmic noon. These events are thought to be the seeds for early-type galaxies observed in the local universe. It is possible then that the physical phenomena associated with the compaction to a star formation phase, a "blue nugget", followed by the central quenching of star formation, a "red nugget",  is leaving an imprint on the galaxies' morphologies. This would line up with our results as the quenching event (transitioning from blue nugget to red nugget) tends to occur near a critical M$_*$ of $10^{10}M\odot$ and leaves the galaxy with a M$_*$ slightly above this critical mass. We see a divide about $M_* = 10^{10.5}M\odot$ which is within the expected most event M$_*$. Additionally, the transition from supernova to AGN feedback described to take place around this M$_*$ might also be leaving physical signatures on the galaxy structure.

\subsubsection{The Role of Morphology in Current M$_*$ and SFR} 

\begin{figure}
    \centering
    \includegraphics[scale=0.35]{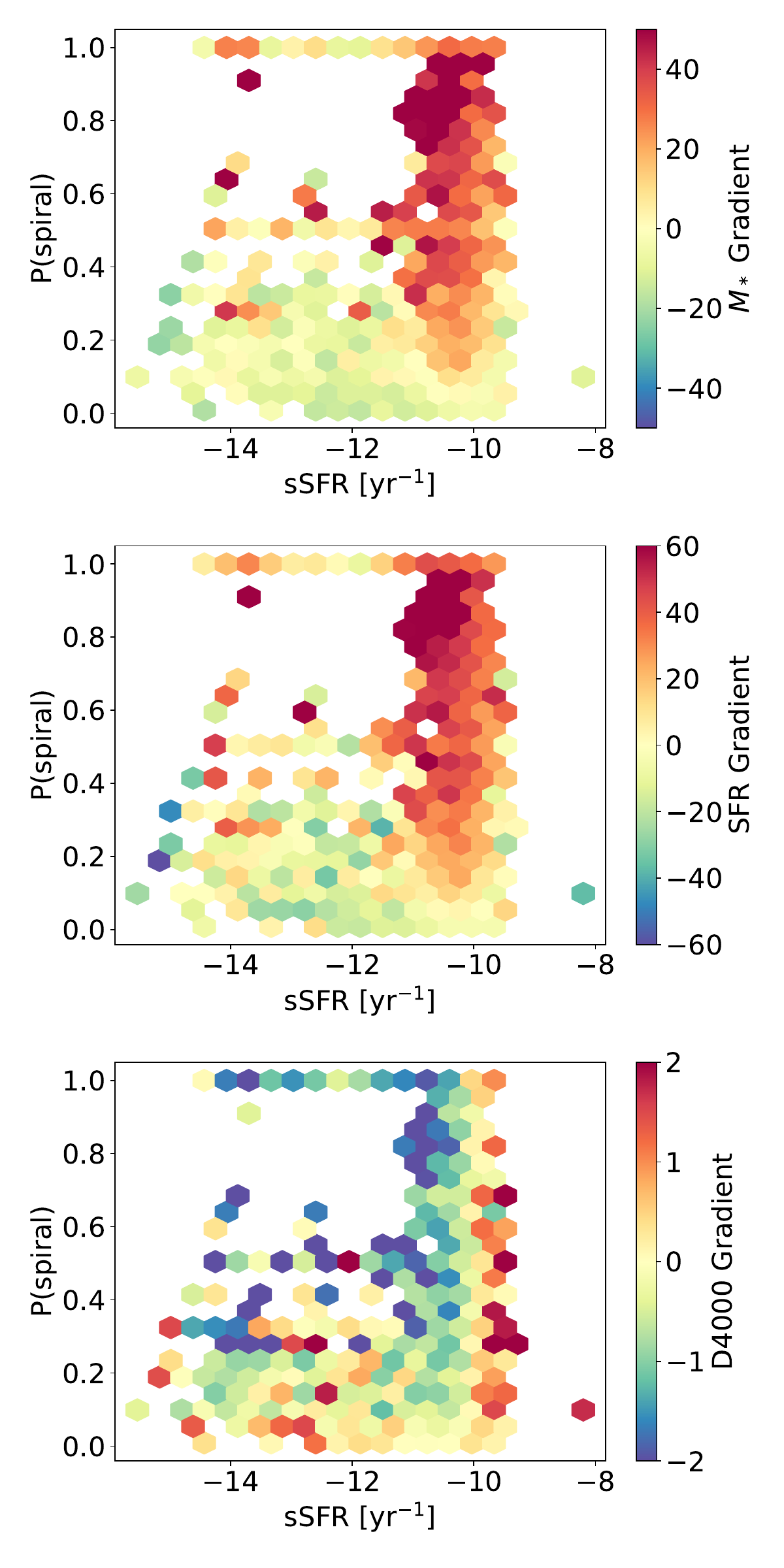}
    \caption{The top plot shows a positive correlation exist between P(spiral) and  $\nabla M_*$ and the bottom plot that a positive correlation exist between P(spiral) and $\nabla$SFR for increasing sSFR. This shows that our method can detect the spiral structures that we expect of SFMS galaxies. The values for sSFR in this figure come from the PIPE3D VAC. The values for P(spiral) come from the Galaxy Zoo VAC.}
    \label{fig:spiral-sfr-mass relation}
\end{figure}

\cref{fig:spiral-sfr-mass relation} shows how galaxies that have been highly voted to contain a spiral in Galaxy Zoo (P(spiral)) are strongly correlated with a positive $\nabla M_*$ and $\nabla$SFR. Moreover, the higher P(spiral) the higher the gradient value tends to be. We further verify this by plotting the galaxies in the sSFR-P(spiral) plane, which allows us to see the trend between spiral arms, star formation, and gradient profiles. In this plot, the galaxies most likely to be spirals are in the upper right, where we have a combination of a high sSFR and high P(spiral) value. \cref{fig:spiral-sfr-mass relation} indicates that for galaxies with clear or large spiral arms the network is focusing on them to make its prediction of both M$_*$ and SFR.

For SFR this trend is exactly what we expect as it is well-known that a large amount of star formation occurs in the spiral arms of spiral-like galaxies. Furthermore, we know that spiral-like galaxies are the morphological type of galaxies that have the most star formation happening (relative to their M$_*$). It is reassuring that the networks also picks up on this morphological feature, and recognize they are important to an accurate prediction of SFR.

The more involved explanation is that of why $\nabla M_*$ follows a similar trend to $\nabla$SFR, because even in spiral galaxies we know most of M$_*$ is concentrated near the center of the galaxy. The reason why the galaxy outskirts are still important morphological features for the network as it predicts M$_*$ might be linked to what was discussed in \cref{sec:mass age divide}. It could be that the network is picking up on important morphological features that lie on the outskirts of galaxies that can help determine on which side of the golden mass divide the galaxy lies. 

\subsubsection{The Age-Morphology Correlation}

\begin{figure*}
    \centering
    \includegraphics[scale=0.7]{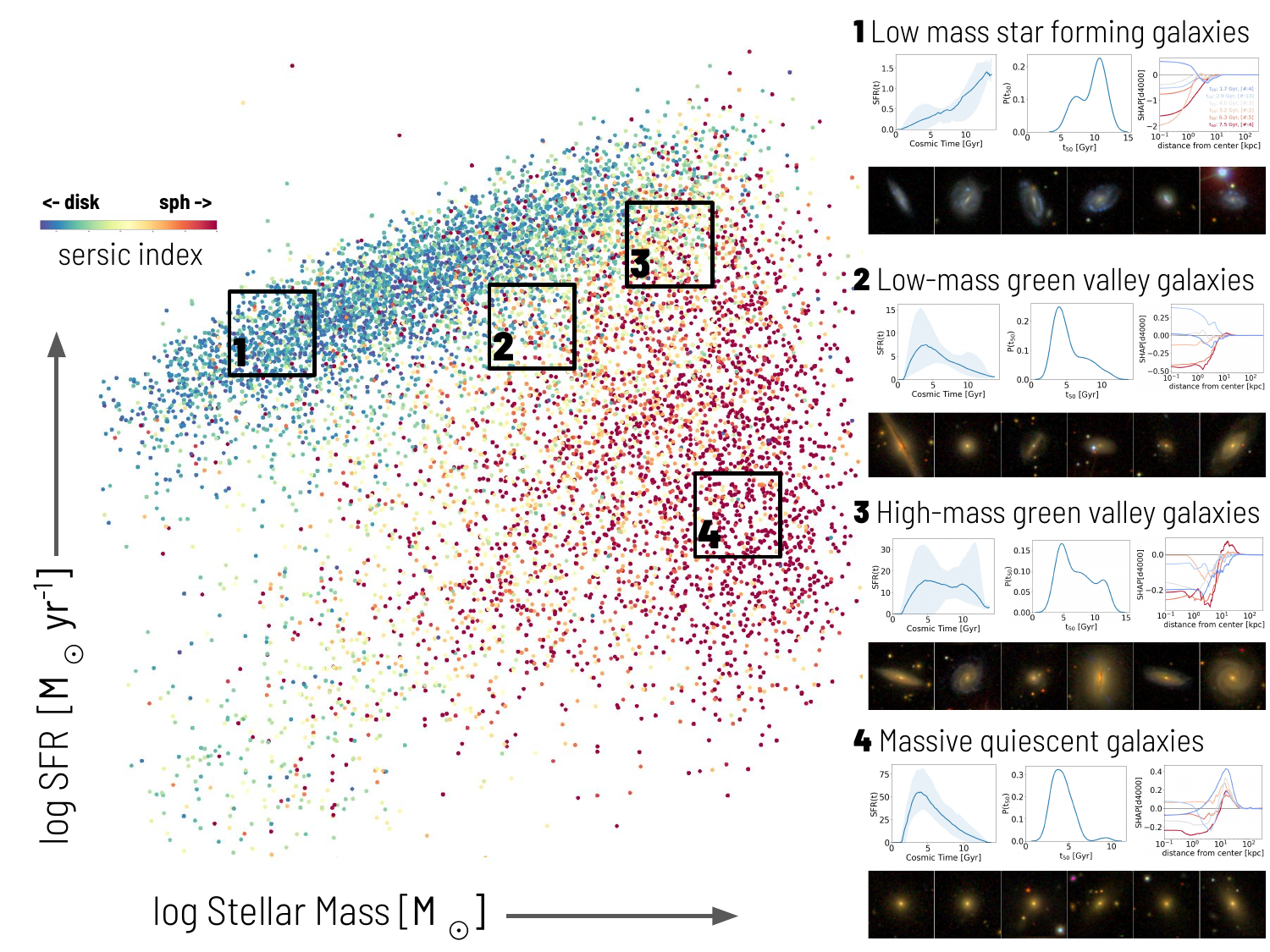}
    \caption{A sample of galaxies from different regions of the M$_*$-SFR plane (shown on the left panel, colored by their sersic index) are chosen. Each region's galaxies have their: RGB images, SFHs, t$_{50}$ values, and radial D4000 SHAP map profiles shown on the corresponding right-hand panel. The leftmost plot shows the SFH of the galaxies in the sub-sample. The blue line on the SFH plot shows the average SFH of all the galaxies chosen in the sub-sample, while the shaded blue region shows the range of all the SFHs in the sub-sample. The center plot shows the distribution of the t$_{50}$ values of the galaxies in the sub-sample (normalized to 1). The rightmost plot shows the radial profiles of the D4000 SHAP maps of the galaxies in the sub-sample binned by their t$_{50}$ value. We see that the D4000 SHAP maps tend to pick out distinct morphological features even within a given bin for galaxies of different ages.
    }
    \label{fig:age_morph}
\end{figure*}

We explore the age-morphology correlation at fixed M$_*$ and SFR in \cref{fig:age_morph}, since a possible concern of our results is that correlations with the SFR and stellar mass are driving the d4000 estimates instead of independent morphological information. To study this, we select samples of galaxies in bins of fixed stellar mass and SFR, and find that (i) there is a sizable spread in t50 values, and (ii) D4000 SHAP maps tend to differ for galaxies of different ages and contain predictive power. A summary of this effect is shown in Figure \ref{fig:age_morph} with a full treatment in \ref{app:age morph}. Figure \ref{fig:age_morph} selects samples of galaxies in different regions of the M$_*$-SFR plane and shows their: RGB images, median SFHs, the distribution of t$_{50}$ values, and median radial D4000 SHAP map profiles in bins of t$_{50}$. The plot shows that the D4000 SHAP maps are sensitive to different morphological features in different bins of t$_{50}$ even at fixed mass and SFR. It also shows how galaxies that are currently star-forming have higher SHAP values near the center when predicting the galaxy's age, meanwhile heavily quenched and older galaxies show higher SHAP values at the outskirts of the galaxy. Across the 4 sub-samples selected we see that generally, galaxies in different bins of t$_{50}$ tend to focus on distinct morphological features to predict the d4000 spectral index. 

The exception to this is seen for current star-forming galaxies. In that bin, we see that all galaxies regardless of t$_{50}$ show similar SHAP radial profiles. This makes sense as we expect very few galaxies in this sample to have old t$_{50}$ values, with similar features being used to determine mass, SFR and D4000. 

For green valley galaxies, we expect a mix of galaxies similar to the `inbetweeners' in Sec. \ref{sec:physical significance of gradients}, which explains why we see a range of behaviors in the SHAP radial profiles for galaxies with different t$_{50}$. Looking at the average SFHs of the galaxies in the sub-sample we see that it even has the shape of a rejuvenated galaxy. This indicates that the galaxies in this sub-sample had events in their recent past that reignited their star formation and likely impacted their morphology. Appendix \ref{app:age morph} shows these trends across the complete sample size of this work. 
 
We know that early-type galaxies are some of the oldest galaxies in the local universe. This could help explain why we also see a divide of gradients for D4000. It could be that the very same physical phenomena that occur at the golden mass give rise to the morphological features of high importance to determining galaxy age. Since the golden mass compaction and quenching event is thought to give rise to early-type galaxies in the local universe, the physical imprints of this event might be a way to infer the galaxy's age from its morphology. This would explain why the network looks at the same morphological features for galaxies with low mass and young age (low D4000) and for galaxies with high mass and old age (high D4000). There is evidence in the literature that bulge growth and disk quenching occur as a galaxy ages, and this might be the morphological imprint that is being picked up from age. \citep{Morph_age1,Morph_age2,Morph_age3,Morph_age4}.

\subsection{Trends with Quiescence and Rejuvenation}

\subsubsection{How do Galaxies Grow Old? Correlations Between Quenching and Morphology} \label{sec: quench and merger}

\begin{figure*}[ht]
    \centering
    \hbox{\hspace{0.5cm} \includegraphics[scale=0.43,trim={0cm 9cm 0cm 9cm},clip]{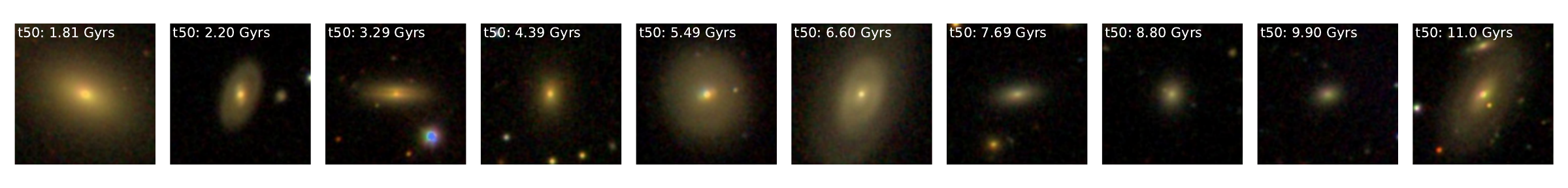}}
    \includegraphics[width=\textwidth]{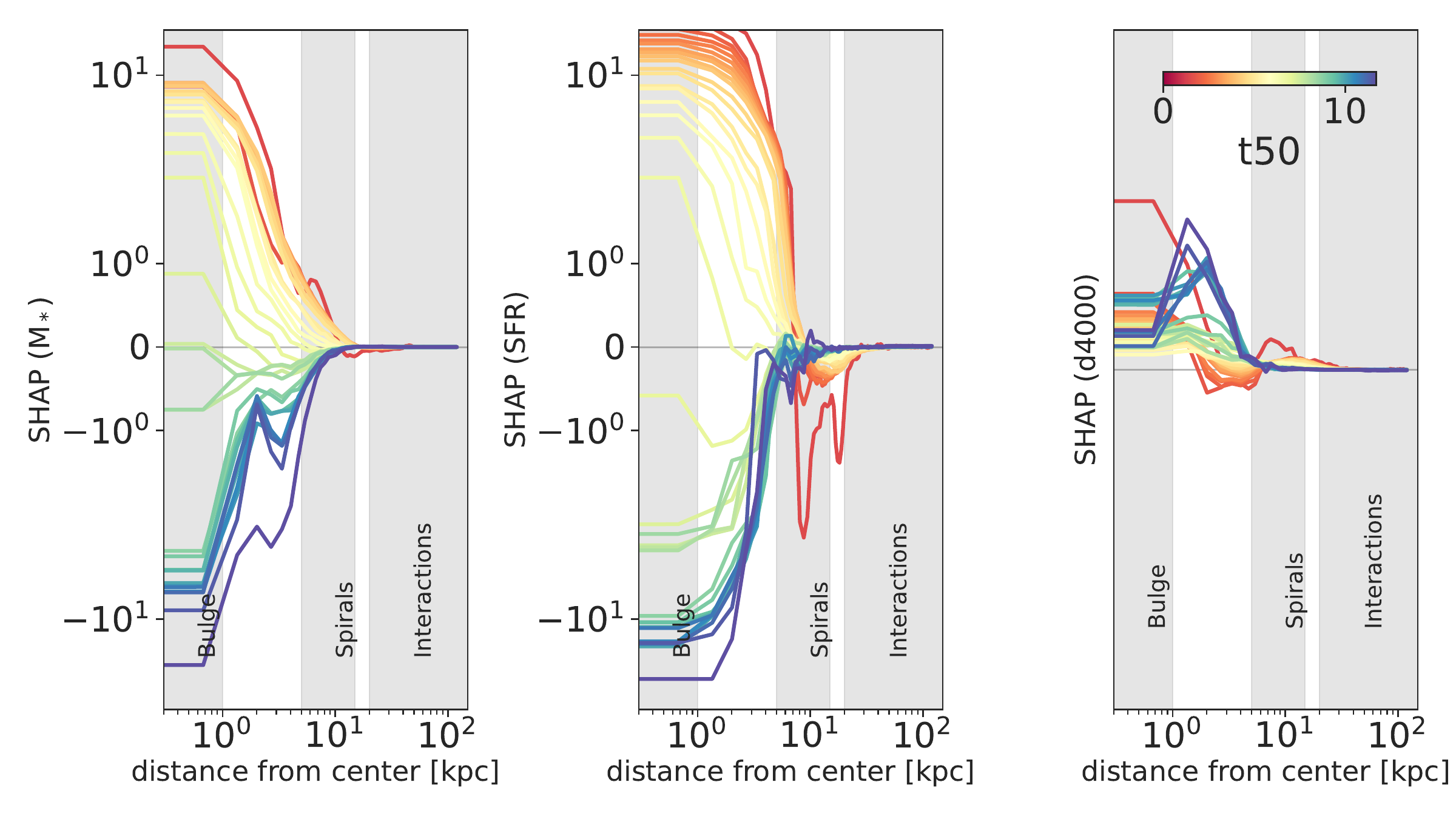}
    \caption{The radial SHAP profile plots for galaxies in the sample (N=5329) which are classified as quiescent (sSFR $<$ -10.8 The galaxies are binned by t$_{50}$ (in cosmic time Gyr). We see that galaxies that quenched earlier have a higher focus in the bulge regions, while galaxies that quenched later have a higher focus in the spiral arm region. These are the types of morphological profiles we expect for these groups of galaxies. The row of galaxies above shows various galaxies from the sample that are quenched (sSFR $<$ -10.8) in increasing t$_{50}$ values from left to right. We see that galaxies that quenched earlier (lower t$_{50}$ values) have pronounced bulges than galaxies that quenched later (higher t$_{50}$ values).}
    \label{fig:quiescent radial plot}
\end{figure*}

The literature shows that as galaxies quench they have their bulges grow and that the bulge grows larger the longer the galaxy has been quenched (\cite{bulge_growth1}, \cite{bulge_growth2}, \cite{bulge_growth3}). In \cref{fig:quiescent radial plot} we examine how our method looks at the morphology of galaxies that are quenched, and how the time since quenching affects the morphology of these galaxies. The plot shows that galaxies with low t$_{50}$ values (quenched earlier on) have a more pronounced bulge structure and that the network considers it to be a significant morphological structure to make its prediction. In a similar vein, for galaxies that have higher t50 values (quenched later) we see that the bulge does not play such an important role in the prediction and instead, the network focuses more on morphological features in the outskirts of the galaxy such as spiral arms. This agrees with the literature as we expect galaxies that quenched earlier to have more pronounced bulges making an important morphological feature to predict the galaxy's M$_*$, SFR, and D4000. 

\begin{figure*}[ht]
    \centering
    \hbox{\hspace{0.5cm} \includegraphics[scale=0.43,trim={0cm 9cm 0cm 9cm},clip]{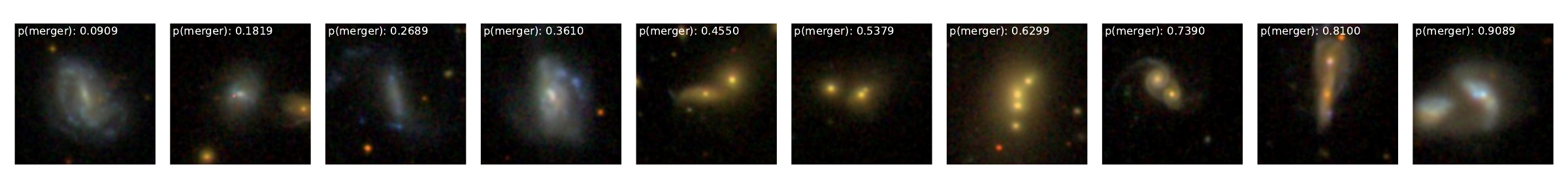}}
    \includegraphics[width=\textwidth]{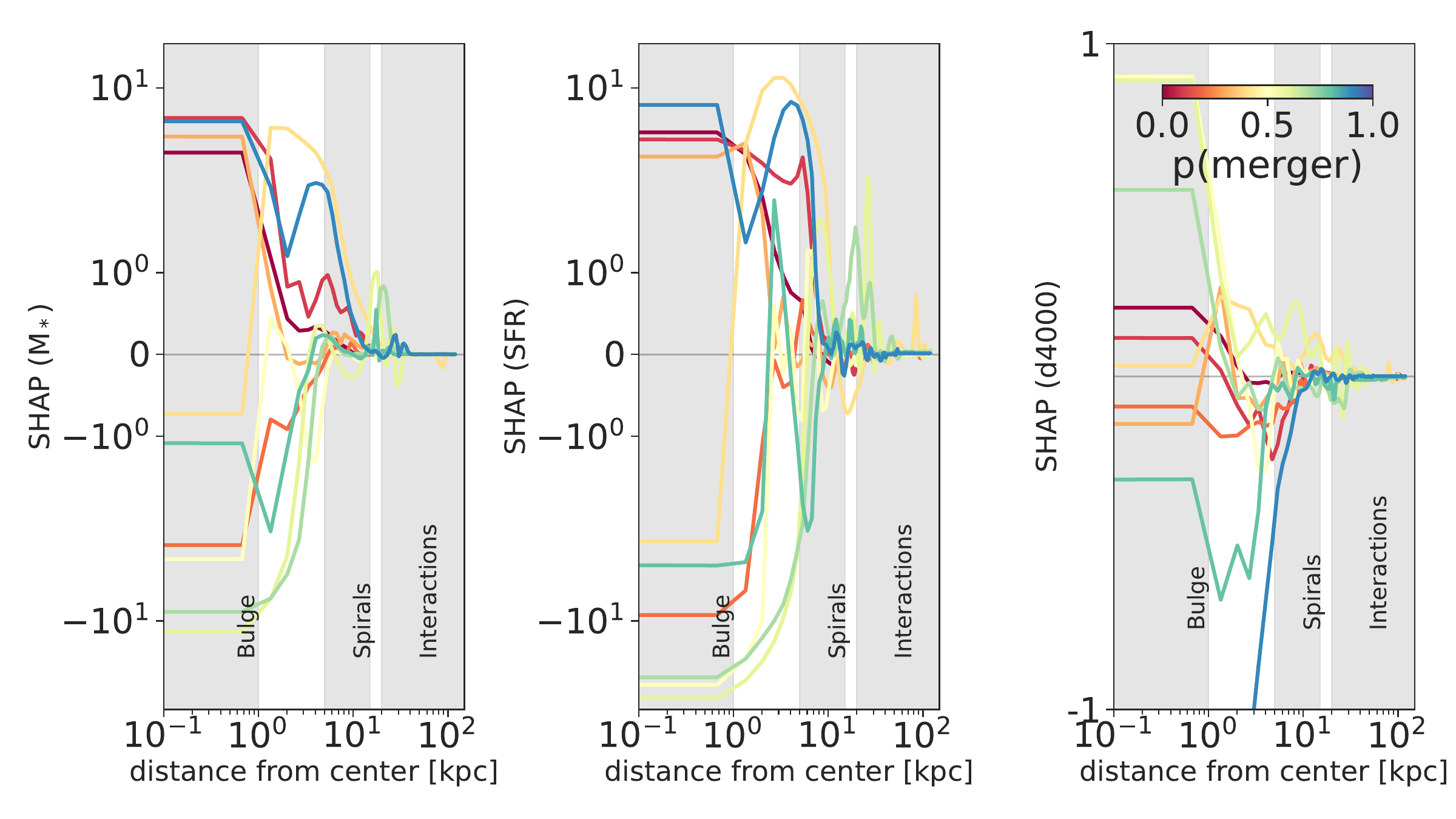}
    \caption{The radial SHAP profile plots for galaxies in the sample have a P(merger) (N=498) value from the PyMorph VAC. We see that galaxies with higher p(merger) values have SHAP profiles that extend out to the interactions sections of the plots. This is due to the network also detecting the interaction between the two galaxies and using this information to inform its prediction. The row of galaxies above the radial profiles shows galaxies from the sample that have a p(merger) value from the PyMorph VAC with increasing p(merger) values from left to right.}
    \label{fig:merger radial plot}
\end{figure*}

We explore galaxies undergoing mergers to see if our method can pick up the secondary object and its morphology to make a more informed decision about the predicted parameters. The radial SHAP profiles for galaxies that have a p(merger) value from the PyMorph VAC (N=498) can be seen in \cref{fig:merger radial plot} where the galaxies are binned by their p(merger) value. For higher values of p(merger), we see that the network picks up on the companion objects as we see a spike in the radial profiles in the interaction zone. The plot shows that galaxies with higher p(merger) values have SHAP profiles that extend out to the interactions sections of the plots. This means that the model is using information from the entire galaxy, including the interacting regions, to make its prediction. This spike is not present in the galaxies with low p(merger) values. Interestingly, there is no spike in the $\nabla$D4000 radial profile, likely because while a merger event would impact a galaxy's M$_*$, SFR it would have a much smaller effect on the age of the stars in the merging pair. If for example the merging pairs are old quiescent galaxies and some young stars are formed as a result they will still be the vast minority among the much older stars both galaxies had before the merger.

\subsubsection{Rejuvenating Galaxies}
\begin{figure*}
    \centering
    \includegraphics[scale=0.7]{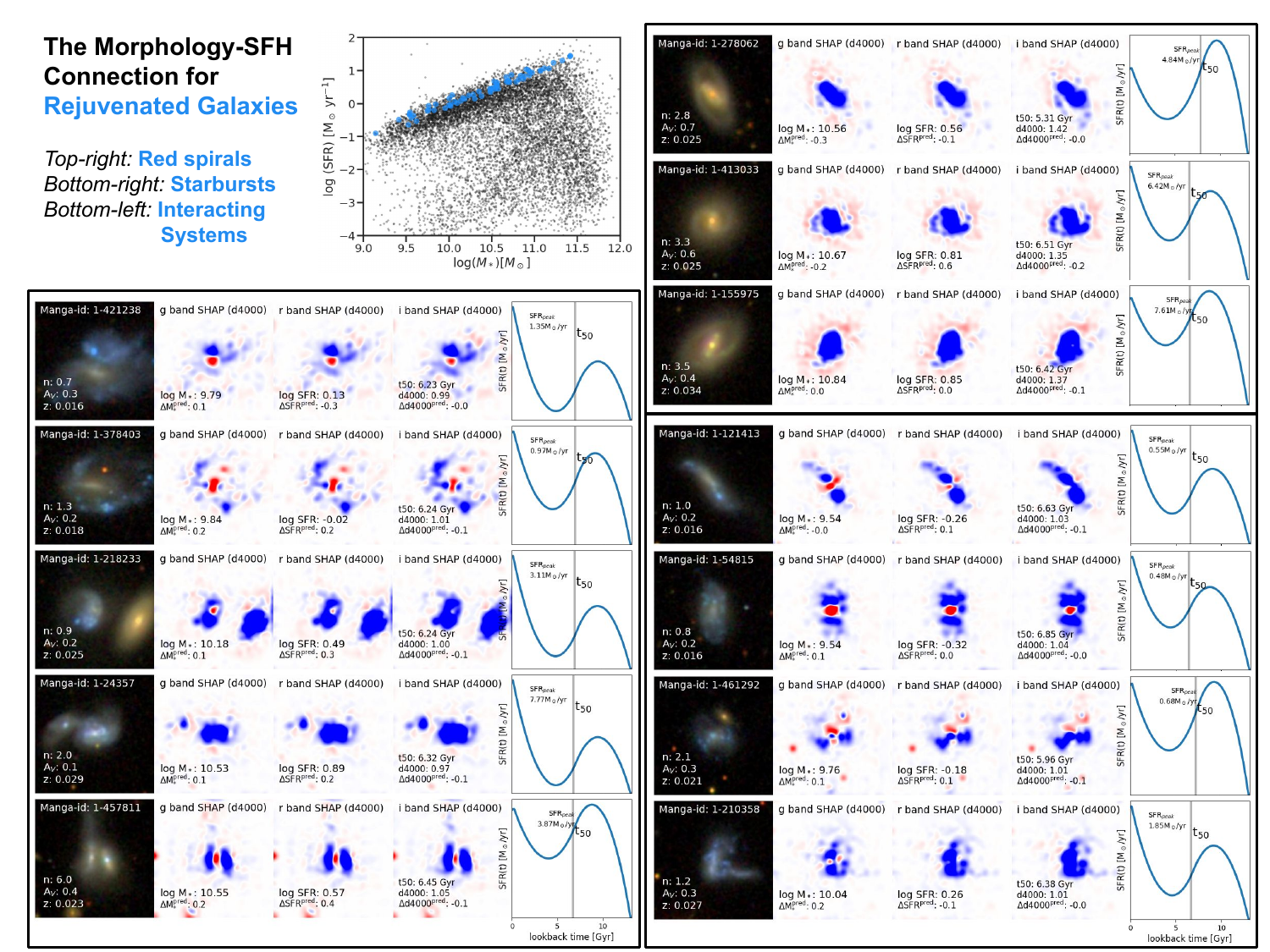}
    \caption{Rejuvenated galaxies from our sample. The top panel shows the location of the rejuvenated galaxies (blue dots), which tend to lie on or above the M$_*$-SFR correlation, with the full sample shown in grey. There are three panels showing 3 distinct populations of rejuvenated galaxies. For each population, the image, D4000 SHAP maps by color, and SFHs are shown for a sub-sample of galaxies. Additionally, the prediction accuracy is overlaid on the SHAP maps. Rejuvenated galaxies of the same population tend to show similar SHAP maps.}
    \label{fig:rejuvination}
\end{figure*}

We explore galaxies that have multiple major episodes of star formation to see if our method can detect morphological signatures of rejuvenation. A sample of rejuvenating galaxies (selected based on their SFHs) can be seen in \cref{fig:rejuvination}, along with their D4000 SHAP maps in each broadband. We find that the rejuvenated galaxies in MaNGA can roughly be grouped into three populations: spirals with bars or other central features that could be triggering central star formation through gas infall, irregular galaxies that are experiencing starbursts, and interacting systems. We find that rejuvenated galaxies of the same population tend to show similar SHAP maps. This leads us to believe that there might be morphological signatures of rejuvenation which the network is picking up on. 

\subsection{Limitations and Caveats of Method}
While we believe the method presented in this paper is a powerful scalable tool to use in future galaxy evolution studies some limitations will need to be addressed. The main limitation is that while the method works well in the local universe we can obtain high SNR images of the galaxies where things like spiral arms or bars can be resolved. This may not be as straightforward when extending the method to studying galaxy morphology at higher redshifts. Considerable work will have to be done to the network architecture and possibly its pipeline to accommodate selection biases and lower SNR data. 

Additionally, galaxy inclination is not corrected for in this paper. In this paper, we mainly seek to develop the Katachi framework and apply it to a sample of galaxies, for which we obtain an extremely data-rich set of morphological predictors. By averaging over color and azimuthal angle we are throwing away some information, however, the power of the framework shows that despite this we recover significant trends that tell us about the morphological imprints left behind by past star formation. This is the main point that we want to highlight in the current paper, and we are working on a follow-up paper that studies these trends in further detail. We believe in the ability of this type of morphological analysis to be able to bring great insight into other data sets. 

A second limitation of the method introduced here is the limitation by the XAI method selected to interpret the network (in this case SHAP values). The field of XAI is constantly and rapidly expanding. What was once thought of as a good interpretability tool is quickly replaced or improved upon by a new method. Working on minimizing the black box aspect of deep learning will aid us in further improving the pipeline presented in this paper. Since this method for studying galaxy morphology is only as good as the XAI method is we will need to keep a constant lookout for new XAI methods that emerge and apply them to our current pipeline. There is a possibility that we are missing out on key insights learned by the network because our interpretability method is simply not good enough. For example, there might be finer morphological features the network is picking up on as features, but SHAP might not be the best way to probe the network to reveal this.

A third issue is the dependence of this analysis on a reliable training set. In the low-z universe, we were able to make use of a well-studied IFU sample to infer `true' values that we used in the training. However, this is still subject to uncertainties in spectral modeling that come from assumptions about stellar population synthesis models, and priors used in modeling dust attenuation, star formation histories, and more. As we develop a more thorough understanding of these effects, it will be important to incorporate them at the training phase.

\section{Conclusion} \label{sec:conclusion}
We presented Katachi \begin{CJK*}{UTF8}{min}(形)\end{CJK*}, a unique, scalable approach to studying galaxy morphology using deep learning that does not rely on discrete classification. This method allows us to move away from traditional methods that fit the light profiles of galaxies or classify them into discrete bins and in doing so misses out on fine details of the galaxy's morphology. We achieve this by training a chain of CNN networks to predict the M$_*$, SFR, and D4000 break strength of galaxies based solely on their RGB images constructed from the SDSS \textit{gri} broad bands. 
The trained network is able to predict the stellar mass (M$_*$) to a RMSE accuracy of 0.22 dex, current star formation rate (SFR) to a RMSE accuracy of 0.31 dex for star forming galaxies, and the half-mass time (t$_{50}$) to a RMSE accuracy of 0.23 dex (in lookback time), comparable to traditional SED fitting methods \citep{2024arXiv240112300W}.
We then interpret what the network is learning to be important morphological features of the galaxy to predict each parameter using SHAP maps. Using these SHAP maps we define a gradient based on the radial profiles of the maps that tells us about the difference in average SHAP value at the center of the galaxy compared to at 20 kpc from its center. The gradients give us a quantitative summary of which morphological areas and features of the galaxy are most tied to the predicted parameters. This allows us to find trends between the current morphology of galaxies and their properties. 

In \cref{sec:color} we briefly touch on how the network gains color information by examining the SHAP maps of galaxies in each of their three color components. We find that the network leverages the color information in a physically driven way to understand the underlying link between color and the predicted parameters. In most cases, it uses color as a first-order approximation of the parameter and then uses morphological structure and features to further fine-tune its prediction. A more thorough examination of this across higher redshifts, where it will become more important, will be examined in future works. 

In \cref{sec:mass age divide} we find evidence that galaxy structure in the local universe is affected by the quenching mechanisms proposed to happen at cosmic noon by \cite{golden-mass-2}. We found a sharp contrast in galaxy morphology for galaxies below and above the golden mass threshold mentioned in that study using our gradient approach. We also find possible evidence that this quenching event might be imprinting a distinct set of morphological features on the galaxy depending on its age. This is seen by our gradient method for which the $\nabla$D4000 behavior is not so ordered or structured as it is for $\nabla M_*$ or $\nabla$SFR. Additionally, we find our method correctly identifies the link we expect of spiral-type galaxies having higher sSFR rates, while more elliptical galaxies have lower sSFR rates. This is a well-understood and studied trend, but we can recreate it using our new method of inferring galaxy morphology directly from the image without the need to classify it into a discreet class. 

Furthermore, in \cref{sec: quench and merger}, our method also being able to track the bulge growth of galaxies which serves both as an indicator of the galaxy's age as well as how long ago it quenched. Additionally, we see that our model can deal with interacting systems and understand the morphology of both objects to fine-tune its prediction. In doing this it can pick up on the merging system as being of morphological importance. 

We extended our new methodology further in \cref{sec: SFH and morphology} to discuss the connection between current galaxy morphology and their SFHs by reconstructing the full SFH of all the galaxies in the sample. Using the full SFH of the galaxies we noted that galaxies with the same morphological profiles clustered together on the M$_*$-SFR plane as we moved through cosmic time, and that there is correlated morphology-t$_{50}$ information even at a fixed stellar mass and SFR. This verifies the idea that galaxies that share the same current-day morphology also share very similar SFHs.  

We acknowledge the limitations and caveats that come with this new methodology which centers around having to tailor the network to different surveys, and the limitations that are imposed by the use of XAI which is still in its infancy. We are optimistic however that further use and improvement of this method will lead to more rich and meaningful explorations of galaxy morphology in the future. This will be especially important with the new generations of telescopes such as Euclid, Rubin, and Roman which will give us unprecedented amounts of high-quality galaxy images and data. 

Future works will make use of this method with JWST data to better understand the morphology and SFH relation across various redshift ranges. Additionally, the method can be used as a public framework to predict galaxy properties on the vast amount of imagining data that will be coming out of Euclid, Rubin, and Roman. Euclid has a resolution of 0.1 arcseconds in the visible spectrum and observes galaxies at $0.2<z<2.2$. Rubin has a resolution of 0.7 arcseconds in the visible spectrum and will observe galaxies at $0.3<z<1$. Roman will have a 1.1 arcsecond resolution and is expected to observe galaxies at $2<z<5$. 

All these future telescopes are expected to generate a plethora of images comparable to those from SDSS-IV MaNGA over a wide range of redshifts. We believe that our method will be expandable to these other telescopes and their surveys given how flexible the use of CNNs and SHAP are. The CNNs will work with images of any resolution, and we expect cutouts from these surveys to contain comparable information to the \textit{gri} images from SDSS-IV MaNGA. 

\section*{Acknowledgements}

We would like to thank the anonymous referee for their helpful comments, and Rachel Somerville, Mike Walmsley, Karen Masters, Ariyeh Maller and Mike Blanton for useful discussions. Support for J.P.A was provided by the Japan Society for the Promotion of Science (JSPS) Core-to-Core Program (grant number: JPJSCCA20210003), the GPPU program at Tohoku University, and by the Japanese Ministry of Education, Culture, Sports, Science and Technology (MEXT) scholarship. Support for K.G.I, was provided by NASA through the NASA Hubble Fellowship grant HST-HF2-51508 awarded by the Space Telescope Science Institute, which is operated by the Association of Universities for Research in Astronomy, Inc., for NASA, under contract NAS5-26555. J.P.A and K.G.I would like to thank the SURP program at the University of Toronto for providing the basis for the start of this project 3 years ago. S.C is supported by the JSPS under Grant No. 21J23611. J.P.A would like to thank his parents, sister and lovely partner for their continuous encouragement, and thank K.G.I and M.A for believing in his research potential. Para ti, Santi.\\   

\textbf{Code Acknowledgements:}
dense basis \citep{dense_basis,dense_basis_2}
Marvin \citep{marvin}
matplotlib \citep{matplotlib}
Numpy \citep{numpy}
PyTorch \citep{Pytorch}
Astropy \citep{astropy1,astropy2,astropy3}
SciPy \citep{scipy}
hickle \citep{hickle}
FSPS \citep{FSPS}
shap \citep{SHAP}

\appendix 
\crefalias{section}{appendix}

\section{Chain Network Comparison} \label{app:chain network appendix}
\begin{figure}[h!]
    \centering
    \includegraphics[scale=0.7]{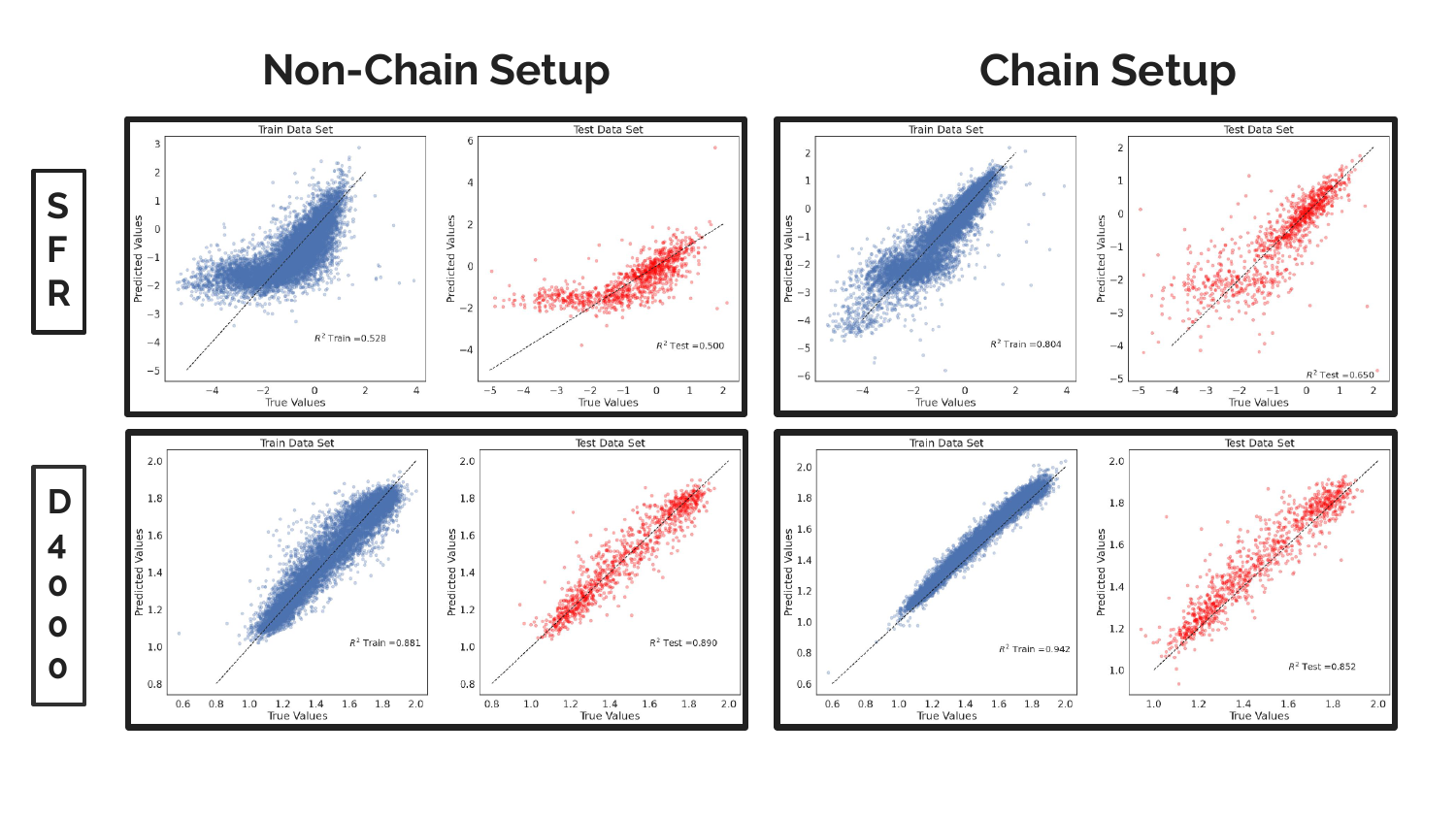}
    \caption{A compression of the accuracy of the ResNet50 architecture networks vs the chain architecture networks. The scatter plots show the true values of the parameters on the x-axis and the values predicted by the networks on the y-axis. A 1 to 1 line is drawn as a reference of $100\%$ accuracy. The plots show the results for the training and test data sets separately. Note that the \say{tail} for low SFR predictions for the ResNet50 network is fixed when using the chain architecture.}
    \label{fig:solo vs chain}
\end{figure}

The top row of \cref{fig:solo vs chain} shows the accuracy of the network for SFR predictions, while the bottom row shows the accuracy of the network for D4000 break strength predictions. The left column shows the accuracy for both parameters when the network is outside of the chain architecture (i.e has no additional information as input except the RBG image), while the right-hand column shows the accuracy for both parameters when the networks are put into the chain architecture as shown in \cref{fig:chain arch}. Of special note for the SFR predictions is how being in the chain architecture (i.e. having M$_*$ as well as the RGB image as an input) allows it to better predict the SFR of quenched galaxies.

\section{t$_{50}$ Predictions Compared to PIPE3D} \label{app:t50 appendix}
\begin{figure}[h!]
    \centering
    \includegraphics[scale=0.45,trim={2cm 0cm 2cm 0cm}]{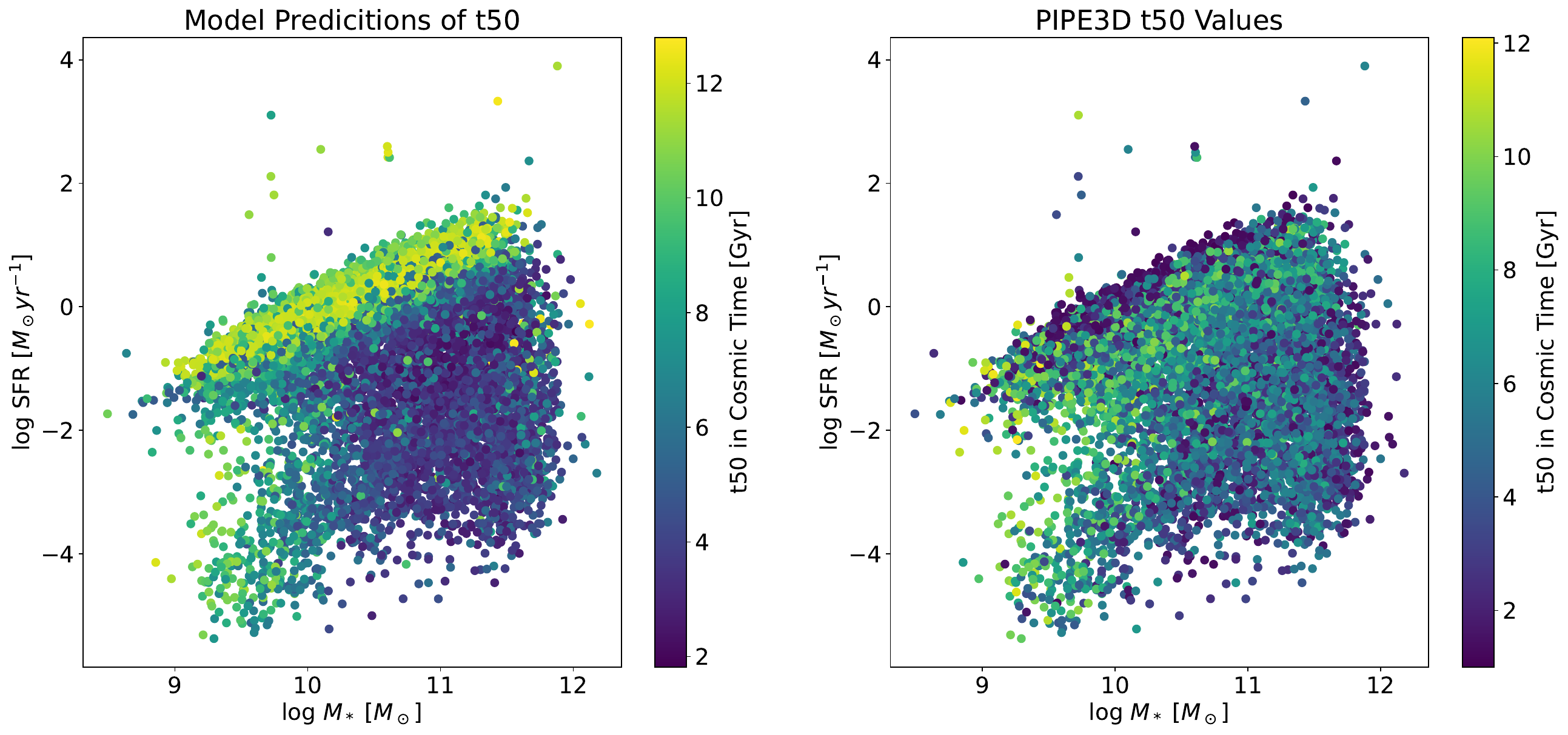}
    \caption{A comparison of our predicted t$_{50}$ values to the t$_{50}$ values found in the PIPE3D VAC.}
    \label{fig:t50 model vs PIPE3D}
\end{figure}

\cref{fig:t50 model vs PIPE3D} compares the t$_{50}$ values we derived using the methods outlined in \cref{sec:t50 model}, and compare it to those found in the PIPE3D VAC. Our method gives us t$_{50}$ values that give us a more logical or expected trend as we see the galaxies on the SFMS have the youngest ages, while the quiescent bulge under the SFMS has the oldest ages, This is in line with what we expect from the literature.  

\vspace{10cm}

\section{Randomly Selected SHAP Maps} \label{app:SHAP map appendix}
\begin{figure}[h!]
    \centering
    \hbox{\hspace{0.6cm}\includegraphics[width=\textwidth, trim={9cm, 0 , 9cm, 0}]{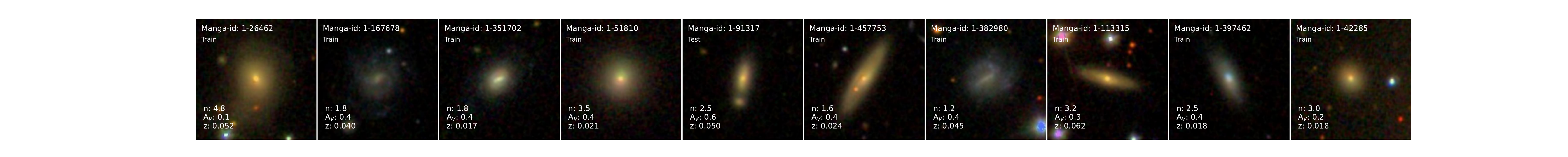}}
    \vspace{1.5cm}
    \hbox{\hspace{0.5cm} \includegraphics[width=\textwidth, trim={9cm, 0 , 9cm, 0}]{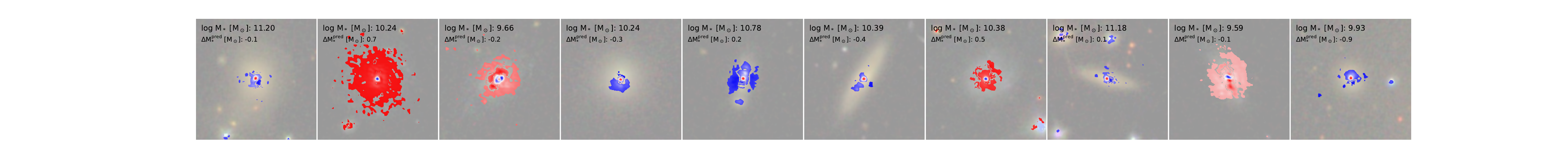}}
    \vspace{1.5cm}
    \hbox{\hspace{0.5cm} \includegraphics[width=\textwidth, trim={9cm, 0 , 9cm, 0}]{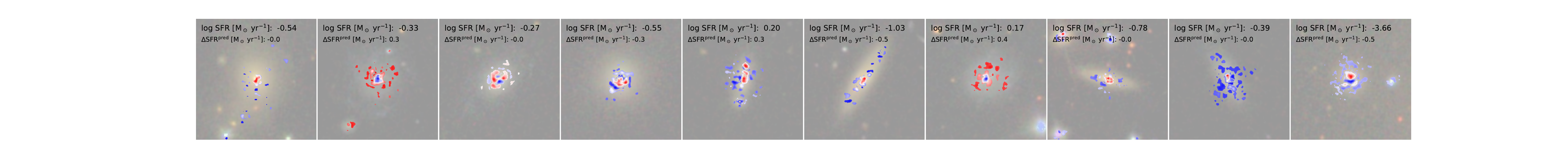}}
    \vspace{1.5cm}
    \hbox{\hspace{0.5cm} \includegraphics[width=\textwidth, trim={9cm, 0 , 9cm, 0}]{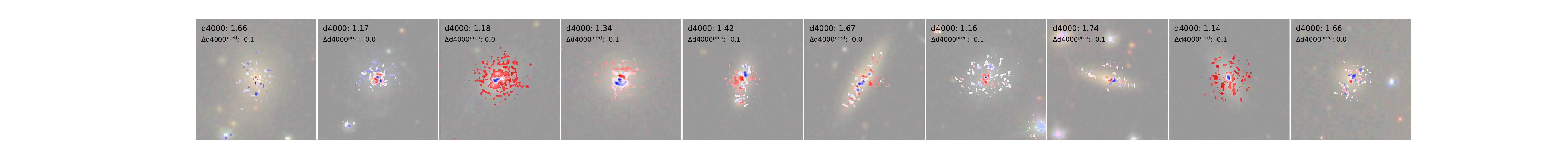}}
    \caption{SHAP maps and images for a random sample of 10 galaxies. The first row shows the RGB images of the galaxies that the chain networks take as input. The second row, third row, and fourth row show the SHAP map generated by the M$_*$, SFR, and D4000 prediction network along with the true and predicted M$_*$, SFR, and D4000 values respectively. The red pixel indicates a positive SHAP value and the blue pixels indicate a negative SHAP value.}
    \label{fig:shap map random}
\end{figure}

\cref{fig:shap map random} shows the RGB images and 3 sets of SHAP maps (one for each predicted parameter) for 10 randomly selected galaxies from the sample. 

\vspace{1cm}

\section{Full Morphology-SFH Plots} \label{app:full sfh morph plots}
\begin{figure}[h!]
    \centering
    \includegraphics[scale=0.07]{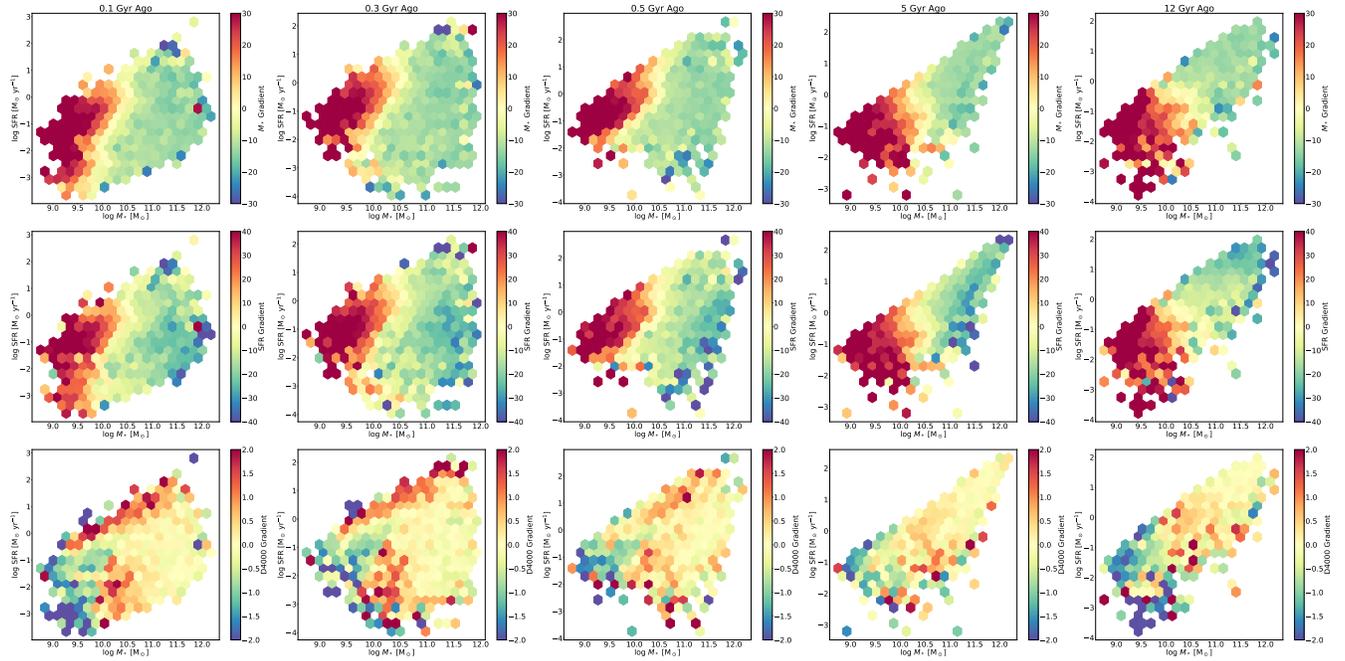}
    \caption{Galaxies in the sample plotted on the M$_*$-SFR plane at different cosmic times. Each column represents a different cosmic time and the rows represent $\nabla M_*$, $\nabla$SFR, and $\nabla$D4000 respectively.}
    \label{fig:morph-sfh-full}
\end{figure}

\cref{fig:morph-sfh-full} shows the same motion of the galaxies across the M$_*$-SFR plane backwards in time. The galaxies are colored by their respective gradient. 

\section{Full Age-Morphology Relationship Plot} \label{app:age morph}
\begin{figure*}[h!]
    \centering
    \includegraphics[width=\textwidth]{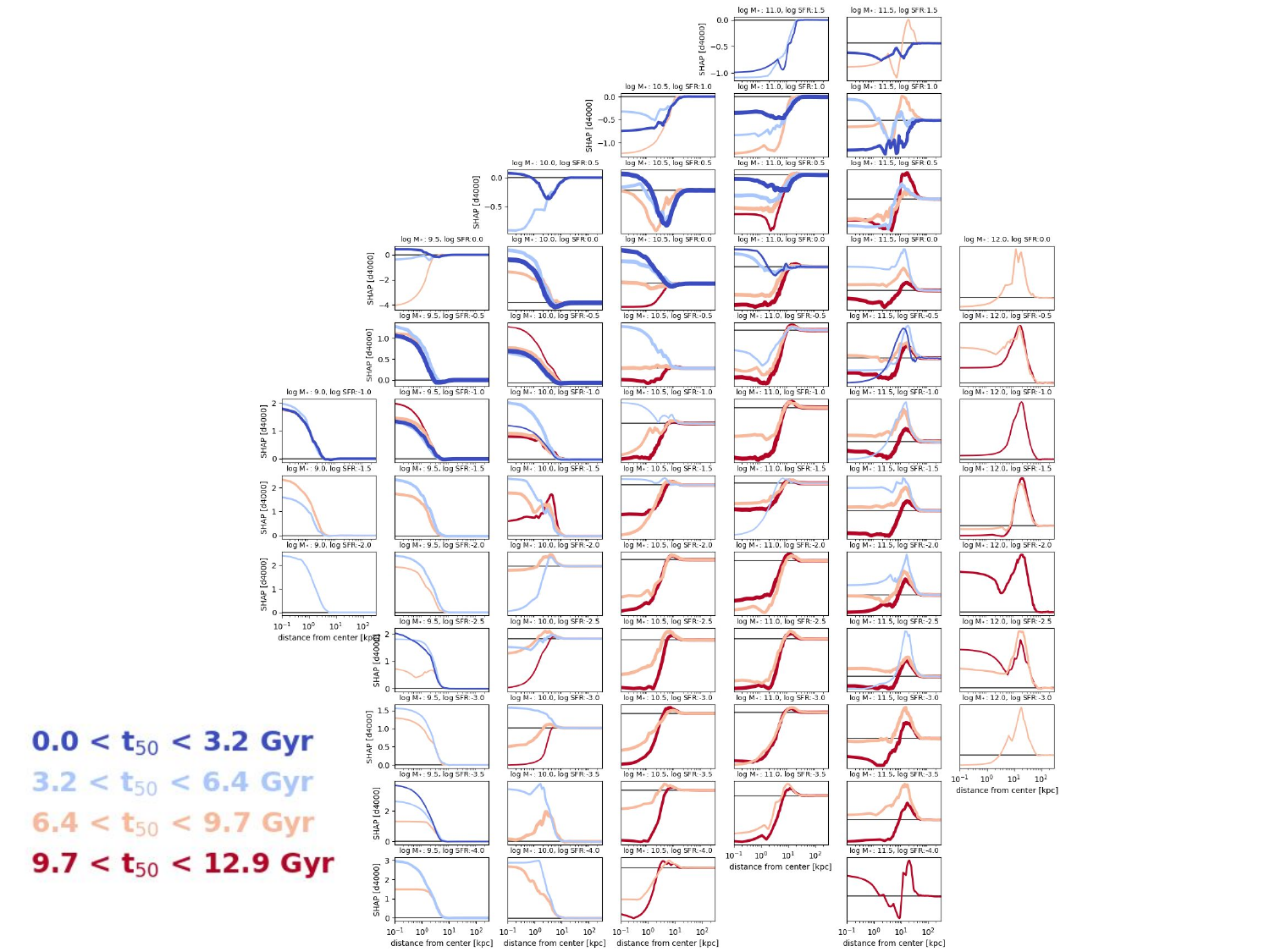}
    \caption{D4000 SHAP Map radial profiles for all the galaxies in the sample, organized by M$_*$ (columns) and SFR (rows). The radial profiles are binned by the t$_{50}$ values of the galaxies.}
    \label{fig:age-morph-full}
\end{figure*}

\cref{fig:age-morph-full} shows the relation of age and morphology across the full sample used in this study.

\section{Dust Attenuation and Metallicity Effects} \label{app:dust metal}

\begin{figure}[h!]
    \centering
    \hbox{\hspace{0.5cm} \includegraphics[scale=0.43,trim={0cm 9cm 0cm 9cm},clip]{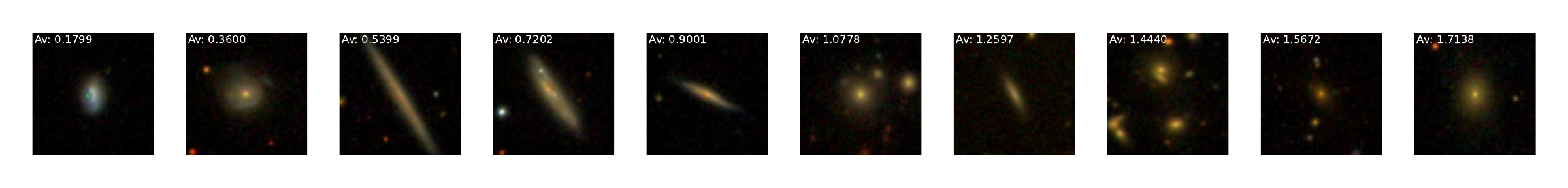}}
    \includegraphics[width=\textwidth]{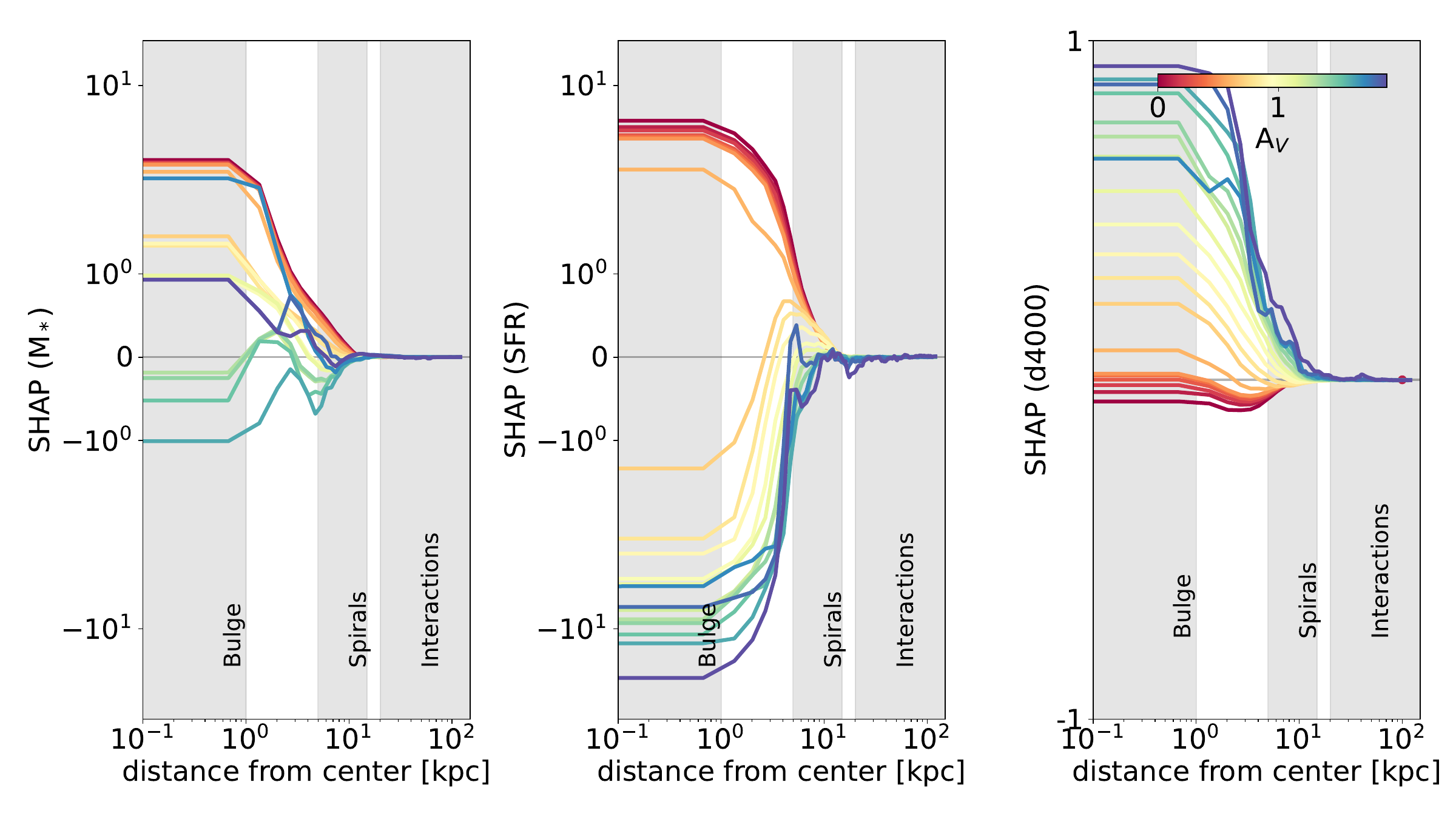}
    \caption{The SHAP radial profile of the galaxies in the sample binned by their Av values. We see more dusty galaxies have a positive correlation with outskirt properties. This is expected as the most dusty galaxies will be star-forming disks and spirals. We see the opposite trend among the least dusty galaxies, where these galaxies have positively correlated morphological features in the bulge of the galaxy. This is also expected as these galaxies are likely to be older quiescent galaxies with pronounced bulges.}
    \label{fig:av radial plot}
\end{figure}

\begin{figure}[h!]
    \centering
    \hbox{\hspace{0.5cm} \includegraphics[scale=0.43,trim={0cm 9cm 0cm 9cm},clip]{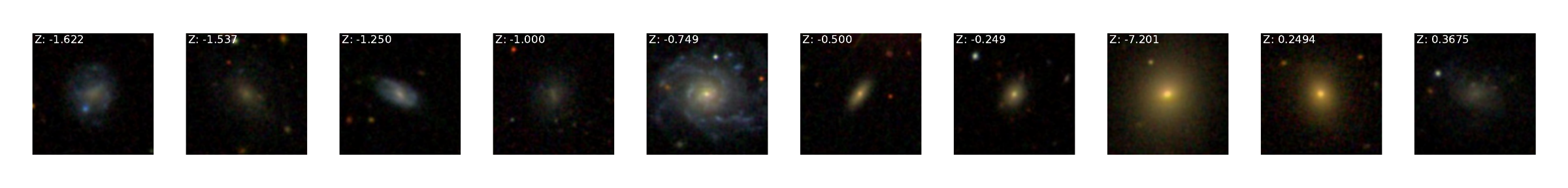}}
    \includegraphics[width=\textwidth]{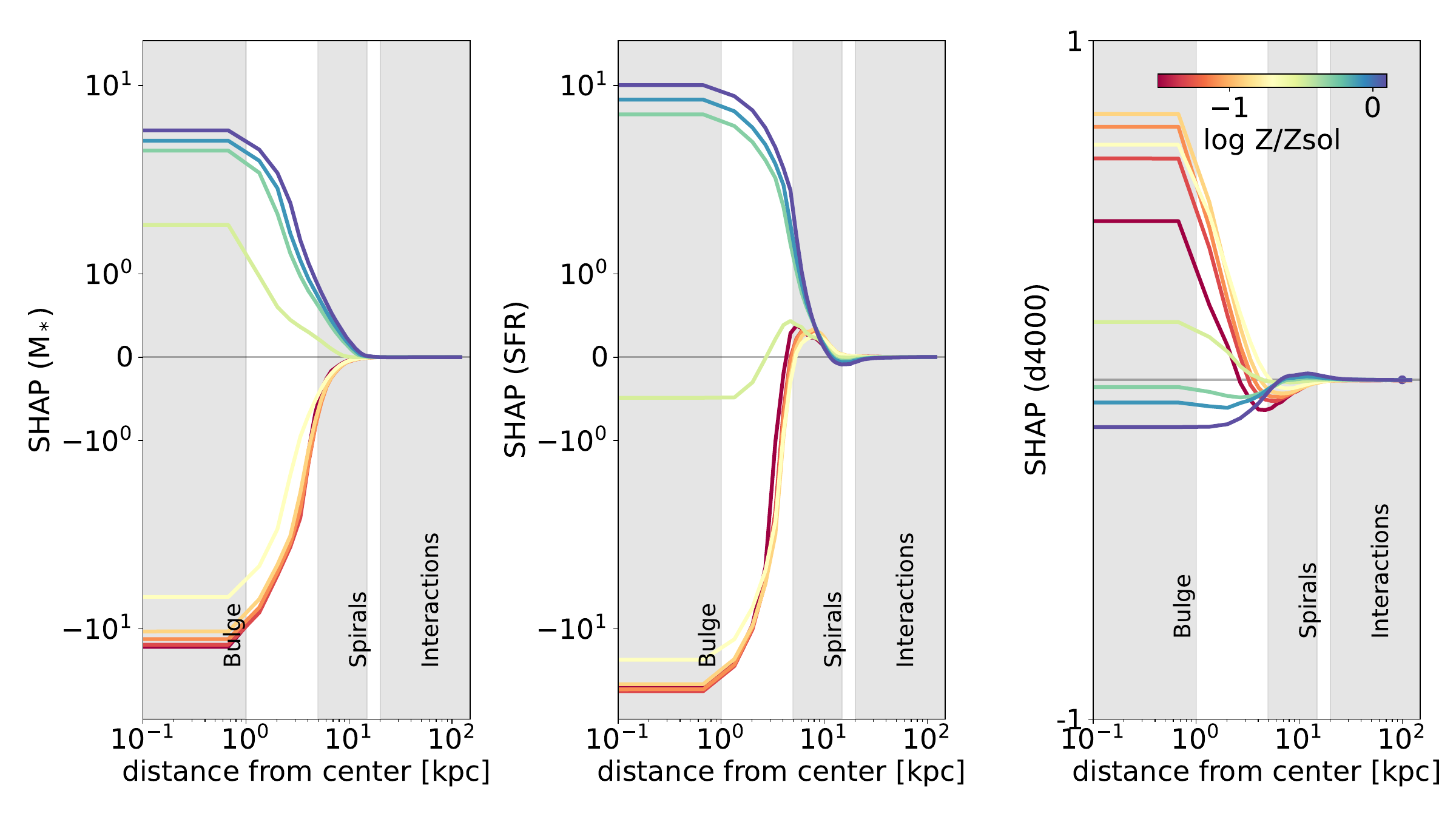}
    \caption{The SHAP radial profile of the galaxies in the sample binned by their metallicity. We see more metal-rich galaxies have a positive correlation with bulge properties. This is expected as the most metal-rich galaxies tend to be massive quiescent galaxies with a significant bulge. We see the opposite trend among metal-poor galaxies, where these galaxies have positively correlated morphological features in the outskirts of the galaxy. This is also expected as these galaxies are likely to be younger and star-forming disks and spirals still undergoing enrichment.}
    \label{fig:z radial plot}
\end{figure}

\cref{fig:av radial plot} shows SHAP radial profiles for the galaxies in the sample, binned by their dust attenuation (Av) values. Similarly, \cref{fig:z radial plot} shows the same radial profiles, but binned by the galaxies' metallicity value (Z). 

\vspace{100cm}

\bibliography{katachi}{}
\bibliographystyle{aasjournal}



\end{document}